\documentclass[aps,pra,superscriptaddress,twocolumn,tightenlines,showpacs,floatfix,amsmath,amssymb]{revtex4-1}

\usepackage{physics}
\usepackage{mathtools}
\usepackage{graphicx}
\usepackage{dcolumn}
\usepackage{bm}
\usepackage{bbold}
\usepackage{xcolor}
\usepackage{hyperref}

\begin{document}

\title{Colored noise driven unitarity violation causing dynamical quantum state reduction}

\author{Aritro Mukherjee}
\author{Jasper van Wezel}
\affiliation{Institute for Theoretical Physics Amsterdam,
University of Amsterdam, Science Park 904, 1098 XH Amsterdam, The Netherlands}

\date{\today}

\begin{abstract}
Spontaneous unitarity violations were recently proposed as a cause of objective quantum state reduction. This complements proposals based on stochastic modifications of Schrodinger's equation, but also differs from them in several aspects. Here, we formalise the description of models for spontaneous unitarity violation, and show that they generically imply models of dynamical quantum state reduction driven by colored noise. We present a formalism for exploring such models as well as a prescription for enforcing explicit norm-preservation, and we show that the resulting pure state dynamics is described by a modified von-Neumann Liouville equation which in a particular limit reduces to the Gorini-Kossakowski-Sudarshan-Lindblad master equations. We additionally show adherence to Born's rule emerging in the same limit from a physical constraint relating fluctuating and dissipating components of the model.
\end{abstract}

\maketitle

\section{Introduction\label{sec:0}}
Despite the fact that quantum physics accurately predicts the dynamics of microscopic objects, reconciling its unitary time evolution with the observation of single measurement outcomes in macroscopic settings remains one of the central foundational problems of modern science~\cite{Bassi_03_PhyRep, leggett2005quantum, overview, arndt2014testing, carlesso2022present,Komar62,Wigner95,Wezel10,aritro1}. Approaches to solving this quantum measurement problem can be roughly divided into three categories, based on decoherence, interpretation, or modifications of Schr\"odinger's equation respectively. Decoherence-based approaches necessarily apply to ensembles of measurements, averaged over all possible states of an environment~\cite{Zurek_1982, Schlosshauer_2005,Zurek2009,theo13}, and therefore do not resolve the problem of finding a single measurement outcome in a single experiment~\cite{aritro1,Dieks_1989,Adler_2003,Stillfried_2008,Fortin2014}. Interpretation-based approaches assume Schr\"odinger's equation is universally applicable, and predict its persistence even to human scales~\cite{Everett_1957,Bohm52A,Bohm52B, rovelli1996relational,Fuchs_2014}. Objective collapse theories finally, introduce small modifications to Schr\"odinger's time evolution that have no noticeable effect on the microscopic scale of elementary particles, but which dominate the dynamics in the macroscopic regime~\cite{BohmBub_66_RevModPhys, Pearle_76, Pearle_89_PRA, Gisin84, Ghirardi_1986, Diosi_87_PLA,Ghirardi_90_PRA,percival95, Penrose_96,Wezel10,diosi_1989}. Since these theories introduce actual changes to the laws of quantum dynamics, they provide experimentally testable predictions for dynamics at the mesoscopic level, which is a field of active and ongoing investigation~\cite{overview, Bouwmeester_2003, underground, vinante_mezzena_falferi_carlesso_bassi_2017, carlesso_bassi_falferi_vinante_2016,Adler_2007_experiment}.

One particular objective collapse theory that was recently introduced focuses on the fact that quantum measurement implies both a broken spatial symmetry in the measurement machine, and a breakdown of the reversibility of time evolution~\cite{Wezel10, Mertens_PRA_21, Mertens22, WezelBerry, aritro1}. The similarity of this phenomenology with the spontaneous breakdown of symmetries during phase transitions prompted a model for Spontaneous Unitarity Violation (SUV), in which time-inversion symmetry breaks down during quantum measurement in a manner directly analogous to that of spontaneous symmetry breaking~\footnote{Notice the difference between the time-inversion symmetry discussed here, and the time-reversal symmetry that is spontaneously broken in for example a magnet. The magnetized equilibrium configuration is static and evolves the same way under time evolution forwards and backwards in time. That is, its dynamics still has time inversion symmetry.}. 

The origin of SUV in spontaneous symmetry breaking yields two distinguishing features as compared to other objective collapse models. Firstly, the preferred basis of quantum measurement in models of SUV is determined by the broken symmetries of the measurement device. Only systems that break a spatial or internal symmetry such as translations or (spin or phase) rotations possess the type of low-energy spectrum that makes them susceptible to arbitrarily small unitarity-breaking perturbations~\cite{Wezel_2008}. Only these systems can undergo spontaneous unitarity violations, and they will always be reduced to the symmetry-broken or ordered states that we customarily identify with thermodynamically stable classical objects~\cite{Wezel_SSBlecturenotes}. Moreover, because the end-states of quantum state reduction are symmetry-broken, ordered states, they are stable under environmental decoherence and are thus also `pointer states' in the general sense~\cite{Zurek_1981}.

Secondly, neither spontaneous unitarity violations nor spontaneous symmetry breaking are spontaneous in the sense of occurring without cause. In both cases, the process is the result of a small but non-zero physical perturbation breaking the underlying symmetries of time or space, respectively. They are spontaneous in the sense that for large, macroscopic systems the perturbation required to break the symmetry is so small that it cannot be measured or controlled even with astronomically large resources~\cite{Wezel_SSBlecturenotes}. Macroscopic systems such as measurement machines, therefore, cannot avoid the effect of even vanishingly small non-unitary perturbations and their behaviour differs qualitatively from that of microscopic particles (see Appendices~\ref{app:SSB},\ref{app:Quantum Strong Measurement} for more details). As the perturbation breaks the unitarity of time evolution, it cannot be represented by a quantum observable. Models for SUV can thus be seen as effective descriptions which alter Schr\"odinger's equation by taking into account the lowest order effect of physics beyond quantum mechanics~\cite{Wezel10}, which dominate the dynamics at the quantum-classical crossover. Although the origin of the perturbation is unknown, it must be physical and thus described by a dynamically evolving state or field configuration of its own. Here, we model the perturbation with a stochastic process, which must necessarily be colored (correlated in time) to result from a physical process. SUV differs in this respect from well-known objective collapse models based on uncorrelated (white) noise~\cite{Bassi_03_PhyRep,Bassi2013Review, Pearle_89_PRA, Gisin84,Diosi_87_PLA,Ghirardi_90_PRA,percival95,diosi_1989}. At the same time, it also stands apart from generalizations of these models to the regime of correlated stochastic processes~\cite{Adler2007_col0,Adler_2008_col1,Bassi_2002_col4,Pearle_1996_col6,Ferialdi_2012_col7}, because of the explicit separation between non-linear and stochastic contributions to the SUV time evolution~\cite{Mertens22}.

These two distinguishing features of SUV are the starting point for constructing a general Markovian framework for its quantum state reduction dynamics in this article. In Sec.~\ref{sec2:DQSR}, we briefly review the quantum measurement problem and motivate why models of SUV form a class of modifications to Sch\"oringer's dynamics yielding quantum state reduction. In Sec.~\ref{sec:3}, we provide a general Markovian framework augmenting the dynamics of the quantum state with a coupled stochastic field. Specifically, in Sec.~\ref{sec:3B} we provide a prescription for obtaining norm-preserving quantum state dynamics in any SUV model, and in Sec.~\ref{sec:3C} we formulate the corresponding modified von-Neumann-Liouville equations for pure states.
In Sec.~\ref{sec:3A} we introduce two examples of analytically tractable models of colored noise and we discuss their mathematical properties. In Sec.~\ref{sec:4} we study SUV dynamics in the specific case of a two-state superposition, comparing it with white noise driven objective collapse models.
Focusing on the limit of small correlation time in Sec.~\ref{sec:4A}, we derive effective white-noise driven equations for quantum state reduction applicable to single shot measurements. This is achieved analytically using a temporal multi-scale noise homogenization analysis, starting from the Kolmogorov backward equations. In Sec.~\ref{sec:4B} we obtain a linear quantum (Markov) semi-group description with an explicit Gorini-Kosakowsky-Sudarshan-Lindblad (GKSL) master equation for the state dynamics after averaging over an ensemble of measurements. The emergence of Born's rule in this limit is shown to be a consequence of a relation between the fluctuating and dissipating contributions of the theory. We summarize in Sec.~\ref{sec:5} and provide a general discussion of the similarities and differences between SUV models and alternative objective collapse theories driven by white noise.

\section{Dynamical Quantum state Reduction via Spontaneous Violations of Unitarity}\label{sec2:DQSR}
In this section, we briefly review how theories of SUV address the quantum measurement problem. In the von-Neumann measurement scheme, the process of the state $\ket{\psi_S}$ of a microscopic quantum system being measured by a macroscopic apparatus described by the state $\ket{\psi_A}$ is separated into two stages (see Appendix~\ref{app:Quantum Strong Measurement} for details)~\cite{Von_Neumann2018-bo}. First, the initial product state of the system and apparatus $\ket{\psi_S}\ket{\psi_A}=\left(\sum_i\alpha_i\ket{S_i}\right)\ket{\psi_A}$ evolves into the macroscopic superposition $\sum_i\alpha_i\ket{S_i}\ket{A_i}$, with $\ket{A_i}$ denoting well-separated pointer states of the measurement device indicative of the system being in state $\ket{S_i}$. This first step does not necessitate any interpretation or physics beyond the standard quantum mechanical interaction between system and apparatus~\cite{Von_Neumann2018-bo}. The second, separate step of the measurement process reduces the macroscopic, entangled superposition state to just one of its components, with the probability of any particular component being selected corresponding to Born's rule. This second step requires either interpretations or proposals modifying quantum mechanical time evolution, since the probabilistic, non-linear, and time-inversion breaking nature of the second step all conflict with the unitary dynamics generated by Schr\"odiner's equation~\cite{Bassi2013Review,Wezel10,Mertens_PRA_21}.

Objective collapse theories attempt to solve the dichotomy between microscopic quantum particles evolving unitarily and macroscopic measurement devices undergoing quantum state reduction by introducing small modifications to Schro\"odinger's equation. In the case of Spontaneous Unitarity Violations (SUV), the changes considered break time-inversion symmetry and hence lead to non-unitary dynamical quantum state reduction (DQSR). Because macroscopic objects that already break a spatial or internal symmetry, such as translations or (spin or phase) rotations, possess a so-called `thin' low-energy spectrum~\cite{Wezel_SSBlecturenotes}, they are susceptible to arbitrarily small unitarity-breaking perturbations~\cite{Wezel_2008}. Such small additions to the usual quantum evolution will not significantly affect objects on the scale of microscopic quantum particles within the lifetime of the universe, while the diverging susceptibility in macroscopic symmetry-breaking objects allows the perturbations to dominate their dynamics. Moreover, since only reduction into a basis of symmetry-breaking states yields a diverging susceptibility, these states form the only possible basis for quantum state reduction to occur in~\cite{Wezel_2008}. The symmetry-breaking end states of the quantum measurement process are pointer states both in the sense of being classical (ordered, symmetry-broken) states associated with any type of actual pointer encountered in quantum measurements, and in the sense of being stable under interactions with an external environment~\cite{Zurek_1981,Wezel_SSBlecturenotes}.
Assuming that unitarity is not a fundamental property of our universe~\cite{Penrose_96,diosi_1987,penrose14,Karolyhazy_66,christian_2005,Adler_2014_grav0,Singh_2015_grav1,Diósi_2009_grav2,Gasbarri_2017_grav3}), the diverging susceptibility to non-unitary perturbations renders classical states the unavoidable fate for any sufficiently macroscopic object.

To be specific, models of SUV consider time evolution described by a modified Schr\"odinger equation of the generic form:
\begin{align}
i\hbar \frac{\partial|\psi_t\rangle}{\partial t} =  [\hat{H_t} + i\epsilon \mathcal{N} \hat{G}_t]|\psi_t\rangle.
\label{Eq:key}
\end{align}
Here $\hat{H_t}$ is the standard, possibly time-dependent, Hamiltonian acting on the joint state $\ket{\psi_t}$ of the microscopic system and measurement device. Notice that the subscript $t$ here denotes time dependence, so that $\ket{\psi_t}\equiv \sum_i \alpha_i(t) \ket{S_i}\ket{A_i}$ (see Appendix~\ref{app:Quantum Strong Measurement}). The unitarity-breaking perturbation is written as $\epsilon \mathcal{N} \hat{G}_t$. Here $\mathcal{N}$ represents the (large) total number of particles (or more generally, degrees of freedom) in the joint system-apparatus setup. It is written explicitly here to emphasize the fact that the non-unitary perturbation must couple to an order parameter of the measurement device, which scales extensively with system size~\cite{Wezel_2008}. Moreover, the strength $\epsilon$ of the perturbation is assumed to be exceedingly small, so that $\epsilon\mathcal{N}$ is small for microscopic systems and has negligible effect on their dynamics, while being large for macroscopic measurement devices and affecting an almost instantaneous evolution in them.

The operator $\hat{G}_t:=\hat{G}(\psi_t, \xi_t)$ is Hermitian, so that $i\hat{G}_t$ is anti-Hermitian and generates non-unitary time evolution. The operator $\hat{G}_t$ must be stochastic and non-linear~\cite{Mertens_PRA_21}, and thus depends on the state $|\psi_t\rangle$ as well as the instantaneous value of a time-dependent stochastic variable $\xi_t$. Crucially, the stochastic variable represents a separate physical process and its value and probability distribution do not, at any time, depend on the quantum state $|\psi_t\rangle$. As detailed in Sec.~\ref{sec:3A} below, the stochastic process described by the real-valued variable $\xi_t:=\xi(t)$ is continuous and Markovian. We will generically refer to this stochastic contribution as `noise' in the remainder of this article. 

Note that $\hat{G}_t$ depends on the instantaneous value of the noise, given by the integral $\xi_t = \xi_0 + \int^t_0 d\xi_t$, but it does not depend on the noise differential $d\xi_t$. This is an important distinction from objective collapse models in which the quantum state evolution itself is modified to resemble a drift-diffusion process of the quantum state, by including stochastic contributions described by a so-called Wiener process, usually denoted as $dW_t$~\cite{Bassi_03_PhyRep,Bassi2013Review, Pearle_89_PRA, Gisin84,Diosi_87_PLA,Ghirardi_90_PRA,percival95,diosi_1989}. Equation~\eqref{Eq:key} instead describes the evolution of a quantum state exposed to a physical noise source, described by the stochastic variable $\xi_t$ as detailed in Sec.~\ref{sec:3A}. Furthermore, $\ket{\psi_t}$ describes the full state of the joint evolution of system and measurement device (and optionally a further local environment) simultaneously undergoing DQSR, and is not an effective model. It differs in this respect from the effective non-unitarity encountered in Gorini-Kossakowski-Sudarshan-Lindblad (GKSL) master equations, obtained for example by tracing out an environment in open quantum systems~\cite{Lindblad1976,GKS76}. 

Finally, notice that as written, Eq.~\eqref{Eq:key} is time-local but non-Markovian, since a temporal integration of Eq.~\eqref{Eq:key} requires specifying the values of the noise $\xi_t$ for all times. In Sec.~\ref{sec:3} we will supplement Eq.~\eqref{Eq:key} with an explicit equation for the time evolution of the noise. Together, the equations for the noise and state evolution are Markovian in their corresponding joint-probability space. Additionally, in the special case of highly correlated noise and fast state dynamics, $\xi_t$ may be considered to be effectively constant ($d\xi_t\approx0$) throughout the measurement process, making Eq.~\eqref{Eq:key} Markovian on its own. This special case has been considered before~\cite{Mertens_PRA_21,Mertens22}, and will be briefly reviewed in Sec.~\ref{sec:4} below.

\section{Colored noise driven unitarity violation}\label{sec:3}
The general class of models whose dynamics is written in the form of Eq.~\eqref{Eq:key} will describe SUV if the operator $\hat{G}_t$ couples to the order parameter associated with the (necessarily present) broken symmetry of a measurement device. The non-unitary dynamics in these models is generically driven by colored (correlated) noise. In contrast to white noise driven drift-diffusion like models~\cite{Bassi_03_PhyRep}, which are inherently Markovian, Eq.~\eqref{Eq:key} by itself is not Markovian because of the dependence on $\xi_t = \xi_0 + \int^t_0 d\xi_t$, where the integration is for all times since the measurement process initiated. Augmenting Eq.~\eqref{Eq:key} with an explicit equation for the dynamics of the noise yields a pair of dynamical equations describing the joint evolution of the pair $\{\ket{\psi_t},\xi_t\}$ which together are time-local and Markovian~\cite{Hanggi94,Luczka2005}. Explicitly, this pair of equations can be written as:
\begin{align}
i\hbar \frac{\partial|\psi_t\rangle}{\partial t} &=  [\hat{H_t} + i\epsilon \mathcal{N} \hat{G}_t]|\psi_t\rangle \notag \\
d\xi_t&=f(\xi_t)dt + g(\xi_t)dW_t
\label{Eq:StateXiPair}
\end{align}
Notice that the first line is Eq~\eqref{Eq:key}, repeated here for convenience. The second line corresponds to a general class of It\^o drift-diffusion processes that are continuous, real-valued stochastic processes. We assume that the drift $f(\xi_t)$ and diffusion $g(\xi_t)$ are both smooth functions, and that their values at time $t$ depend only on the value of the stochastic variable $\xi_t$ at the same time $t$, and no other previous or future time. That is, $\xi_t$ is a continuous Markovian stochastic process (It\^o process). The stochastic nature of the process $\xi_t$ is effected by the increments $dW_t$ denoting a standard Wiener process (corresponding to the Brownian motion $W_t=\int dW_t$)~\cite{revuz1999continuous,oksendal2003stochastic,gardiner2004handbook}. 

As usual, $W_t$ (with $W_0\equiv0$) is a real-valued, continuous, but non-differentiable process with independent, zero-mean Gaussian increments $dW_t$ taken from the Gaussian probability distribution with mean $0$ and standard deviation, $\sqrt{dt}$, denoted as $\mathbb{N}\left(0,\sqrt{dt}\right)$. This implies the standard time independent expectation values $\mathbb{E}\left[dW_t\right] = \mathbb{E}\left[W_t\right]=0$ and $\mathbb{E}\left[dW_t^2\right]=dt$ while $\mathbb{E}\left[dW_tdW_s\right]=0$ for $t\neq s$. Here $\mathbb{E}$ indicates the expectation value with respect to an ensemble of $dW_t$ values sampled from the Gaussian distribution function. These relations lead to the so-called It\^o multiplication rules, with $dt^2 \sim 0$, $dtdW_t \sim 0$ and $dW_t^2\sim dt$~\cite{gardiner2004handbook, oksendal2003stochastic}. 
Because the dynamical equation for $\xi_t$ is time-local and Markovian, together, the pair of relations in Eq.~\eqref{Eq:StateXiPair} is rendered Markovian in the augmented state space $\{\ket{\psi_t},\xi_t\}$. A detailed general description of such augmented spaces in the context of Markovianization of noise-driven stochastic differential equations may be found for example in Refs.~\cite{Luczka2005,Hanggi94}. 

In the following, we describe a procedure for ensuring that the dynamics defined by Eq.~\eqref{Eq:StateXiPair} is norm-preserving. This will allows us to obtain dynamics modelling single measurement events as well as formulate modified von-Neumann-Liouville equations for the statistics of an ensemble of measurements. Further, we introduce two explicit and physically motivated examples of correlated noise processes relevant to SUV models.

\subsection{Normalization prescription}\label{sec:3B}
Although the norm of a quantum state is unobservable and does not affect the evolution of observable quantities under the dynamics of Eq.~\eqref{Eq:StateXiPair}~\cite{Mertens_PRA_21,Mertens22,aritro1}, it is convenient to define the unitarity-breaking term in such a way that it is norm-conserving. It is possible to make any unitarity-breaking operator of the general form of Eq.~\eqref{Eq:StateXiPair} norm preserving. To do so, we first rewrite the set of equations governing the pair $\{\ket{\psi_t},\xi_t\}$ in It\^o's differential notation:
\begin{align}
d|\psi_t\rangle &= \frac{1}{\hbar}\left[-i\hat{H}_t +  \hat{G}_t\right]|\psi_t\rangle \,\,dt \nonumber \\
d\xi_t&=f(\xi_t)dt + g(\xi_t)dW_t
\label{Eq4:key_ito_norm}
\end{align}
For convenience, here we absorbed the factor $\epsilon \mathcal{N}$ into the definition of $\hat{G}_t$. 

Formally, given a noise trajectory $\xi_t$, the state at any time $T$ can be found from an integral of the form  $\ket{\psi_T}=\ket{\psi_0}+\frac{1}{\hbar}\int^T_0 \left[-i\hat{H}_t +  \hat{G}_t\right]\ket{\psi_t}dt$. In the integrand, the operator $\hat{G}_t$ depends functionally on the stochastic variable $\xi_t$, but it does not contain the stochastic differential $d\xi_t$. This makes it an ordinary Riemann integral, independent of any Stratonovich or It\^o interpretation, even in the case of multiplicative noise~\cite{Luczka2005,Hanggi94}. In other words, the integral over stochastic dynamics can be interpreted in either the It\^o or Stratonovich sense, and the usual rules of calculus persist. To see this explicitly, consider the so-called quadratic variations of the quantum state dynamics. The quadratic variation for a general stochastic process $X_t$ is defined as $\mathcal{Q}_T[X_t,X_t]:=\int^T\mathbb{E}_X\left[dX_t^2\right]$. Here, the expectation value $\mathbb{E}_X$ is taken over an ensemble of stochastic trajectories $X_t$. Formally, this definition can also be written in differential notation as $d\mathcal{Q}_t[X_t,X_t]:=\mathbb{E}_X\left[dX_t^2\right]$ (see Appendix~\ref{app:QuadVar} for details). For a twice-differentiable function $K(X_t)$ of the stochastic process, we can then use It\^o's lemma to find 
$dK(X_t) = \left[ dX_t\partial_x K(x) 
+\frac{1}{2}d\mathcal{Q}_t[X_t,X_t]\,\partial_x^2 K(x)\right]_{x=X_t}$. From this expression it is clear that if $d\mathcal{Q}_t[X_t,X_t]$ is zero for all times, It\^o's correction term vanishes and the calculus of functions $K(X_t)$ of the process $X_T$ is the same as that of regular functions. For the specific stochastic process of the quantum state $\ket{\psi_t}$ evolving according to Eq.~\eqref{Eq4:key_ito_norm}, we find a vanishing quadratic variation $d\mathcal{Q}_t(\bra{\psi_t},\ket{\psi_t}):=\mathbb{E}_{\psi}\left[\langle d\psi_t|d\psi_t\rangle\right]$. Here, $\ket{d\psi_t}$ is short for $d\ket{\psi_t}$ and the expectation value $\mathbb{E}_{\psi}$ is averaged over realizations of $\ket{\psi_t}$ corresponding to an ensemble of continuous noise processes $\xi_t$. Because $\langle d\psi_t|d\psi_t\rangle$ is of order $dt^2$, the quadratic variation evaluates to zero, and the usual rules of calculus can be applied to functions of the quantum state, evolving via Eq.~\eqref{Eq4:key_ito_norm}. In particular, this implies that the quantum state evolution is both continuous and (once) differentiable, in contrast to white-noise driven objective collapse theories~\cite{Bassi_03_PhyRep,Bassi2013Review, Ghirardi_90_PRA, Pearle_89_PRA}.

The vanishing of the scalar quadratic variation, $\mathbb{E}_{\psi}\left[\langle d\psi_t|d\psi_t\rangle\right]=0$, moreover yields the variation of the norm $N_t:=\langle \psi_t |\psi_t\rangle$, for which the It\^o-Leibniz product rule and Eq.~\eqref{Eq4:key_ito_norm} implies $dN_t = \langle d\psi_t|\psi_t\rangle + \langle \psi_t|d\psi_t\rangle +\mathbb{E}_{\psi}\left[\langle d\psi_t|d\psi_t\rangle\right]\,\,
= \frac{2 dt}{\hbar} \langle\psi_t|\hat{G}_t|\psi_t\rangle$. This  shows that in general, normalization is not preserved under the dynamics of Eq.~\eqref{Eq4:key_ito_norm}. Denoting the un-normalized states as $|\psi^{U}_t\rangle$, a corresponding normalized state $|\psi^{N}_t\rangle$ can be obtained at any time from the expression $|\psi^N_t\rangle={|\psi^{U}_t\rangle} / {\sqrt{\langle\psi^{U}_t|\psi^{U}_t\rangle}}$. Assuming that the state was normalized at some time $t$, we can use Eq.~\eqref{Eq4:key_ito_norm} and compute $d|\psi^N_{t}\rangle$, yielding 
$
d|\psi^N_{t}\rangle = \frac{1}{\hbar}\left[-i\hat{H}_t+\hat{G}_t-\langle\hat{G}\rangle_t\right]|\psi^N_{t}\rangle \,\,dt+\mathcal{O}\left(dt^2\right)
$.
This expression for the time evolution of the normalized state is the same as that in the first line of Eq.~\eqref{Eq4:key_ito_norm}, except for the addition of the correction term $\langle\hat{G}\rangle_t={\langle\psi_t|\hat{G}_t|\psi_t\rangle}/{\langle\psi_t|\psi_t\rangle}$. Because the correction is proportional to the identity operator, it does not influence the relative weights or relative phases of the evolving quantum state, instead can be seen as a geometrical constraint on the state dynamics. It can thus be added to Eq.~\eqref{Eq4:key_ito_norm} without changing any physically observable quantities to yield the general norm-preserving quantum state dynamics:
\begin{align}
d|\psi_t\rangle &= \frac{1}{\hbar}\left[-i\hat{H}_t +  \left(\hat{G}_t-\langle\hat{G}\rangle_t\right)\right]|\psi_t\rangle \,\,dt \nonumber \\
d\xi_t&=f(\xi_t)dt + g(\xi_t)dW_t
\label{Eq:ito_normalized}
\end{align}
From here on, we will denote the norm-preserving unitarity breaking term in the generator for time evolution as $\hat{\mathcal{G}}_t :=\hat{G}_t-\langle\hat{G}\rangle_t$. It again depends on the stochastic process $\xi_t$ but not on the noise differential $d\xi_t$, and leads to stochastic yet continuous, differentiable, and norm-preserving quantum state dynamics.

\subsection{Modified von-Neumann Liouville  equations}\label{sec:3C}
Having found a general norm-preserving form for the SUV dynamics, we can consider the dynamics of the pure state density operator $\hat{\mathbb{P}}_t=|\psi_t\rangle\langle\psi_t|$. Using the It\^o-Leibniz rule, $
d\hat{\mathbb{P}}_t=|d\psi_t\rangle\langle\psi_t|+|\psi_t\rangle\langle d\psi_t|+\mathbb{E}_{\psi}\left[|d\psi_t\rangle\langle d\psi_t|\right]
$, and the vanishing of the operator-valued quadratic variation, $d\mathcal{\hat{Q}}_t(\ket{\psi_t},\bra{\psi_t}):=\mathbb{E}_{\psi}\left[\ket{ d\psi_t}\bra{d\psi_t}\right]=O(dt^2)=0$, we obtain:
\begin{align}
\frac{\partial\hat{\mathbb{P}}_t}{\partial t}&=\frac{-i}{\hbar}\,\left[\hat{H}_t\,,\,\hat{\mathbb{P}}_t\right]+\frac{1}{\hbar}\,\left\{\hat{\mathcal{G}}_t\,,\,\hat{\mathbb{P}}_t\right\}
\label{Eq_ProtoMaster0}
\end{align}
This expression is the modified von-Neumann-Liouville equation describing the dynamics of a pure state, given a trajectory for the noise $\xi_t$. Note that it is not a master equation, which would be obtained by averaging over an ensemble of noise trajectories. This means that, as before, Eq.~\eqref{Eq_ProtoMaster0} by itself is non-Markovian. It needs to be augmented with the prescription for the noise dynamics in Eq.~\eqref{Eq:ito_normalized} in order to obtain a Markovian process in the joint space $\{\mathbb{P}_t,\xi_t\}$. 

We can obtain a formal expression for the noise-averaged master equation by labelling each possible noise trajectory within a given time interval $0<t<T$ with a (continuous) index $\omega$, denoted as $\xi_t(\omega)$. The particular stochastic operator corresponding to one single trajectory $\xi_t(\omega)$ may then be labelled $\mathcal{\hat{G}}_t(\omega)$, and the time evolution of a given initial pure state is obtained by integrating Eq.~\eqref{Eq_ProtoMaster0}, which yields $\hat{\mathbb{P}}_T(\omega)=\hat{\mathbb{P}}_0 +\frac{-i}{\hbar} \,\int^T_0 dt\left[\hat{H}_t\,,\,\hat{\mathbb{P}}_t(\omega)\right]
+\frac{1}{\hbar}\,\int^T_0 dt\left\{\hat{\mathcal{G}}_t(\omega)\,,\,\hat{\mathbb{P}}_t(\omega)\right\}
$

Formally averaging over all possible trajectories of the noise requires a probability measure, $d\mu(\omega)$, such that the total probability of finding any noise trajectory is normalized, $\int _\Omega d\mu(\omega)=1$. The noise averaged pure state then becomes $\mathbb{E}_\omega[\hat{\mathbb{P}}_T(\omega)]=\int d\mu(\omega)\hat{\mathbb{P}}_T(\omega):=\hat{\rho}_T$, where $\hat{\rho}_T$ is the usual quantum mechanical density matrix describing a statistical mixture of time-evolving quantum states. Computing this average for Eq.~\eqref{Eq_ProtoMaster0} yields a formal expression for the master equation:
\begin{align}
\hat{\rho}_T= \hat{\mathbb{P}}_0 &-\frac{i}{\hbar}\,\int^T_0 dt\left[\hat{H}_t\,,\,\hat{\rho}_t\right] \notag \\
&+\frac{1}{\hbar}\, \int d\mu(\omega)\int^T_0 dt \,\left\{\hat{\mathcal{G}}_t(\omega)\,,\,\hat{\mathbb{P}}_t(\omega)\right\}
\label{Eq_FormalMaster}
\end{align}
Here, $\hat{\mathbb{P}}_0=\hat{\rho}_0$ because we assumed the initial state to be pure. The integrals in the expectation value cannot be commuted, i.e. the time integral must be performed for each stochastic trajectory separately, and only then can we average over the ensemble of trajectories.Eq.~\eqref{Eq_FormalMaster} is a formal expression and further analytical evaluation for general noise processes is highly non-trivial~\cite{Hanggi94,Luczka2005}. We will examine its behaviour and numerically evaluate the evolution of specific initial states for two specific implementations of the noise dynamics in Sec.~\ref{sec:4} below.

\subsection{Colored Noise models}\label{sec:3A}
Having found a formal expression for the SUV dynamics driven by a general noise process, we will now focus on two particular (physically motivated) models for the evolution of the noise. Both fall in the class of It\^o drift-diffusion models, and are continuous Markov stochastic processes. 

The first stochastic process we consider is the so-called Ornstein-Uhlenbeck (OU) process~\cite{OU_OG1930,Doob42,gardiner2004handbook,Risken1996}, given by:
\begin{align}
d\xi_t = - \xi_t \,\frac{dt}{\tau} + \sqrt{\frac{2}{\tau}} dW_t
\label{Eq:OU SDE}
\end{align}
The OU process has correlations that decay exponentially within the characteristic time $\tau$, and it evolves towards a steady-state Gaussian distribution centered on $\xi=0$, independent of initial conditions and given by $\rho^{\infty}(\xi)=\frac{1}{\sqrt{2\pi}}e^{\frac{-\xi^2}{2}}$~\cite{gardiner2004handbook,Risken1996,Pavliotis2008}. A well-known application of the OU process in statistical physics is modelling the velocity of a particle undergoing damped Brownian motion in a fluid~\cite{Zwanzig2001}. In the setting of SUV dynamics considered here, there is no thermal environment inducing random motion, and $\xi_t$ instead represents an effective description for the influence of a physical field outside of quantum mechanics driving stochastic dynamics in Hilbert space~\cite{Wezel10}. 

In the OU process, $\xi_t$ can take values on the entire real line. The trajectory of an individual noise realisation with sharp initial condition $\xi_0$ is analytically described by~\cite{gardiner2004handbook}:
\begin{align}
\xi_t = \xi_0 e^{-t/\tau} +  \sqrt{\frac{2}{\tau}} e^{-t/\tau} \int_0^t e^{s/\tau}dW_s
\label{Eq:OU_gen_sol_main}
\end{align}
Here integrating a time-dependent function with the Wiener measure represents a standard It\^o integral. From Eq.~\eqref{Eq:OU_gen_sol_main}, we can compute the mean and the autocorrelation, given initial conditions. Sampling $\xi_0$ from its steady state distribution, and using Eq.~\eqref{Eq:OU_gen_sol_main}, we find:
\begin{align}
    \mathbb{E}_{\xi}[\xi_0] = \mathbb{E}_{\xi}[\xi_t]  &=0 \notag \\
    \mathbb{E}_{\xi}[\xi_t \,\xi_s] -\mathbb{E}_{\xi}[\xi_t] \mathbb{E}_{\xi}[\xi_s]  &= e^{- |t-s|/\tau}
\end{align}
These results may also be obtained by considering the (adjoint) infinitesimal generator of It\^o processes, solving the corresponding Fokker-Planck-Kolmogorov (FPK) equations, and computing the mean and autocorrelation from given initial conditions for an ensemble of noise processes (as described in more detail in Appendix~\ref{app:Noise_Characterization})~\cite{gardiner2004handbook,Risken1996,oksendal2003stochastic}.

The second process we consider for driving SUV dynamics, arises from Brownian motion on a spherical manifold. The position of a Brownian walker on a sphere is given by two coordinates, its polar angle ($\theta_t$) and the azimuthal angle ($\phi_t$). Both coordinates evolve stochastically, but because the circumference of circles at constant latitude depend on the value of the latitude, the $\phi_t$ dynamics depends on the instantaneous value of $\theta_t$, while the dynamics of $\theta_t$ is independent of longitude (see Appendix~\ref{app:Noise_Characterization}). 

Rather than the position on the sphere itself, we take the cosine of its latitude as the noise variable in the unitarity-breaking term of Eq.~\eqref{Eq:ito_normalized}, i.e. $\xi_t=\cos(\theta_t)$. This definition has the advantage that it yields a stochastic variable on the bounded domain between $\xi_t=-1$ (the south pole of the sphere) and $\xi_t=1$ (the north pole). Moreover, the steady state distribution of the Brownian motion covers the sphere uniformly, which results in the steady state distribution of $\xi_t=\cos(\theta_t)$ being uniform across the range $[-1,1]$. These features of the Brownian motion on the sphere have previously been shown to enable the emergence of Born's rule in the limit of infinite correlation time or time-independent noise in models of SUV~\cite{Mertens_PRA_21,Mertens22}.

The stochastic dynamics resulting from the spherical Brownian motion (SBM) process is given by:
\begin{align}
d\xi_t = - \xi_t \,\frac{dt}{\tau} + \sqrt{\frac{1-\xi_t^2}{\tau}} dW_t
\label{Eq:SBM SDE}
\end{align}
Here, the final term indicates SBM is a multiplicative noise process, and is interpreted in the It\^o sense. The time scale $\tau$ again corresponds to the correlation time of the noise process, while the drift term proportional to $dt$ ensures the noise is mean-reverting with mean zero.

As for the OU process, and as described in Appendix~\ref{app:Noise_Characterization}, the mean and autocorrelation for the SBM process can be found from considering its infinitesimal (adjoint) generator and the corresponding Fokker-Planck-Kolmogorov (FPK) equations~\cite{Risken1996,CNOSPearsonDiffPaper2008,CNOStimelocalPearsontoJacobiAscione2021,CNOSWong1964}. Sampling $\xi_0$ from the steady state distribution $\rho^{\infty}(\xi)=\frac{1}{2}$ within the interval $[-1,1]$, we have $\mathbb{E}_{\xi}[\xi_0]=0$ and we find:
\begin{align}
    \mathbb{E}_{\xi}[\xi_t] = \mathbb{E}_{\xi}[\xi_0]  &=0 \\
    \mathbb{E}_{\xi}[\xi_t \,\xi_s] -\mathbb{E}_{\xi}[\xi_t] \mathbb{E}_{\xi}[\xi_s] &=\frac{1}{3}e^{-|t-s|/\tau}
\end{align}

Both the OU and SBM processes thus have a symmetric steady state distribution and a correlation time $\tau$. In the limit of vanishing correlation time, both processes flow smoothly to an approximate white noise limit, with Gaussian white noise for OU, and uniform white noise for SBM. The consequences for SUV dynamics in this limit are explored in the next section in the context of an initial superposition over two pointer states.

\section{Two-state superposition}\label{sec:4}
Having obtained a norm-preserving generic model for SUV, its corresponding modified von-Neumann-Liouville equation and pure state evolution, and having defined two stochastic processes of interest, we will now apply all of these to the specific situation of an initial configuration for the combined system and measurement device being superposed over two pointer states. A non-linear and stochastic generator for the SUV dynamics of an initial two-state superposition was introduced in Refs.~\cite{Mertens_PRA_21,Mertens22,aritro1}. It was shown there to yield spontaneous unitarity violations, resulting in a reduction to symmetry-breaking pointer states, with final-state probabilities obeying Born's rule in the limit of infinitely long correlation time in the noise. The explicit form used was given by
$
\hat{G}(\psi_t,\xi_t) = \hat{\sigma}_3 (J \langle\hat{\sigma}_3\rangle_t+G\xi_t)/2$.
Here $J$ is the strength of the non-linear contribution, $G$ determines the average magnitude of the stochastic fluctuations, and $\langle\hat{\sigma}_3\rangle_t := \bra{\psi_t}\hat{\sigma}_3\ket{\psi_t}/{\langle\psi_t\ket{\psi_t}}$, which is the quantum expectation value of the Pauli z-matrix in the basis of the superposed pointer states. 

The combined state of the system and measurement device can be written as $\ket{\psi_t}=\alpha_t\ket{0}+\beta_t\ket{1}$. Under the dynamics defined by Eq.~\eqref{Eq4:key_ito_norm}, the state evolves stochastically towards either of the two attractive fixed points $\ket{0}$ or $\ket{1}$ at the poles of the Bloch sphere spanned by the two superposed pointer states. Since the operator $\hat{G}_t$ is purely real, it will not affect the phases of the coefficients $\alpha_t$ and $\beta_t$. The unitary contribution $\hat{H}_t$ to the time evolution on the other hand, causes Rabi oscillations encircling the Bloch sphere. These oscillations do not introduce any attractive fixed points in the quantum state evolution, nor affect the fixed points introduced by $\hat{G}_t$. With regard to the final state distribution, $\hat{H}_t$ can thus be ignored without loss of generality, while in the large-$\mathcal{N}$ limit (strong $\hat{G}_t$) or for cases with $[\hat{H}_t,\hat{G}_t]=0$, it can be ignored altogether. We therefore set $\hat{H}_t=0$ from here on. In that case, the time evolution does not affect the total or relative phase of the state coefficients, and again without loss of generality, we may consider $\alpha_t$ and $\beta_t$ to be real for all time~\cite{Mertens_PRA_21,Mertens22}.

Using the normalization prescription of Sec.~\ref{sec:3A}, i.e. $\hat{\mathcal{G}}_t=\hat{G}_t-\langle\hat{G}\rangle_t$, we obtain the generator for norm-preserving SUV dynamics appearing in Eq.~\eqref{Eq:ito_normalized}:
\begin{align}
\hat{\mathcal{G}}_t = \frac{1}{2}\left( \hat{\sigma}_3-\langle\hat{\sigma}_3\rangle_t \right) \left( J \langle\hat{\sigma}_3\rangle_t+G\xi_t\right)
\label{Eq:curlyG2}
\end{align}
At this point, it is instructive to compare the colored noise driven SUV state dynamics with the well-known stochastic Schr\"odinger equation (SSE), which forms the basis for the time evolution encountered in various so-called continuous spontaneous localization (CSL) models~\cite{Pearle_89_PRA, Bassi_03_PhyRep,Ghirardi_90_PRA}.
\begin{align}
\text{SUV:  }i\hbar \, d|\psi_t\rangle = & \hat{H}_t\ket{\psi_t}dt+ \frac{i}{2}\bigg[J\langle\hat{\sigma}_3\rangle_t\left(\hat{\sigma}_3-\langle\hat{\sigma}_3\rangle_t\right) dt \bigg. \notag \\ 
&+ \bigg. G\left(\hat{\sigma}_3-\langle\hat{\sigma}_3\rangle_t\right) \xi_t(\omega)dt \bigg]|\psi_t\rangle \notag \\
\text{SSE:  }i\hbar \, d|\psi_t\rangle= &\hat{H}_t\ket{\psi_t}dt+ \frac{i}{2} \bigg[{-\gamma}\left(\hat{\sigma}_3-\langle\hat{\sigma}_3\rangle_t\right)^2 dt \bigg. \notag \\
&+ \bigg. 2 \sqrt{\gamma}\left(\hat{\sigma}_3-\langle\hat{\sigma}_3\rangle_t\right)dW_t\bigg]|\psi_t\rangle
\label{Eq:Suv_compare}
\end{align}
As before, $\xi_t(\omega)$ here denotes a particular realization of the stochastic noise process $\xi_t$ labelled by the (continuous) index $\omega$. Notice that both the SUV and SSE equations can here be expressed in terms of Pauli operators, owing to the fact that we focus on two-state dynamics. In more general CSL models based on the SSE, the Pauli operator is replaced by some form of local density operator.

Comparing the two expression in Eq.~\eqref{Eq:Suv_compare}, several differences may be noticed. First of all, the SUV process includes a stochastically evolving external field $\xi_t$, whose evolution is defined independently from the quantum mechanical state vector. SUV thus requires the noise process to be physical (in the sense of existing independent from the quantum state), and describing the state dynamics necessarily encompasses defining a pair of evolution equations (see Sec.~\ref{sec:3}). The SSE on the other hand, is given by a single It\^o stochastic differential equation, in which the $dW_t$ directly modifies the definition of the quantum mechanical time increment, making it stochastic without specifying any external source.

\begin{figure}[t]
\includegraphics[width=\columnwidth]{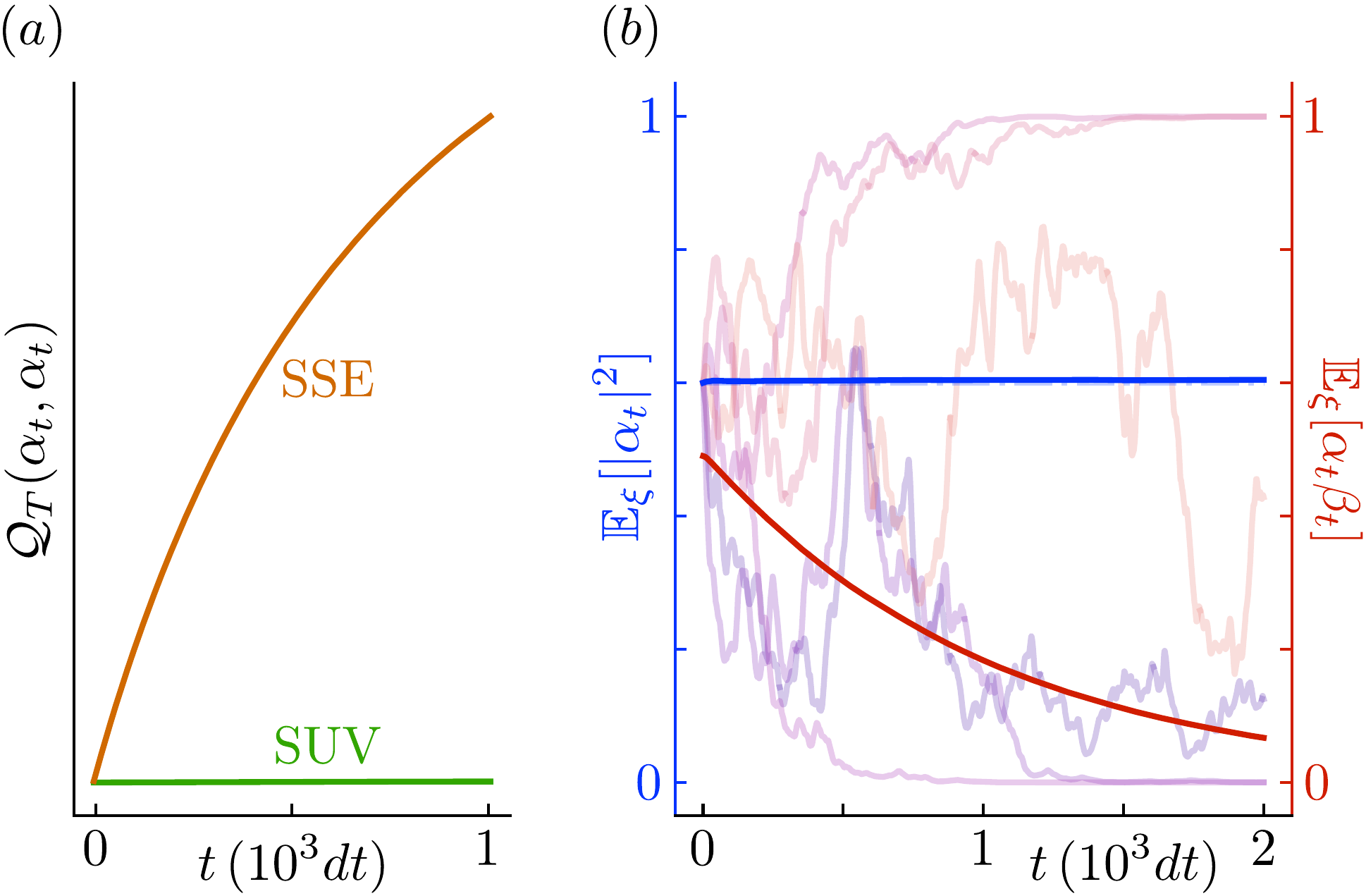}
\caption{\label{fig1}State dynamics. (a) The numerically evaluated quadratic variation $\mathcal{Q}_T(\alpha_t,\alpha_t)$ as a function of time for both the SUV and SSE dynamics defined in Eq.~\eqref{Eq:Suv_compare}. The quadratic variation is seen to be identically zero for the SUV process, while it grows monotonically in time for the SSE. Here, we used the parameter values $G=1$, $J=2$, $\tau=1$, $\gamma=0.5$ and the numerical timestep $dt=0.001$. The expectation value in $\mathcal{Q}_T(\alpha_t,\alpha_t)$ is averaged over $20,000$ noise realizations. (b) State evolution obtained by numerically evaluating Eqs.~\eqref{Eq_FormalMaster} and~\eqref{Eq:curlyG2}  with the state initialized at $|\alpha_0|^2=0.6$ (Several implementations of $|\alpha_0|^2$  shown in shaded lines) using OU noise (similar results are obtained for SBM noise). Also shown are the noise averaged density matrix elements $\mathbb{E}_{\xi}[|\alpha_t|^2]$ (diagonal element, blue line) and $\mathbb{E}_{\xi}[\alpha_t \beta_t]$ (off-diagonal element, red line), for which the average is over an ensemble of noise realizations. Here, we used the parameter values $G=10$, $J=2$, $\tau=0.01$, and the numerical time step $dt=0.001$. The expectation values are averaged over $50,000$ OU noise realizations. For these parameter values, $|\alpha_t|^2$ follows Born's statistics and is seen to be a martingale, $\mathbb{E}_{\xi}[|\alpha_t|^2]=|\alpha_0|^2$.
}
\end{figure}

Secondly, the state dynamics in SUV is continuous and differentiable, since all terms on the right hand side are proportional to $dt$. The state dynamics generated by the SSE on the other hand, is continuous but not differentiable, because of the $dW_t$ appearing in its diffusion term. The smoothness properties of the SUV and SSE state dynamics may be made explicit by considering the quadratic variation $\mathcal{Q}_T(\alpha_t,\alpha_t)$, where $\alpha_t := \bra{0}\ket{\psi_t}$. While the quadratic variation is proportional to $dt^2$ and thus vanishes for SUV dynamics, it is proportional to $dt$ and thus non-vanishing for the SSE process. This is shown for an ensemble of stochastic realizations of both equations in Eq.~\eqref{Eq:Suv_compare} (see Fig.~\ref{fig1}a).

Thirdly, the modifications to Schr\"odinger's equation introduced by the SUV equations can be seen not to affect the expectation value of energy, if energy is defined as the operator generating time translation (i.e. according to Noether's theorem). Both the SUV and SSE dynamics can be written in the form $i \hbar\,d|\psi_t\rangle = \hat{\mathcal{H}}_t\ket{\psi_t}dt=(\hat{H}_t + i\hat{\Delta}_t)\ket{\psi_t}dt$ with $\hat{H}_t$ the standard quantum mechanical Hamiltonian and $\hat{\Delta}_t$ the modifications in Eq.~\eqref{Eq:Suv_compare}. Using Noether's theorem and the fact that time translation symmetry (in contrast to time inversion) is preserved by Eq.~\eqref{Eq:Suv_compare}, `energy' can in general be defined as the expectation value of the generator of time evolution, so that $E=\langle\hat{\mathcal{H}}_t\rangle$. Within SUV, this energy retains the same form as under the unperturbed Scr\"odinger equation, i.e. $\langle\hat{\mathcal{H}}_t\rangle=\langle\hat{H}_t\rangle$ regardless of the unitarity-breaking terms in the SUV dynamics. In SSE on the other hand, the energy defined this way obtains the correction $\langle \hat{\mathcal{H}}_t \rangle = \langle \hat{H}_t \rangle - i\gamma \langle ( \hat{\sigma}_3 - \langle\hat{\sigma}_3\rangle_t)^2\rangle /2$. SSE thus predicts a contribution of non-unitary fluctuations to the expectation value $E$, while these are absent in models of SUV.

Finally, both the deterministic and stochastic terms in the SUV and SSE expressions have subtly different forms. In particular, the stochastic term in the SSE has $dW_t$ multiplied directly by the factor $\hat{\sigma}_3-\langle\hat{\sigma}_3\rangle$, which in that case ensures both normalization and Born's rule~\cite{Bassi_03_PhyRep}. This form is obtained from a set of un-normalized quantum state diffusion equations (the so called raw ensemble) via a Girsanov transformation, which explicitly introduces the multiplicative factor $\hat{\sigma}_3-\langle\hat{\sigma}_3\rangle$ in the noise process in order to ensure adherence to Born's statistics into the normalized equations (physical ensemble)~\cite{Pearle_89_PRA,Ghirardi_90_PRA,Bassi_03_PhyRep}. The full stochastic increment in the time evolution of the physical ensemble thus depend on the instantaneous state of the system in precisely such a way as to ensure adherence to Born's rule. In the SUV dynamics, on the other hand, the factor $\hat{\sigma}_3-\langle\hat{\sigma}_3\rangle$ appears as an overall (geometric) factor constraining the state dynamics, as shown in Sec.~\ref{sec:3B}. It does not influence the evolution of the physical noise $\xi_t$, whose stochastic increments are given by a dynamical law independent of the quantum state. In this case, adherence to Born's rule emerges from the dynamics rather than being hard-wired in its construction, and it depends on a physical relation between the deterministic and stochastic terms in the SUV process (as made explicit below)~\cite{Mertens22,aritro1}.

Returning to the dynamics for SUV of an initial two-state superposition generated by Eqs.~\eqref{Eq:ito_normalized} and~\eqref{Eq:curlyG2}, we note that this has been previously analysed in the limit of highly correlated noise, i.e. $\tau\to\infty$ for the SBM noise model~\cite{Mertens_PRA_21,Mertens22}. This limit is encountered when the dynamics due to measurement, result in the quantum state evolving within a much shorter time than the correlation time $\tau$ of the noise. The noise can then be approximated to be constant, $d\xi_t\approx0$ and thus $\xi_t(\omega)=\xi_0(\omega)$, which is the noise sampled at initial times. This extreme limit of fast measurement yields Markovian evolution with the trivial noise dynamics $d\xi_t\approx0$, i.e. the quantum state dynamics samples the noise only once during its evolution. It was shown that quantum state reduction to the pointer state $\ket{0}$ or $\ket{1}$ occurs with Born's rule probabilities in this limit whenever $J=G$~\cite{Mertens22} for uniformly distributed noise in the range $[-1,1]$. The requirement of Born's rule emerging thus gives a way of probing the relation between the physical fluctuation and dissipation processes driving the SUV dynamics. In the same $\tau\to\infty$ limit, it was furthermore shown that SUV models yielding Born's rule can be defined for arbitrary initial state superpositions~\cite{aritro1}.

Here, we will instead be interested in the opposite limit of vanishing correlation time, $\tau\to 0$. Homogenizing the fast-varying colored noise will then allow us to approach a white-noise limit, as described below. In this limit, the existence of a linear quantum (Markov) semi-group can be shown, and explicit master equations for both the OU and SBM noise processes can be computed.

\subsection{Fast noise limit}\label{sec:4A}
From here on, we will focus on the SUV dynamics starting from an initial configuration for the combined system and measurement device that is superposed over two pointer states. Moreover, we will consider the OU and SBM noise processes in the limit of vanishing (but not identically zero) correlation time, $\tau\to 0$. This limit corresponds to the noise fluctuating much faster than any other time scale in the dynamics. Using Eq.~\eqref{Eq:curlyG2}, we consider the pair of equations: 
\begin{align}
\frac{\partial|\psi_t\rangle}{\partial t}&= \frac{1}{2}\left(\hat{\sigma}_3-\langle\hat{\sigma}_3\rangle_t\right)\left[J\langle\hat{\sigma}_3\rangle_t+G\xi_t\right]|\psi_t\rangle,\nonumber\\
d\xi_t&=-\xi_t\frac{dt}{\tau} + g(\xi_t)dW_t.
\label{Eq4:smalltaupair}
\end{align}
Here, we use $g(\xi_t)=\sqrt{{2}/{\tau}}$ for the OU noise process, while $g(\xi_t)=\sqrt{{1-\xi_t^2}/{\tau}}$ for SBM (see Sec.~\ref{sec:3A} and Appendix.~\ref{app:Noise_Characterization}). In both cases, the initial value for the noise is taken from the corresponding steady-state distribution, which implies that $\mathbb{E}[\xi]=0$ for all times in both cases.
Together, the two expressions in Eq.\eqref{Eq4:smalltaupair} constitute a Markovian evolution for the combined state and noise dynamics.

As in the case of infinite correlation time~\cite{Mertens_PRA_21,Mertens22}, we expect that also for general $\tau$, the condition of finding Born's probabilities for an ensemble of quantum measurements will yield a (state-independent) relation between the model parameters $J$ and $G$. To find this relation, we would need to average over the noise in the state dynamics, and study the statistics of $\ket{\psi}$ alone. In general, this is analytically intractable. It is in principle possible numerically (as shown in Fig.~\ref{fig1}), but the computational complexity of sweeping the complete parameter range grows rapidly with decreasing $\tau$ and increasing $G$ or $J$. We therefore focus on the vanishing correlation time limit, where we can analytically perform a multi-scale temporal analysis to homogenize over the noise, given that the noise is sufficiently ergodic within the considered time scale~\cite{Pavliotis2008,HorsthemkeBook2006}. 

Following this approach, we will find that the coupled dynamics of the quantum state and noise becomes weakly equivalent to a process driven by white noise in the vanishing $\tau$ limit. The weak equivalence implies agreement in terms of time-dependent probabilities, i.e. agreement at the level of averages, but not on individual trajectories.  Schematically, this will be shown subsequently to imply (under temporal integrals) that: \begin{align}
\lim_{\tau\small\to0} G \left(\hat{\sigma}_3-\langle\hat{\sigma}_3\rangle_t\right) |\psi_t\rangle \,\xi_t dt\,\to \mathcal{D} \left(\hat{\sigma}_3-\langle\hat{\sigma}_3\rangle_t\right) |\psi_t\rangle \circ dW_t,\nonumber
\end{align} where `$\circ$' denotes the Stratonovich product and the diffusion parameter $\mathcal{D}$ is the effective coupling constant of the stochastic term in the limit of vanishing correlation time. $\mathcal{D}$ depends on a combination of the $\tau$ and the original coupling constant $G$, such that $\mathcal{D}$ is held constant as the correlation time is taken to zero. The white-noise driven state dynamics is described by a single Markovian equation, similar to (but distinct from) the SSE. This can be used to obtain a master equation for the state dynamics, which in turn yields the condition under which Born's rule expectation values are obtained. 

~\\
\emph{Kolmogorov Backward equation} --
Since we are interested in obtaining a master equation for the state evolution which indicates adherence to Born's rule, we will consider the time evolution of the squared component $z_t\equiv|\braket{0}{\psi_t}|^2$, rather than that of the state vector itself. Using the conservation of normalization and writing $\bra{0}\ket{\psi_t}=\alpha_t$ we find from Eq.~\eqref{Eq4:smalltaupair} that:
\begin{align}
d \alpha_t =  dt\, \alpha_t \left(1-\alpha_t ^2\right) \left[G \xi_t+J (2\alpha_t^2 -1 )\right]
\end{align}
We can use this equation to compute $dz_t$, using It\^o's lemma and noting the vanishing of the quadratic variation $\mathbb{E}_{\alpha}[d\alpha^2_t]=0$, to obtain:
\begin{align}
    dz_t &= J_{z} dt +  \mathcal{G}_z \xi_t \frac{dt}{\sqrt{\tau}},\nonumber\\
    d\xi_t &= -\xi_t\frac{dt}{\tau} + g(\xi_t) dW_t.
    \label{Eq:ZPair}
\end{align}
Here, $J_z = 2J \,z(1-z)(2z-1)$ and $\mathcal{G}_z = 2 D \,z(1-z) $, where the time dependence of $z_t$ is kept implicit. Note the introduction of the scaling $1/\sqrt{\tau}$ and the variable $D=\sqrt{G^2\tau}$ in the state dynamical equations. These will be used to homogenize over the noise, as explained in  Refs.~\cite{Pavliotis2008,HorsthemkeBook2006,BoninTraversaSmallTau}. We will find that $D$ is proportional to a diffusion constant in the final master equation, and should thus be interpreted as a physical parameter. We will therefore consider $\tau$ and $D$, rather than $\tau$ and $G$, to be independent variables from here on. 

The transition probability density, $\mathbb{T}(z_t,\xi_t,t|z_s,\xi_s,s)$ is defined for $t>s$ as the probability of finding the quantum state and the noise to be in an infinitesimal region around $(z_t,\xi_t)$ at time $t$, conditional on sharp initial conditions at $(z_s,\xi_s)$ at time $s$. Notice that $\mathbb{T}(z_2,\xi_2,t|z_1,\xi_1,t)=\delta(z_2-z_1)\delta(\xi_2-\xi_1)$, and that the Chapman-Kolmogorov equation $\mathbb{T}(z_3,\xi_3,t_3|z_1,\xi_1,t_1)=\int d\xi_2 dz_2\,\mathbb{T}(z_3,\xi_3,t_3|z_2,\xi_2,t_2)\,\mathbb{T}(z_2,\xi_2,t_2|z_1,\xi_1,t_1)$ holds for all times because the combined state and noise dynamics is a Markovian process. 

Under these conditions, we can define the so-called Kolmogorov Backward equation (see Appendix.\ref{app:FP_BK}) describing the evolution of likelihoods for having a probability density $\rho(z,\xi,t)$ at time $t<T$ given a terminal condition $\rho(z_T,\xi_T,T)$\cite{oksendal2003stochastic,Pavliotis2008,gardiner2004handbook}. This uses the relation $\rho(z,\xi,t)=\int dz_T\,d\xi_T \, \rho(z_T,\xi_T,T) \, \mathbb{T}(z_T,\xi_T,T|z,\xi,t)$ and results in the system of equations:
\begin{align}
-\partial_t \rho(z,\xi,t) &= \left(\Lambda_J + \frac{1}{\sqrt{\tau}} \Lambda_G + \frac{1}{\tau} \Lambda_{\xi}\right) \rho(z,\xi,t) \notag \\
\text{with}~~~ \Lambda_J &:= J_z \partial_z \notag \\
\Lambda_G &:= \xi\,\mathcal{G}_z\partial_z \notag \\
\Lambda_{\xi} &:= -\xi\partial_\xi +\frac{\tau g^2(\xi)}{2}\partial^2_\xi
\label{Eq:Op defs BK}
\end{align}
Here, we used that $\tau g^2(\xi)$ is independent of $\tau$ for both the OU and SBM noise processes. Note that while in Eq.~\eqref{Eq:ZPair} the functions $J_z$ and $G_z$ are interpreted as functions of time through their dependence on the stochastic process $z_t$, in Eq.~\eqref{Eq:Op defs BK} they are interpreted directly as functions of $z$ itself (see also Appendix~\ref{app:Noise_Characterization}). The differential operator $\Lambda_J$ depends on $z$ alone, while $\Lambda_G$ depends on $z$ and $\xi$, and $\Lambda_\xi$ depends on $\xi$ alone. This means the quantum state dynamics depends on the noise, but the noise dynamics is independent of the state, as noted before. Notice that the dynamics of the noise can be seen to be faster than the state dynamics in the small-$\tau$ limit, due to the coefficient of $\Lambda_\xi$. We do not, however, assume the noise to always be in its steady-state or late-time distribution $\rho^{\infty}(\xi)$, since this would yield a trivial, effectively noiseless situation that does not capture the noise-induced time-dependence of the probability density~\cite{HorsthemkeBook2006,Pavliotis2008}.

~\\
\emph{Multi-scale Noise Homogenization} ---
The Kolmogorov Backward equations allow a systematic construction of the probability density $\rho(z,\xi,t)$ by expressing it as a power series in the small parameter $\tau^{1/2}$:
\begin{align}
\rho(z,\xi,t)= \sum^{\infty}_{k=0} \tau^{k/2} \rho_k = \rho_0 + \sqrt{\tau}\rho_1 + \mathcal{O}[\tau]
\label{Eq:rho_tau_exp}
\end{align}
Using Eq.~\eqref{Eq:Op defs BK} we then obtain separate equations for each power of $\tau$:
\begin{align}
    &\tau^{-1}: & -\Lambda_{\xi}\rho_0 &=0 \notag\\
    &\tau^{-1/2}: & -\Lambda_{\xi}\rho_1& =\Lambda_G\rho_0\notag \\
    &\tau^{0}: & -\Lambda_{\xi}\rho_2&=(\Lambda_J+\partial_t)\rho_0 + \Lambda_G\rho_1 \notag\\
    &\tau^{p/2}: & -\Lambda_{\xi}\rho_{p+2} &=(\Lambda_J+\partial_t)\rho_p + \Lambda_G\rho_{p+1} 
    \label{Eq:Pert4}
\end{align}
Here, the expression in the final line is valid for $p>-2$ with the condition that $\rho_{c}=0$ for all ${c<0}$. Since $\Lambda_{\xi}$ is a differential operator only in $\xi$, the first line implies that $\rho_0$ depends only on the state vector and time, $\rho_0 := \rho_0(z,t)$. 

The further system of equations consists of expressions of the form $\Lambda_{\xi}\rho_k=f_k(z,\xi)$. Solutions to these expressions can be constructed using Fredholm's Alternative theorem~\cite{Pavliotis2008,HorsthemkeBook2006,BoninTraversaSmallTau}. While applicable in more general settings, in a finite dimensional Hilbert space, this theorem states that
the operator equation $A {x_i}= {y_i}$ with operator $A$ and vectors ${x_i}$ and ${y_i}$  has a solution if $n\cdot y_i=0$ for vectors $n$ in the null subspace of the adjoint operator $A^*$ (i.e. $A^{*}{n}=0$). In other words, ${y_i}$ must be orthogonal to the null subspace of $A^{*}$ for a solution of $A {x_i}= {y_i}$ to exist, and the solvability condition ${n}\cdot {y_i}=0~\forall i$ can then be used to explicitly construct the solution.

This procedure can be applied to the expressions of the form $\Lambda_{\xi}\rho_k=f_k(z,\xi)$ in Eq.~\eqref{Eq:Pert4}, because the adjoint operator $\Lambda_{\xi}^*$ describes the forward evolving Fokker-Planck-Kolmogorov equations of the noise alone~(see Appendix~\ref{app:Noise_Characterization},  \ref{app:FP_BK} and Refs.~\cite{Risken1996,gardiner2004handbook,HorsthemkeBook2006,Pavliotis2008}). Since both SBM and OU processes admit a long time steady state probability distribution, $\rho^{\infty}(\xi)$, we have $\Lambda_{\xi}^{*}\rho^{\infty}(\xi)=0$, which defines the null subspace of the adjoint operator. Furthermore, orthogonality of a generic function $y(z,\xi)$ with the null subspace is then given by the (function space) inner product, $\int d\xi\,\rho^\infty(\xi) y(z,\xi) =\mathbb{E}^{\infty}_{\xi}[y(z,\xi)]=0$.

Written this way, the solvability condition for the $O(\tau^{-1/2})$ equation becomes $\mathbb{E}^{\infty}_{\xi}[\Lambda_G\rho_0]=0$. This can be recognised as a condition on the nature of the coupling of the state and noise dynamics, and is known as the centering condition~\cite{Pavliotis2008,HorsthemkeBook2006,BoninTraversaSmallTau}. In this case, the ergodicity properties of the OU and SBM processes guarantees the centering condition to hold, since $\mathbb{E}^{\infty}_{\xi}[\xi]=0$. 
We thus find that a solution of the first order component $\rho_1$ exists, and that given $\rho_0$, it can be written as $\rho_1 = - \Lambda^{-1}_{\xi}\Lambda_G\rho_0$. Substituting this expression for $\rho_1$ in the equation for $\Lambda_{\xi}\rho_2$, we then obtain the next solvability condition:
\begin{align}
    \mathbb{E}^{\infty}_{\xi}\left[ (\Lambda_J+\partial_t)\rho_0 -\Lambda_G\Lambda^{-1}_{\xi}\Lambda_G\rho_0 \right]=0
    \label{Eq:Final_Solvbl0}
\end{align}
Only the final term on the right hand side depends on $\xi$ so that this expression reduces to:
\begin{align}
    (\Lambda_J+\partial_t)\rho_0 -\mathbb{E}^{\infty}_{\xi}\left[\Lambda_G\Lambda^{-1}_{\xi}\Lambda_G\rho_0 \right]=0
    \label{Eq:Final_Solvbl}
\end{align}
The remaining expectation value can be computed using the so-called cell problem ansatz~\cite{Pavliotis2008,HorsthemkeBook2006,BoninTraversaSmallTau}. This starts from noticing that $\Lambda_G = \xi\,\mathcal{G}_z\partial_z$, so that $\Lambda^{-1}_{\xi}\Lambda_G\rho_0$ can be written as $\Lambda^{-1}_{\xi} (\xi F(z))$ for some unknown function $F(z)$. The cell problem ansatz now assumes that there is a function $r(z,\xi)$ such that $\Lambda^{-1}_{\xi} (\xi F(z)) = r(z,\xi)$. If this function exists, it implies $\Lambda_{\xi} r(z,\xi) =  \xi F(z)$. Finding a function $r(z,\xi)$ that satisfies this relation is then equivalent to finding an expression for $\Lambda^{-1}_{\xi} (\xi F(z))$. In our case, $\Lambda_{\xi} = -\xi\partial_\xi +\frac{\tau g^2(\xi)}{2}\partial^2_\xi$, and the function $r(z,\xi)=-\xi F(z)$ solves the differential equation $\Lambda_{\xi} r(z,\xi) =  \xi F(z)$ for both the OU and SBM noise processes. Inserting this into the expectation value finally yields:
\begin{align}
\mathbb{E}^{\infty}_{\xi}\left[\Lambda_G\Lambda^{-1}_{\xi}\Lambda_G\rho_0 \right]=-\mathbb{E}^{\infty}_{\xi}\left[ \xi^2\right]\mathcal{G}_z  \partial_z \mathcal{G}_z  \partial_z \rho_0
\label{Eq:Final_Solvbl2}
\end{align}

Together with Eq.~\eqref{Eq:Final_Solvbl}, Eq.~\eqref{Eq:Final_Solvbl2} yields the solvability condition for the dynamics up to order $\tau^0$. That is, the stochastic variable has been homogenized over to order $\tau^0$, yielding an expression for the time evolution of probabilities for $z$ alone:
\begin{align}
\left(\partial_t+ \Lambda_J +\mathbb{E}^{\infty}_{\xi}\left[ \xi^2\right]\mathcal{G}_z \partial_z \mathcal{G}_z \partial_z\right) \rho_0(z,t) =0
\label{eq:stratBK}
\end{align}
This equation is the solvability condition on the system of Eq.~\eqref{Eq:ZPair}, and the solutions of Eq.~\eqref{eq:stratBK} represent are weakly equivalent (agreeing on ensemble averaged probabilities $\rho(z,t)$) to those of Eq.~\eqref{Eq:ZPair} in the limit $\tau\to 0$, when the probability distribution $\rho(z,\xi,t)$ equals $\rho_0(z,t)$. Furthermore, Eq.~\eqref{eq:stratBK} constitutes a Kolmogorov Backward equation (in the Stratonovich representation) for the time evolution of the probabilities of $z$ alone (after homogenizing over the noise)~\cite{gardiner2004handbook,Hanggi94,oksendal2003stochastic}. By inspection, it can be recognized that the same Kolmogorov Backward equation would be (straightforwardly) obtained from the white-noise driven process~\cite{Pavliotis2008,HorsthemkeBook2006,BoninTraversaSmallTau, gardiner2004handbook,Risken1996,oksendal2003stochastic}:
\begin{align}
d z_t = J_z dt + \sqrt{2\mathbb{E}^{\infty}_{\xi}\left[ \xi^2\right]}\,\mathcal{G}_z \circ dW_t
\label{Eq:zStrat}
\end{align}
Here, $J_z$ and  $\mathcal{G}_z$ are defined as in Eq.~\eqref{Eq:ZPair}, and the symbol $\circ$ denotes the Stratonovich (rather than It\^o) product. The fact that Eqs.~\eqref{Eq:ZPair} and~\eqref{Eq:zStrat} imply the same Kolmogorov Backward equations in the $\tau\to 0$ limit, means that they are weakly equivalent in that limit. That is, the processes defined by Eqs.~\eqref{Eq:ZPair} and~\eqref{Eq:zStrat} yield the same probability distribution functions $\rho(z,t)$ for the quantum state amplitudes $z$ after averaging over an ensemble of noise (or Wiener process) realisations, even though individual trajectories differ. Indeed, such weak equivalence is not unexpected, in light of the Wong-Zakai theorems showing (weak) convergence of random ordinary differential equations to white noise driven stochastic differential equations~\cite{WongZakai1965relation,wongZakai1965convergence,WongZakai1969,WongZakaiReview}. Explicitly writing out Eq.~\eqref{Eq:zStrat} in terms of the state vector dynamics yields the (norm preserving) stochastic differential equation:
\begin{align}
d|\psi_t\rangle= &\frac{J}{2} \langle\hat{\sigma}_3\rangle_t \left(\hat{\sigma}_3-\langle\hat{\sigma}_3\rangle_t\right)   \ket{\psi_t}\,dt \nonumber\\
&+\frac{\mathcal{D}}{2} \left(\hat{\sigma}_3-\langle\hat{\sigma}_3\rangle_t\right)  \ket{\psi_t}\circ dW_t
\label{Eq:SmalltauFinal}
\end{align}
Here, we introduced the effective diffusion parameter, $\mathcal{D}=\sqrt{2D^2\mathbb{E}^{\infty}_{\xi}[ \xi^2]}$. Note that $\mathbb{E}^{\infty}_{\xi}[ \xi^2]=1$ for OU and $\mathbb{E}^{\infty}_{\xi}[ \xi^2]=1/3$ for SBM (see Appendix~\ref{app:Noise_Characterization}). The white-noise driven process of Eq.~\eqref{Eq:SmalltauFinal} constitutes approximate solutions of the coupled noise and state dynamics of Eq.~\eqref{Eq:ZPair} in the limit of vanishing correlation time (weak equivalence); it is similar to, but distinct from the SSE of Eq.~\eqref{Eq:Suv_compare}; and it can be used to construct master equations for the state dynamics in the $\tau\to 0$ limit. 

Notice that in the context of white noise driven quantum state reduction processes, such as SSE and CSL, the Fokker-Planck-Kolmogorov system for the quantum state dynamics alone, may be straightforwardly obtained~\cite{Pearle_89_PRA,Pearle_76,Pearle1984_bookRef}. In Appendix~\ref{app:FDR_Emergent_Born_Pearle_comarison}, we compare this approach to the analysis of SUV dynamics, and show that the FPK equations of Eq.~\eqref{Eq:SmalltauFinal} adhere to the same set of constraints as the FPK equations obtained from the SSE or CSL models.

\subsection{Master Equations}\label{sec:4B}
Starting from the noise-homogenized dynamical equations for the state evolution in Eq.~\eqref{Eq:SmalltauFinal}, we can convert the Stratonovich product into It\^o form using the conversion formula $X_t\circ dW_t=X_t dW_t+\frac{1}{2}dX_t dW_t$\cite{oksendal2003stochastic}. Using $X_t=\frac{\mathcal{D}}{2}\big(\hat{\sigma}_3-\langle\hat{\sigma}_3\rangle_t\big)\ket{\psi_t}$ this yields:
\begin{align}
d|\psi_t\rangle= \frac{J}{2}&\langle\hat{\sigma}_3\rangle_t\left(\hat{\sigma}_3-\langle\hat{\sigma}_3\rangle_t\right)|\psi_t\rangle\,dt\nonumber\\
&+\hat{C}_s|\psi_t\rangle dt+ \frac{\mathcal{D}}{2}\left(\hat{\sigma}_3-\langle\hat{\sigma}_3\rangle_t\right)|\psi_t\rangle\, dW_t.
\label{Eq:ItoWhiteOG}
\end{align}
Here, the multiplicative noise term is interpreted in It\^o's sense and $\hat{C}_sdt$ is the Stratonovich drift correction term, which reads:
\begin{align}
\hat{C}_s=\frac{\mathcal{D}^2}{4} \left( \frac{1}{2}\left(\hat{\sigma}_3-\langle\hat{\sigma}_3\rangle_t\right)^2 -\left[\langle\hat{\sigma}_3^2\rangle_t -\langle\hat{\sigma}_3\rangle_t^2 \right]\right)
\label{Eq:Strat Correction}
\end{align}
Note that Eqs.~\eqref{Eq:ItoWhiteOG} and~\eqref{Eq:Strat Correction}, which describe a white-noise driven process weakly equivalent to the SUV dynamics in the $\tau\to 0$ limit, are still different from the SSE. 

From the dynamics in It\^o form, we can compute the evolution of the pure state statistical operator $\hat{\mathbb{P}}_t=\ket{\psi_t}\bra{\psi_t}$, by employing It\^o's lemma, $d\hat{\mathbb{P}}_t=|d\psi_t\rangle\langle\psi_t|+|\psi_t\rangle\langle d\psi_t|+\mathbb{E}_{\psi}\left[|d\psi_t\rangle\langle d\psi_t|\right]$. In contrast to previous sections discussing colored noise, the quadratic variation in the white-noise driven evolution is non-vanishing. From Eqs.~\eqref{Eq:ItoWhiteOG} and~\eqref{Eq:Strat Correction} we thus find:
\begin{align}
 d\hat{\mathbb{P}}_t&= \frac{dt}{2}\left[(J-\mathcal{D}^2) \left\{\left( \hat{\sigma}_3\langle\hat{\sigma}_3\rangle_t-\langle\hat{\sigma}_3\rangle_t^2\right),\mathbb{\hat{P}}_t\right\}\right. \,     \nonumber\\
&\left. +\frac{\mathcal{D}^2}{2}\hat{\sigma}_3\mathbb{\hat{P}}_t\hat{\sigma}_3-\frac{\mathcal{D}^2}{4} \left\{\hat{\sigma}_3^2 , \mathbb{\hat{P}}_t \right\} \right]  \nonumber\\
&+\frac{\mathcal{D}}{2}\left\{\left(\hat{\sigma}_3-\langle\hat{\sigma}_3\rangle_t\right),\mathbb{\hat{P}}_t\right\}dW_t \label{Eq:PurestateAlmostMaster}
\end{align}

As explained in Sec.~\ref{sec:3B} the noise averaged expectation value of the pure state statistical operator is the usual quantum mechanical density matrix, $\mathbb{E}_{\psi}[\hat{\mathbb{P}}_t]=\hat{\rho}_t$, thus we now consider an ensemble of pure state evolutions for different realisations of the stochastic dynamics and obtain an expression for the time evolution of the density matrix. This corresponds to averaging over an ensemble of  measurements. Because the multiplicative coefficient in front of the $dW_t$ term in Eq.~\eqref{Eq:PurestateAlmostMaster} is interpreted in It\^o's convention, it will average to zero: $\mathbb{E}[F_t dW_t]=0$ for generic functions $F_t$, which is not true for a Stratonovich product (i.e. $\mathbb{E}[F_t\,\circ dW_t]$ is not generically zero)~\cite{Bassi_03_PhyRep,oksendal2003stochastic,gardiner2004handbook}. 
We are then left with:
\begin{align}
 \frac{\partial\hat{\rho}_t}{\partial t} &= \frac{1}{2}\left[(J-\mathcal{D}^2) \,\mathbb{E}_{\psi}\left[\left\{\left( \hat{\sigma}_3\langle\hat{\sigma}_3\rangle_t-\langle\hat{\sigma}_3\rangle_t^2\right),\hat{\mathbb{P}}_t\right\}\right]\right. \,     \nonumber\\
&\left. +\frac{\mathcal{D}^2}{2}\hat{\sigma}_3\hat{\rho}_t\hat{\sigma}_3-\frac{\mathcal{D}^2}{4} \left\{\hat{\sigma}_3^2 , \hat{\rho}_t \right\} \right]
\label{Eq:PurestateAlmostMaster2}
\end{align}
In this expression, all non-linearites are multiplied by the factor $(J-\mathcal{D}^2)$. In order to obtain a linear quantum (Markov) semi-group description of the density matrix evolution, guaranteeing Born's Rule, we are then forced to demand $J=\mathcal{D}^2$. The realisation that there should be a relation between $J$ and $\mathcal{D}$ in the SUV dynamics is similar to the relation between model parameters required in the theoretical description of the velocity of a particle undergoing damped Brownian motion~\cite{Einstein1905, Smoluchowski1906, Kubo_1966, gardiner2004handbook, Risken1996,Zwanzig2001}. In Eq.~\eqref{Eq:PurestateAlmostMaster}, as in that classic case, it is found that the fluctuation ($\mathcal{D}$) and the dissipative ($J$) terms, which were at first assumed to be independent physical effects, in fact must be related to one another if the system is to obtain a well-defined ensemble-averaged equilibrium at long times. In the present case, the final state statistics are encoded in Born's rule, whereas for the velocity of a particle undergoing damped Brownian motion the equilibrium configuration is imposed to be the Maxwell-Boltzmann distribution. In both cases, the requirement of reaching equilibrium exposes a physical relation between model parameters, and the condition $J=\mathcal{D}^2$ can be seen as an application of the fluctuation-dissipation theorem to the colored-noise driven SUV model in the vanishing correlation time limit. 

Imposing the fluctuation-dissipation relation, the dynamics of the density matrix is given by:
\begin{align}
\frac{\partial\hat{\rho}_t}{\partial t} = -i\,\left[\hat{H}_t\,,\,\hat{\rho}_t\right]+
\frac{\mathcal{D}^2}{4} \left(\hat{\sigma}_3\hat{\rho}_t\hat{\sigma}_3- \hat{\rho}_t \right).
\label{Eq:MasterQbit}
\end{align}
This final expression is a Gorini-Kossakowski-Sudarshan-Lindblad (GKSL) type master equation for the evolution of the density matrix. Notice that it  is linear in $\hat{\rho}$ and  does not allow for super-luminal signalling~\cite{Gisin:1989sx,Bassi2015}. In fact, Eq.~\eqref{Eq:MasterQbit} is of the same form as the Linbladian evolution obtained for a dephasing qubit~\cite{Breur_Petr02}, and for other objective collapse theories such as the SSE~\cite{Bassi_03_PhyRep}. This implies that during time evolution, the off-diagonal (interference) terms of the density matrix decay while the diagonal elements are preserved, thus guaranteeing the emergence of Born's rule.  

Notice however that this does \emph{not} imply SUV is equivalent to or indistinguishable from dephasing or alternative objective collapse theories. Rather, the density matrix encountered in the dephasing Linbladian, $\rho_t$ in Eq.~\eqref{Eq:MasterQbit} is the full density matrix for the combined system and measurement device averaged over an ensemble of measurements for the closed system-device setup, rather than an ensemble of environments in decoherence.Moreover, the underlying pure state dynamics was shown above to differ significantly from the SSE. Although there is a weak equivalence between the ensemble averages of SUV, SSE, and dephasing, their predictions for the actual pure state evolution in each individual measurement are fundamentally different. This can be exploited to construct explicit protocols allowing the predictions of SUV to be distinguished from those of other theories in mesoscale experiments~\cite{Wezel_2012}. 

Although the dynamics of Eq.~\ref{Eq:key} takes pure states to pure states, the entropy within an ensemble of measurements increases over time, as pure ensembles become mixed. The role of entropy production in objective collapse models~\cite{Entropy_production_article}, as well as the irreversible (semi-group) dynamics of Eq.~\eqref{Eq:MasterQbit} with Born's statistics emerging only in the presence of a fluctuation-dissipation relationship, present possible connections to the notions of irreversibility encountered in foundations of non-equilibrium statistical theory~\cite{BISHOP20041_AB0,PRIGOGINE1999_AB1,Morozov1998_Zubarev_AB2,Jou1996_zubarev_AB3,LuzziVasconcellosRamos2006_AB4,Eu1998_AB5}. We leave this potential route of enquiry for future work.

\section{Conclusions}\label{sec:5}
In this article, we proposed a general Markovian framework for describing dynamical quantum state reduction as a consequence of spontaneous unitarity violations. It was shown to yield a class of SUV models obeying normalized state evolution with vanishing quadratic variation, driven by a physical noise field with state-independent dynamics. We gave formal expressions for the modified Von-Neumann Liouville equations describing pure state evolution, which can be numerically evaluated for any particular implementation of the noise. 

Focusing on two specific noise processes, Ornstein-Uhlenbeck and Spherical Brownian Motion, we analysed the dynamics of a two-state superposition. In the fast-noise limit, we homogenized over the noise to obtain a Kolmogorov Backward equation of the quantum state dynamics alone, which shows the equivalence of ensemble averaged predictions of the SUV dynamics with those of a white-noise driven process. Notice that the white noise limit does not require fundamentally white noise. It is obtained whenever the noise fluctuates on time-scales shorter than all other time-scales in the dynamics. For the white-noise limit, we explicitly formulated the master equation describing the time dependence of the ensemble-averaged density matrix corresponding to an ensemble of measurements. The master equation is shown to be of the GKSL (i.e. a linear quantum Markov semi-group) type, guaranteeing adherence to Born's rule and the absence of super-luminal communication, if the parameters obey a fluctuation-dissipation relation. 

We thus establish colored noise driven spontaneous unitarity violations as a class of models for dynamical quantum state reduction. They complement existing objective collapse theories, but differ from them in several crucial aspects.

First, only objects that already break a spatial or internal symmetry are susceptible to SUV, and the final states of the quantum state reduction are always the states of broken spatial symmetry. The measurement device thus determines the basis in which measurement outcomes are registered. 

Secondly, spontaneous symmetry breaking in general, and SUV in particular, requires a physical perturbation. The stochastic dynamics encountered in SUV models must therefore be caused by a physical (non-Hermitian) field rather than a stochastic re-definition of the time increment in quantum state evolution. Constructing models this way, using physical (correlated) noise has the additional effect that the energy, defined as the Noether charge associated with time translations, always equals the expectation value of the unmodified Hamiltonian.

Moreover, the diverging susceptibility to physical perturbations leading to SUV (or any other type of spontaneous symmetry breaking) implies that even vanishingly small perturbations will dominate the dynamics of sufficiently large objects, while at the same time having vanishing effect at the microscopic scale. We leave identifying possible origins of the unitarity-breaking perturbation to future research.

Finally, because the noise driving the quantum state reduction evolves independently of the quantum state, Born's statistics for ensemble averaged measurement outcomes are not implied by any property of the noise.
Instead, both the Born distribution and the absence of superluminal signalling are seen to emerge as a consequence of a fluctuation-dissipation relation being upheld in the SUV dynamics.

\subsection*{Acknowledgement}
The authors gratefully acknowledge illuminating discussions with H. Maassen, K. R. Parthasarathy, L. Diosi, A. Bassi, J. Veenstra, L. Mertens, L. Lenstra, and Sumita Mukherjee.


\appendix
\section{Characterization of Stochastic Processes}\label{app:Noise_Characterization}
In this appendix, we provide a brief reminder of the formal description of stochastic processes in general, and of the two types of real-valued stochastic processes employed in the main text in particular. A more complete treatment is available in Refs.~\cite{oksendal2003stochastic,revuz1999continuous,Pavliotis2008,HorsthemkeBook2006}, while more physically motivated characterizations may be found in Refs.~\cite{Breur_Petr02,gardiner2004handbook,Risken1996}. 

We will use It\^o's calculus for continuous-time stochastic processes, assuming an underlying filtered probability space $(\Omega,\mathcal{F},\mathcal{F}_t,\mathbb{Q})$, endowed with a natural filtration. Here, $\Omega$ is the sample space containing all possible events (of the stochastic process). $\mathcal{F}$ denotes the space of all collections of events (i.e. all possible subsets of $\Omega$), and $\mathbb{Q}$ is the probability measure associating probabilities with abstract events. The total probability obeys $\mathbb{Q}[\mathcal{F}]=1$. The space of all possible collections of events, $\mathcal{F}$, is a $\sigma$-algebra \cite{oksendal2003stochastic}. The (natural) filtration, $\{\mathcal{F}_{0\leq t<\infty}\}\subset\mathcal{F}$, is a family of sub-$\sigma$-algebras with a causal ordering. It is determined by all possible histories of a process leading up to time $t$, such that for $0<t_1<t_2..<t$, we have $\mathcal{F}_{0}\subset\mathcal{F}_{t_1}\subset\mathcal{F}_{t_2}..\subset\mathcal{F}_{t}\subset\mathcal{F}$. Here, each $\mathcal{F}_s$ is the set of all possible collections of events forming a history leading up to time $0<s<t$.  The process $Z_t$ ($t>0$) is termed an adapted process if $Z_t$ is $\mathcal{F}_t$ measurable, i.e. $Z_t$ admits a well defined probability for all possible histories leading up to time $t$, without `looking' into the future (non-anticipating). Intuitively, this condition implies considering causal processes only, and all processes considered in this article are adapted. 

An It$\mathrm{\hat{o}}$ process $Z_t$ is an adapted stochastic process, which can be written in the from of either a stochastic integral or a stochastic differential equation (SDE):
\begin{align}
Z_t &= Z_0 + \int^{t}_0A(Z_s,s)ds + \int^{t}_0 B(Z_s,s) dW_s \notag \\
\Leftrightarrow ~ dZ_t &= A(Z_t,t)dt +B(Z_t,t) dW_t
\label{Eq:App_Ito}
\end{align}
Here, the drift term, $A(Z_t,t)$, and the diffusion term, $ B(Z_t,t)$, are assumed to be both smooth functions and to be $\mathcal{F}_t$ measurable for all $t$ (for a more rigorous discussion see Refs.~\cite{oksendal2003stochastic,Pavliotis2008}). The symbol $W_t$ represents the standard Wiener (Brownian) process with independent increments and continuous paths, defined by the properties $W_0=0$ and $W_{t+dt} - W_{t}\sim \mathbb{N}(0, \sqrt{dt})$. Its mean is always zero, $\mathbb{E}_\mathbb{Q}[W_t]=0$, and its variation equals $\mathbb{E}_\mathbb{Q}[W^2_t]=t$. Furthermore, the Gaussian increments being independent imply the expectation values $\mathbb{E}_\mathbb{Q}[dW_t]=0$ and $\mathbb{E}_\mathbb{Q}[dW_tdW_s]=0$ for $t\neq s$ while $\mathbb{E}_\mathbb{Q}[dW_t^2]=dt$. These give rise to the informal It$\mathrm{\hat{o}}$ multiplication rules $dt^2=dW_t dt=0$  and $dW_t^2=dt$. Finally, notice that (smooth) functions of It\^o processes are also It\^o processes themselves. 

The evolution of probabilities within an ensemble of stochastic processes can be described by its transition probability density $\mathbb{T}(z_f,t_f|z_0,t_0)$, which evolves either forward in time from given initial conditions according to the Fokker-Planck-Kolmogorov (FPK) equations, or backward in time from given terminal conditions according to the Kolmogorov Backward (KB) equations. Both can be written in terms of an infinitesimal generator $\Lambda$ (also called characteristic operator), with $\Lambda(z,t)$ propagating the transition probability density in the backward direction and its adjoint $\Lambda^{*}(z,t)$ propagating it forward:
\begin{align}
\frac{\partial}{\partial t}\mathbb{T}(z_f,t_f|z,t) &= -\Lambda(z,t)\mathbb{T}(z_f,t_f|z,t) \notag \\
\frac{\partial}{\partial t}\mathbb{T}(z,t|z_0,t_0) &= \Lambda^{*}(z,t)\mathbb{T}(z,t|z_0,t_0)
\label{eqApp:FP}
\end{align}
We may integrate out the terminal (initial) conditions from the Kolmogorov Backward (Fokker-Planck-Kolmogorov) equations to find the one-time probability density $\rho(z,t)$, using $\int \mathbb{T}(z_T,t_T|z,t)\rho(z_T,t_T)dz_T = \rho(z,t)$ or $\int \mathbb{T}(z,t|z_0,t_0)\rho(z_0,t_0)dz_0 = \rho(z,t)$. The one-time probability density can then be seen to also be a solution of Eq.~\eqref{eqApp:FP}. The generator $\Lambda$ and its adjoint can be written explicitly in operator form for the It\^o process defined in Eq.~\eqref{Eq:App_Ito} as:
\begin{align}
\Lambda(z,t) &= A(z,t)\frac{\partial}{\partial z} + \frac{B^2(z,t)}{2}\frac{\partial^2}{\partial z^2 } \notag \\
\Lambda^{*}(z,t) &= -\frac{\partial}{\partial z} \,A(z,t)\, + \frac{\partial^2}{\partial z^2 } \frac{B^2(z,t)}{2}\label{Eq.App.FPKOp}
\end{align}
Note that in deriving these forms, an integration by parts is used to obtain one from the other. The associated surface term vanish assuming compact support. Further details may be found in Refs.~\cite{Pavliotis2008,Risken1996}.

\subsection{Ornstein Uhlenbeck}
The Ornstein Uhlenbeck (OU) process is the only stochastic process defined on the sample space $\Omega=\mathbb{R}$, which is Markovian, Gaussian, and stationary, up to scaling or time translation~\cite{Risken1996,gardiner2004handbook}. 
It is defined as:
\begin{align}
dX_t = - X_t \,\frac{dt}{\tau} + \sqrt{\frac{2}{\tau}} dW_t
\end{align}
Here, $\tau$ is the correlation time as well as the average time required by the process to return to the mean (with the mean being $X=0$ in this case). The stochastic trajectory $X_t$ admits a closed formal solution for sharp initial condition $X_0$ of the form:
\begin{align}
X_t = X_0 e^{-t/\tau} +  \sqrt{\frac{2}{\tau}} e^{-t/\tau} \int_0^t e^{s/\tau}dW_s
\label{Eq:OU_gen_sol}
\end{align}
By It$\mathrm{\hat{o}}$'s Isometry, $X_t \sim \mathbb{N}\left[ X_0 e^{-t/\tau} ,  \sqrt{\left(1 - e^{-2t/\tau}\right)} \right]$, i.e. the probability distribution for $X_t$ is a normal distribution with mean $\mathbb{E}[X_t]= X_0 e^{-t/\tau}$ and variance $\mathbb{E}[X_t^2]-\mathbb{E}^2[X_t]=\left(1 - e^{-2t/\tau}\right)$. The (temporal) auto-correlation function is given by:
\begin{align}
    \mathbb{E}[X_t X_s] -\mathbb{E}[ X_t]\mathbb{E}[X_s] &= (e^{- |t-s|/\tau} - e^{-(t+s)/\tau})
\end{align}
For for long times, $t,s \rightarrow \infty$, but keeping $|t-s| \equiv \delta$ constant, the auto-correlation reduces to $e^{-\delta/\tau}$. 

The OU process is also fully characterized by its transition probability densities. For a stochastic process, the probability density to have the value $X_t=x$ at time $t$, conditioned on the process having a value $x_0$ at $t=t_0$, is expressed by the transition probability density $\mathbb{T}(x,t|x_0,t_0)$. It follows the FPK equation given by Eqs.~\eqref{eqApp:FP} and~\eqref{Eq.App.FPKOp}. In case of the OU process, the FPK equation yields:
\begin{align}
\partial_t \mathbb{T}\left(x, t|x_0, t_0\right) &= \frac{1}{\tau}\partial_x \left[ x \, \mathbb{T}\left(x, t|x_0, t_0\right) \right] \notag \\
&~~~~~~~~ +  \frac{1}{\tau}\partial^2_x\mathbb{T}\left(x, t|x_0, t_0\right)
\end{align}
For the Ornstein Uhlenbeck process, the transition probability density can be expressed in closed form for all times, by:
\begin{align}
\mathbb{T}\left(x, t|x_0,0\right) &= \frac{1}{\sqrt{2 \pi \sigma^2}}\, \exp\left[-\frac{(x-x_0e^{-t/\tau})^2}{2 \sigma^2}\right] \notag \\
\text{with} ~~~ \sigma^2 &=  \left( 1- e^{-2t/\tau}\right)
\end{align}
This expression is equivalent to Eq.~\eqref{Eq:OU_gen_sol}. 

At long times ($t\to\infty$), the process $X_t$ will `forget' about the initial value $X_0$ and reach a steady state probability density, $\rho^{\infty}(x)$, which is given regardless of the initial conditions by a zero-mean Gaussian distribution of unit variance:
\begin{align}
    \rho^{\infty}(x)=\frac{1}{\sqrt{2 \pi }}\, \exp\left[-\frac{x^2}{2 }\right]
\end{align}

\subsection{Spherical Brownian Motion}
The second stochastic process considered in the main text is Brownian motion confined to the surface of a sphere (SBM). It falls in the category of Pearson Diffusions and is also a Jacobi diffusion processes (see Refs.~\cite{CNOSPearsonDiffPaper2008,CNOStimelocalPearsontoJacobiAscione2021,CNOSJacobiDEMNI2009518} and references therein for further details). The generator of the SBM process is half the Laplace-Beltrami operator expressed on the sphere (see Chap.~8 in~\cite{oksendal2003stochastic}). For a sphere of unit radius, the Laplace-Beltrami operator is given by:
\begin{align}
    \Delta(\theta,\phi)  = \frac{1}{\sin\theta}\frac{\partial}{\partial \theta}\bigg(\sin\theta\frac{\partial}{\partial\theta}\bigg) + \frac{1}{\sin^2\theta}\frac{\partial^2}{\partial\phi^2}
    \label{Eq:Laplace Belt}
\end{align}
The KB  equation for the stochastic process on the surface of a sphere can then be written in terms of its transition probability density $\mathbb{T}(\theta,\phi,t|\theta_0,\phi_0,t_0)$ using Eq.~\eqref{eqApp:FP} and Eq.~\eqref{Eq.App.FPKOp} as:
\begin{align}
    \frac{\partial}{\partial t}\mathbb{T}(\theta_T,\phi_T,T|\theta,\phi,t)=\frac{-1}{2\tau}\Delta(\theta,\phi)\mathbb{T}(\theta_T,\phi_T,T|\theta,\phi,t) \notag
\end{align}
Integrating out the terminal condition to obtain the one-point probability density, $\rho(\theta,\phi,t)=\int\mathbb{T}(\theta_T,\phi_T,T|\theta,\phi,t)\rho(\theta_T,\phi_T)d\theta_T d\phi_T$, then yields: 
\begin{align}
\frac{\partial}{\partial t}\rho(\theta,\phi,t)&=\frac{-1}{2\tau}\left[\cot\theta\frac{\partial}{\partial \theta}+\frac{\partial^2}{\partial\theta^2} \right. \nonumber\\
&~~~~~~ +\left. \,\frac{1}{\sin^2\theta}\frac{\partial^2}{\partial\phi^2}\right]\rho(\theta,\phi,t)
\end{align}
This expression implies two It\^o processes, one for $\theta_t\in[0,\pi]$ and one for $\phi_t\in[0,2\pi]$. The corresponding stochastic differential equations read:
\begin{align}
    d\theta_t &= \frac{1}{2\tau} \cot(\theta_t) \,dt + \frac{1}{\sqrt{\tau}}dW^{\theta}_t
     \notag\\ 
    d\phi_t &=  \frac{1}{\sqrt{\tau}\sin(\theta_t)}dW^{\phi}_t 
    \label{Eq:theta_t}
\end{align}
This shows that the $\theta_t$ process is independent of the $\phi_t$ process, but not the other way around. The SBM noise considered in the main text is defined by the stochastic process $X_t=\cos\theta_t$. Its sample space is given by $\Omega=[-1,1]$. The drift term in Eq.~\eqref{Eq:theta_t} can be understood as a consequence of the fact that circles of constant latitude (constant $\theta$) on the sphere have the largest circumference at the equator, and tend to zero circumference at the poles. Stepping with equal probability in any direction from a generic point on the sphere is therefore more likely to evolve the latitude towards the equator than away from it. The effective drift causes the $X_t$ process to be temporally correlated. 

\begin{figure}[t]
\includegraphics[width=\columnwidth]{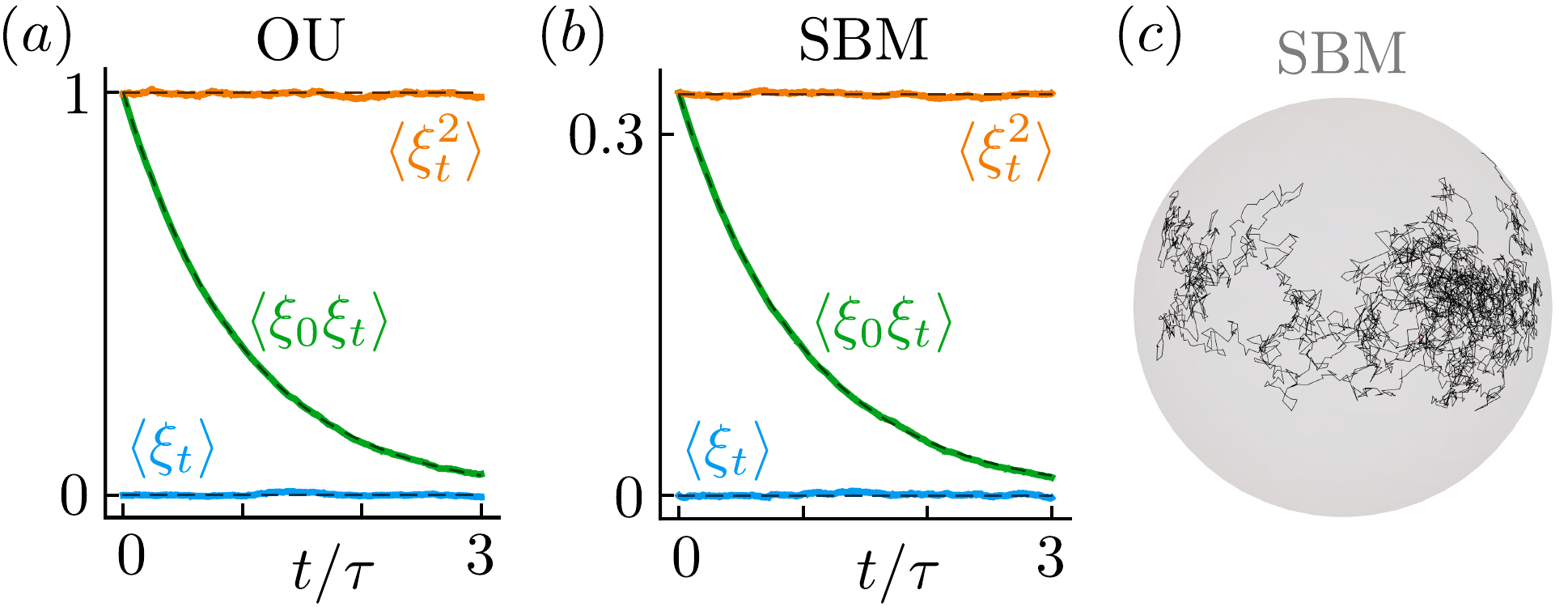}
\caption{\label{fig2} Noise dynamics. (a) The mean value $\langle \xi_t \rangle$ (blue line) of the noise in the OU process, starting from the initial condition $\langle \xi_0 \rangle= 0$ (blue dashed line), the correlation function $\langle \xi_0 \xi_t \rangle$ (green line) compared to the function $\exp(-t/\tau)$ (green dashed line), and the mean variation $\langle \xi_t^2 \rangle$ (orange line) as well as its steady state expectation value $\mathbb{E}^{\infty}[\xi_t^2]$ (orange dashed line). All averages are taken over 50000 realisations of the noise. (b) The same quantities for the SBM process, where the correlation function is now compared to $\exp(-t/\tau)/3$. Again, all averages are taken over 50000 realisations of the noise. (c) The trajectory of a single noise realisation in the SBM process, plotted on the surface of a sphere.
}
\end{figure}

The stochastic differential equations for $X_t$ may be deduced from those for $\theta_t$ using It\^o's lemma, which yields:
\begin{align}
 dX_t = - X_t \,\frac{dt}{\tau} + \sqrt{\frac{1-X_t^2}{\tau}}\, dW_t. \label{Eq:CNOS}
\end{align}
Unlike the OU process, the SBM process described by Eq.~\eqref{Eq:CNOS} has a multiplicative diffusion term interpreted in an It\^o sense. The dynamics of its transition probabilities, can be found from the FPK equation:
\begin{align}
 \partial_t \mathbb{T}\left(x, t|x_0, t_0\right) &=  \frac{1}{\tau} \partial_x \left[x \, \mathbb{T}\left(x, t|x_0, t_0\right)\right]  \nonumber\\ 
 &~~~ + \frac{1}{2 \tau} \partial^2_x \left[ (1-x^2) \mathbb{T}\left(x, t|x_0, t_0\right)\right]
\end{align}
The above FPK equation has an exact formal solution in the form of an infinite sum~\cite{CNOSWong1964}:
\begin{align}
\mathbb{T}\left(x, t|x_0, t_0\right) &= \frac{1}{2} \sum^{\infty}_{n=0} (2 n+1) (n!)^2 P_n(x) P_n(x_0) \nonumber\\
&~~~ \cross\exp \left[\frac{-n (n+1) |t-t_0|}{2 \tau }\right]
\label{Eq:CNOS sol}
\end{align}
Here, $P_n$ are the standard Legendre polynomials, given by the Rodrigues formula $P_n(x) = \frac{1}{n!\,\,2^n} \, \, \frac{d^n}{dx^n} \left(x^2-1\right)^n$.

At long times ($t-t_0\to\infty$), the SBM process `forgets' about its initial state, and approaches a steady state probability density $\rho^{\infty}(x)$. It can be found by taking the limit $t_0\rightarrow-\infty$ in Eq.~\eqref{Eq:CNOS sol}. Only the $n=0$ term then survives, and the steady state distribution is found to be completely flat:
\begin{align}
\rho^{\infty}(x)=\lim_{t_0\rightarrow-\infty}\mathbb{T}\left(x, t|x_0, t_0\right)=1/2
\end{align}

Likewise, the unconditioned mean $\mathbb{E}[x]$ and the expected value $\mathbb{E}[x|x_0]$ conditioned on sharp initial conditions, $x=x_0$ at $t=t_0$ may be obtained by direct integration of the transition probability density:
\begin{align*}
\mathbb{E}[x] &=\int^1_{-1}\int^1_{-1} x \, \mathbb{T}\left(x, t|x_0, t_0\right) \rho^{\infty}(x_0) \, dx_0 dx=0 \notag \\
\mathbb{E}[x|x_0] &= \int^1_{-1} x \, \mathbb{T}\left(x, t|x_0, t_0\right) dx = x_0 \exp\left[-\frac{t-t_0}{\tau}\right]
\end{align*}
In the final line, we used the identity $\int_{-1}^1  x P_n(x)dx \, = 2\sin(n\pi) / (n^2 +n-2)\pi$. Similar to the OU process, we find that the SBM process has exponentially decaying correlations.

Finally, the (temporal) auto-correlation $\mathbb{E}[x_t x_s]- \mathbb{E}[x_t]\mathbb{E}[x_s]$ of the SBM process starting from the stationary state distribution $\rho^{\infty}$ is given by:
\begin{align}
    &\int^1_{-1}\int^1_{-1} x_t \,x_s \, \mathbb{T}\left(x_t, t|x_s, s\right) \rho^{\infty}(x_s)\, dx_t dx_s - 0 \notag \\
    &= \frac{1}{3} \exp\left[-\frac{t-t_0}{\tau}\right]
\end{align}
Here again, we used the identity for $\int  x P_n(x)dx\,$ in the final line. Similar to the OU process, the SBM process has exponentially decaying mean and auto-correlation. Unlike the OU process, however, the SBM process is not Gaussian, and higher moments will not vanish.

\section{Quadratic Variations}
\label{app:QuadVar}
The quadratic variation is a generalization of the total variation, where instead of summing (absolute) changes, the squares of the changes are summed (see Refs.~\cite{oksendal2003stochastic,revuz1999continuous} for further details). The quadratic variation of a stochastic process $Z_t$, defined on an appropriate probability space, is expressed by discretizing time $t\in [0,T]$ into $n$ small intervals $\delta_k = t_k - t_{k-1}$ (possibly of different sizes), such that $T=\sum_{k=1}^n\delta_k$. Then, the quadratic variation is defined by:
\begin{align}
\mathcal{Q}_T[Z_t,Z_t]   &:= 
\lim_{n\rightarrow\infty}
\lim_{||\delta||\rightarrow0} \sum^{n}_{k=1} \left(Z_{t_k} - Z_{t_{k-1}}\right)^2 \notag \\
&:=\int^T \mathbb{E}_{\mathbb{Q}}\left[dZ_t^2\right]
    \label{Eq:QuadVarDef2}
\end{align}
Here, the first line is an operational definition using the partitioning, where $||\delta||$ is the largest partition size. The double limit yielding vanishing partition sizes throughout the interval $[0,T]$ is assumed to exist and the sum to converge in probability. In the second line, the expectation value (with measure $\mathbb{Q}$) is written explicitly to connect to the definitions used in the main text, while it is left implicit in the first line following convention in the mathematics literature.

For the specific case that $Z_t$ is the Wiener process $W_t$, the quadratic variation $\mathcal{Q}_t[W_t,W_t]=t$ follows directly from the the fact that the average (quadratic) variation accumulated by the Wiener process in any interval $\delta$ is equal to $\sqrt{\delta}$. Its differential is denoted $d\mathcal{Q}_t[W_t,W_t]=dt$, so that we find the usual rule employed in It\^o's calculus: $d\mathcal{Q}_t[W_t,W_t]=\mathbb{E}_{\mathbb{Q}}[dW_t^2]=dW_t^2=dt$. 

Next, consider an It\^o drift-diffusion process for $Z_t$, such that $dZ_t=Z_t -Z_{t-dt}=A(Z_{t-dt}) dt + B(Z_{t-dt}) dW_t$, with continuous and differentiable drift function $A(z)$ and diffusion $B(z)$. The quadratic variation for $Z_t$ is then defined as $\mathcal{Q}_T[Z_t,Z_t]=\int^T\mathbb{E}_{\mathbb{Q}}\left[ dZ_t^2\right]=\int_0^T B^2(Z_t)dt$. This implies the rules of calculus for functions of the stochastic process, $K_t=K(t,Z_t)$, obtain a correction (the so-called It\^o's correction) with respect to the usual rules of calculus:
\begin{align}
    dK_t = &dt\frac{\partial K(t,z)}{\partial t}|_{z=Z_t} + dZ_t\frac{\partial K(t,z)}{\partial z}|_{z=Z_t}  \nonumber\\
    &+\frac{d\mathcal{Q}_t[Z_t,Z_t]}{2}\frac{\partial^2 K(t,z)}{\partial z^2}|_{z=Z_t} \, \,,
\end{align}
 Here, $d\mathcal{Q}_t[Z_t,Z_t]=B(Z_t)^2dt$. This result is known as It$\mathrm{\hat{o}}$'s lemma (or an It$\mathrm{\hat{o}}$-Taylor expansion).
Clearly, if $\mathcal{Q}_t[Z_t,Z_t]=0$ happens to vanish for all $t\in \mathbb{R}^+$, the usual rules of calculus can be applied to functions $K(t,Z_t)$. Similarly, in the calculus of non-stochastic functions, quadratic variation does not play a major role, because it is zero for all continuously differentiable functions. 

Turning now to modified Schr\"odinger equations, consider a general modified evolution defined by:
\begin{align}
    d|\psi_t\rangle = dt&\left(\hat{A}(\psi_t)\xi_t+\hat{B}(\psi_t)f(\xi_t) \right. \notag \\
    &~~~~ + \left. \hat{C}(\psi_t)\hat{F}(\xi_t)\right)|\psi_t\rangle
    \label{Eq.A1}
\end{align}
This equation is formally defined by the integral:
\begin{align}
|\psi_t\rangle &= |\psi_0\rangle +\int^t_0 ds\, \hat{\mathcal{G}}_s|\psi_s\rangle \notag \\
\text{with} ~~ \hat{\mathcal{G}}_s &=\hat{A}(\psi_s)\xi_s +\hat{B}(\psi_s)f(\xi_s)+\hat{C}(\psi_s)\hat{F}(\xi_s) 
\label{Eq.A2}
\end{align}
Here $\hat{A}(\psi_t)$, $\hat{B}(\psi_t)$ and $\hat{C}(\psi_t)$ are well-behaved (continuous and differentiable), possibly non-linear operators, which depend only on the state $|\psi_t\rangle$ at time $t$. In the following, we assume there exists a well-defined measure for the $\xi_t$ and $\ket{\psi_t}$ processes, which we denoted $\xi$ and $\psi$ respectively. The functions $f(\xi_t)$ and $\hat{F}(\xi_t)$ (with $\hat{F}$ being an operator valued function) depend on the real valued stochastic variable $\xi_t=\xi_0+\int^t_0d\xi_s$. Its dynamics is specified by an It\^o process of the form $d\xi_t = X_a(\xi_t)dt+X_b(\xi_t)dW_t$ with quadratic variation  $d\mathcal{Q}_t[\xi_t,\xi_t]=\mathbb{E}_\xi\left[d\xi_t^2\right]=X^2_b(\xi_t)dt$. The operator valued function $\hat{F}(\xi_t)$ is assumed to have a spectral decomposition, $\hat{F}(\xi_t)=\sum_iF^i(\xi_t)|\psi^i_t\rangle\langle\psi^i_t|$, where the diagonal elements $F^i(\xi)$ are functions of $\xi_t$ and $\{|\psi^i_t\rangle\}$ denote a complete set of basis vectors, spanning the quantum mechanical Hilbert space. 

Note that the right hand side of Eq.~\eqref{Eq.A1} contains only terms of order $dt$ and not $d\xi_s$. Moreover, the stochastic variable $\xi_t=\int^t \xi_s ds$ is continuous and its integral and hence also $|\psi_t\rangle$ is continuous and once differentiable. We thus expect the quadratic variations of the state in Eq.~\eqref{Eq.A2} to be zero. To explicitly show this, we consider $\mathcal{Q}_T[\bra{\psi_t},|\psi_t\rangle]$ and $\hat{\mathcal{Q}}_T[|\psi_t\rangle,\bra{\psi_t}]$, which are the scalar valued and operator valued analogues of the quadratic variation:
\begin{align}
&\mathcal{Q}_T[\bra{\psi_t},|\psi_t\rangle]:=
\int^T\mathbb{E}_{\psi}(|d\psi_t|^2)\nonumber\\
&=\lim_{n\rightarrow\infty} \lim_{||\delta||\rightarrow0}\sum^{n}_{k=1} \left(\bra{\psi_{t_k}}-\bra{\psi_{t_{k -1}}}\right)\left(|\psi_{t_k}\rangle - |\psi_{t_{k -1}}\rangle\right)\nonumber\\
&= \lim_{n\rightarrow\infty}\lim_{||\delta||\rightarrow0} \sum^{n}_{k=1} \left(t_k-t_{k-1}\right)^2\bra{\psi_{t_{k-1}}}\hat{\mathcal{G}}^{\dagger}_{t_{k-1}}\hat{\mathcal{G}}_{t_{k-1}}|\psi_{t_{k-1}}\rangle\nonumber\\
&=0
\end{align}
As expected, the quadratic variation is of the order of $(t_k - t_{k-1})^2=\delta_k^2$ and hence vanishes for any time $t>t_0$. 

Similarly, the operator valued quadratic variation is:
\begin{align}
&\hat{\mathcal{Q}}_T[|\psi_t\rangle,\bra{\psi_t}]:=
\int^T\mathbb{E}_{\psi}\left[\ket{d\psi_t}\bra{d\psi_t}\right]\nonumber\\
&=\lim_{||\delta||\rightarrow0}\lim_{n\rightarrow\infty} \sum^{n}_{k=1} \left(|\psi_{t_k}\rangle - |\psi_{t_{k -1}}\rangle\right)\left(\bra{\psi_{t_k}}-\bra{\psi_{t_{k -1}}}\right)\nonumber\\
&= \lim_{n\rightarrow\infty}\lim_{||\delta||\rightarrow0} \sum^{n}_{k=1} \left(t_k-t_{k-1}\right)^2 \notag \\
&~~~~~~~~~~~~~~~~~~~~~~ \times \left(\hat{\mathcal{G}}_{t_{k-1}}|\psi_{t_{k-1}}\rangle\bra{\psi_{t_{k-1}}}\hat{\mathcal{G}}^{\dagger}_{t_{k-1}}\right)\nonumber\\
&=0
\end{align}
Again, the quadratic variation is found to vanish for all times. Any calculations involving functions of just $|\psi_t\rangle$ (and not joint functions of both $|\psi_t\rangle$ and $\xi_t$) thus require no correction to the usual rules of calculus, and the quadratic variations vanish for the entire class of models described by Eq.~\eqref{Eq.A1}. This is the result quoted in the main text, which has also been discussed in the context of colored noise driven Langevin-like equations in Refs.~\cite{Hanggi94,Luczka2005}.
Note that the vanishing of the quadratic variations depends crucially on the state evolution being strictly proportional to $dt$, and the stochastic variable not being Gaussian white noise ($\xi_t dt \neq dW_t$).

\section{Fokker-Planck-Kolmogorov equations for It\^o noise driven dynamical systems}\label{app:FP_BK}
In this appendix we formulate the Fokker-Planck-Kolmogorov (FPK) and Kolmogorov-Backward (KB) equations for Eq.~\eqref{Eq:ZPair}, reproduced here for convenience:
\begin{align}
dz_t &= J_z dt +  \mathcal{G}_z \xi_t \,dt\nonumber\\
d\xi_t &= -\xi_t\frac{dt}{\tau} + g(\xi_t) dW_t
\label{Eq:ZPairAppendix}
\end{align}
This pair of equations describes Markovian dynamics for all times $t>0$ in the augmented phase space of $\{z_t,\xi_t\}$. Specific forms for $\mathcal{G}_z$, $J_z$, and $g(\xi_t)$ were used in the main text, but we consider the general case here, restricting them to be atleast twice differentiable functions of $z_t$. Thus, this formalism applies to a large class of noise ($\xi_t$) driven non-linear dynamical systems ($z_t$).

Ensembles of evolutions for the joint state $z_t$ and the stochastic variable $\xi_t$, yield continuous and differentiable probability densities, which evolve according to the KB and FPK equations~\cite{Risken1996,Pavliotis2008,Luczka2005,Hanggi94,gardiner2004handbook,oksendal2003stochastic}. The dynamics of the ensemble averages can be formulated in terms of (joint) transition probabilities $\mathbb{T}(z_f,\xi_f,t_f|z_i,\xi_i,t_i)$, which provide the conditional probability of obtaining at time $t=t_f$ both a state in the interval $(z_f,z_f+dz)$ and a noise value in the interval $(\xi_f,\xi_f+d\xi)$, given initial conditions $(z_i,\xi_i)$ at time $t=t_i$. The Chapman-Kolmogorov equations, $\mathbb{T}(z_3,\xi_3,t_3|z_1,\xi_1,t_1)=\int d\xi_2 dz_2\,\mathbb{T}(z_3,\xi_3,t_3|z_2,\xi_2,t_2)\,\mathbb{T}(z_2,\xi_2,t_2|z_1,\xi_1,t_1)$, hold at all times.The transition probabilities are solutions to the FPK equations which evolve densities in the forward time direction given initial conditions. They also solve the KB equations, which evolve densities in the backward time direction given terminal conditions.  

The infinitesimal generators introduced in Eq.~\eqref{Eq.App.FPKOp} define the FPK and KB equations:
\begin{align}
&\frac{\partial}{\partial t}\mathbb{T}(z_T,\xi_T,t_T|z,\xi,t) = -\Lambda(z,\xi,t)\mathbb{T}(z_T, \xi_T,t_T|z,\xi,t) \notag \\
&\frac{\partial}{\partial t}\mathbb{T}(z,\xi,t|z_0,\xi_0,t_0) = \Lambda^{*}(z,\xi,t)\mathbb{T}(z,\xi,t|z_0,\xi_0,t_0)
\label{eq:FPZ}
\end{align}
For the joint evolution of $z_t$ and $\xi_t$ defined by Eq.~\eqref{Eq:ZPairAppendix}, the infinitesimal generators are given by:
\begin{align}
\Lambda^{*}(z,\xi,t) &= -\frac{\partial}{\partial z}\,\left(J_z + \mathcal{G}_z\xi\right)\, +\frac{1}{\tau}\frac{\partial}{\partial\xi} \xi + \frac{1}{2}\frac{\partial^2}{\partial \xi^2 } g(\xi)^2  \notag \\ \Lambda(z,\xi,t) &= \left(J_z + \mathcal{G}_z\xi\right) \frac{\partial}{\partial z} -\frac{\xi}{\tau}\frac{\partial}{\partial\xi} + \frac{g^2(\xi)}{2}\frac{\partial^2}{\partial \xi^2 }
\end{align}
Note that in the above equations, $J_z$ and $\mathcal{G}_z$ retain only the functional forms, which are functions of $z$, while in the stochastic equations of Eq.~\eqref{Eq:ZPairAppendix}, they are functions of $z_t$. Using the transition probability density and integrating out the density at the terminal time $T$, yields the one-point (backward) probability density, $\rho(z,\xi,t)$ and shows that the Kolmogorov Backward equation also applies directly to it:
\begin{align}
\rho(z,\xi,t)&=\int_\Omega \rho(z_T,\xi_T,t_T) \mathbb{T}(z_T,\xi_T,t_T|z,\xi,t)d\xi_T\,dz_T \notag \\
\frac{\partial}{\partial t}\rho(z,\xi,t)&= -\Lambda(z,\xi,t)\rho(z,\xi,t) \label{eq:BKZ_fin}
\end{align}
Here, $\Omega$ is the appropriate sample space supporting the terminal conditions of joint state and stochastic variable. Similarly, for fixed initial conditions, we find that the Fokker-Planck-Kolmogorov equation applies to  one-point (forward) density function, $\rho(z,\xi,t)$ as:
\begin{align}
\rho(z,\xi,t) &=\int_\Omega \rho(z_0,\xi_0,t_0) \mathbb{T}(z,\xi,t|z_0,\xi_0,t_0)d\xi_0\,dz_0 \notag \\
\frac{\partial}{\partial t}\rho(z,\xi,t) &= \Lambda^{*}(z,\xi,t)\rho(z,\xi,t)
\label{eq:FPZ_fin}
\end{align}
Here again, $\Omega$ is the appropriate sample space supporting the initial conditions of joint state and stochastic variable.
This is the starting point for the analysis of Sec.~\ref{sec:4A}.

\section{Fluctuation-Dissipation relations, the Martingale condition and emergent Born statistics\label{app:FDR_Emergent_Born_Pearle_comarison}}
Here we discuss the properties of Eqs.~\eqref{Eq:SmalltauFinal}-\eqref{Eq:Strat Correction} obtained after the multi-scale noise homogenization procedure outlined in Sec.\ref{sec:4A}. 

After converting to the corresponding It\^o convention, we find for $z_t=|\langle 0\ket{\psi_t}|^2$:
\begin{align}
d z_t = \tilde{J}_z dt + \mathcal{D}\,\tilde{\mathcal{G}}_z  dW_t.
\label{Eq:App:zIto1}
\end{align}
Here, $\tilde{J}_z = 2(J-\mathcal{D}^2) \,z_t(1-z_t)(2z_t-1)$, $\tilde{\mathcal{G}}_z = z_t(1-z_t) $ and $\mathcal{D}=\sqrt{2D^2\mathbb{E}^{\infty}_{\xi}[ \xi^2]}$ is the effective diffusion constant. Given the fluctuation-dissipation relation (FDR) in Sec.\ref{sec:4B}, $J=\mathcal{D}^2$, the first term vanishes and we are left with:
\begin{align}
d z_t =  \mathcal{D}\,z_t(1-z_t)  dW_t.
\label{Eq:App:zIto2}
\end{align}
The expectation $\mathbb{E}_{z}[dz_t]$ is then seen to vanish due to the vanishing expectation value of $dW_t$. The expectation value of $dz_t$ being zero is known as the Martingale condition, and it guarantees adherence to Born's statistics~\cite{Pearle_89_PRA,Pearle1984_bookRef,Ghirardi_90_PRA,Bassi_03_PhyRep}. 

Furthermore, $z_t=1$ and $z_t=0$ are both steady solutions of Eq.~\eqref{Eq:App:zIto2} that have $dz_t=0$ even for individual trajectories. For an ensemble of $z_t$ processes this implies that the long time average $\lim_{t\rightarrow\infty}\mathbb{E}_z\big[z_t(1-z_t)\big]$ vanishes. 

The FPK equation (see Appendix~\ref{app:FP_BK}) corresponding to Eq.~\eqref{Eq:App:zIto2} reads:
\begin{align}
    \partial_t\,\rho(z,t) = \frac{\mathcal{D}}{2}\partial^2_z\,\bigg[z(1-z)\rho(z,t)\bigg].
    \label{Eq:App:zFPK}
\end{align}
This equation is the same as Eq.~(6.6) of Ref.~\cite{Pearle1984_bookRef}, and it guarantees that all the criteria imposed on models of 
quantum state reduction in that reference are upheld. The GKSL master equation obtained in Sec.~\ref{sec:4B} is a more general result, which also implies adherence to these criteria.

\section{Spontaneous Symmetry Breaking \label{app:SSB}}
In this appendix, we review some of the central concepts in the theory of spontaneous symmetry breaking and summarise their use in models of spontaneous unitarity violation. This appendix is based on appendix C of Ref.~\cite{aritro1}, the main points of which are reproduced here for sake of completeness. For a more detailed treatment of spontaneous symmetry breaking, see Ref.~\cite{Wezel_SSBlecturenotes}.

Spontaneous symmetry breaking refers to the situation in which the actually realised state of a system has a lower symmetry than the Hamiltonian governing it. The symmetry breaking is considered ``spontaneous'' if it is unavoidable and impossible to influence in practice. A concrete example is that of a harmonic crystal, described by the Hamiltonian:
\begin{align}
\hat{H} = \sum_j \frac{\hat{P}^2_j}{2m} + \sum_{\langle i,j\rangle} \frac{1}{2}m \omega^2 \left(\hat{X}_i-\hat{X}_{j}\right)^2.
\label{eq:xtal}
\end{align}
Here, $i$ and $j$ label neighbouring sites of an atomic lattice in which $m$ is the atomic mass and $\omega$ the a frequency characterising forces between neighbouring atoms.  

The Hamiltonian of Eq.~\eqref{eq:xtal} is invariant under global translations of all atoms, and all of its eigenstates share this translational symmetry. In fact, all eigenstates of $\hat{H}$ are simultaneously eigenstates of the total momentum $\hat{P}_{\text{tot}}$, which generates global translations.

The Fourier transform of Eq.~\eqref{eq:xtal} can be written as:
\begin{align}
\hat{H} = \frac{\hat{P}_{\text{tot}}}{2mN} + \sum_{k\neq 0} \hat{H}_k.
\label{eq:xtalk}
\end{align}
Here, $N$ is the number of atoms in the harmonic crystal, and $k$ denotes the internal crystal momentum. The internal physics described by $\hat{H}_k$ can be ignored when considering only the global centre of mass and total momentum of the crystal, or when focusing on temperatures below the lowest possible phonon excitation energy.

The ground state of the Hamiltonian in Eq.~\eqref{eq:xtalk} is non-degenerate and has zero total momentum. Global excitations with non-zero total momentum are separated from the ground state by energies of order $1/N$. In the \emph{thermodynamic limit} $N\to \infty$, these globally excited states become degenerate with the ground state, and it becomes possible for a wave packet of degenerate ground states to be formed, with a localised centre of mass. Such a wave packet breaks translational symmetry. 

In reality, $N$ may be large but not infinite, and forcing the crystal into localised state requires a small but non-zero external force:
\begin{align}
\hat{H}_{\text{SB}} = \frac{\hat{P}_{\text{tot}}}{2mN} + \epsilon N \left(\hat{X}_{\text{com}} - x_0\right)^2.
\label{eq:xtalSB}
\end{align}
Here, $\hat{X}_{\text{com}}$ is the centre of mass position, $x_0$ is the centre of the externally applied potential, and $\epsilon$ is its strength. The factor $N$ multiplying $\epsilon$ keeps the energy extensive, and arises from the applied potential coupling to an \emph{order parameter} of the system~\cite{vanwezelAmJPhys}. The non-degenerate ground state $\ket{\psi_{\text{gs}}}$ of the crystal in the presence of the external potential is a Gaussian wave packet obeying the limiting behaviour:
\begin{align}
    \lim_{{N}\to \infty} \lim_{\epsilon\to 0} \ket{\psi_{\text{gs}}} &= \ket{P_{\text{tot}}=0} \notag \\
    \lim_{\epsilon\to 0} \lim_{{N}\to \infty}  \ket{\psi_{\text{gs}}} &= \ket{X_{\text{com}}=x_0}. 
    \label{eq:limits}
\end{align}
That is, if the external force is strictly zero, the ground state of the crystal is delocalised and symmetric. If there is even an infinitesimally small (but non-zero) perturbation $\epsilon$, however, the crystal ground state breaks the symmetry and becomes fully localised in the thermodynamic limit. 

Neither of the limits in Eq.~\eqref{eq:limits} are physical. What their failure to commute shows, is the \emph{diverging susceptibility} of the crystal to symmetry-breaking perturbations. That is, large crystals may be forced into a symmetry-breaking state by an external potential whose strength scales as $1/N$. In practice, the potential required to localise classical objects in our everyday world is so weak that we can never detect or control it, regardless of any future technological progress. For all practical purposes therefore, the ground states of human-sized harmonic crystals are always rendered fully localised in the real world. Because the localisation is unavoidable, unpredictable, and uncontrollable, we say the symmetry is spontaneously broken.

Notice that microscopic harmonic crystals consisting of only a few atoms are not spontaneously localised by weak external forces, and that the perturbation required to localise a macroscopic crystal will have only a negligible effect on microscopic systems. That is, symmetry breaking is an emergent effect that becomes unavoidable as the thermodynamic limit is approached. Moreover, as long as it couples to the order parameter, the precise shape and strength of the external potential are irrelevant to the final state obtained by the crystal. The localised symmetry broken state is universal and independent of details of the symmetry breaking field.

\subsection*{Spontaneous unitarity breaking}\label{app:SSBsub}
As shown in Refs.~\cite{vanwezelprb,Wezel10}, the unitarity of quantum mechanical time evolution can be spontaneously broken in the same way as any other symmetry. That is, the diverging susceptibility to external forces in the thermodynamic limit that causes the usual types of spontaneous symmetry breaking \emph{additionally} causes symmetry-breaking systems to escape the unitarity of Schr\"odinger's time evolution.

As in other types of symmetry breaking, unitarity breaking require a non-zero perturbation. In this case, the perturbation must correspond to a non-Hermitian addition to the Hamiltonian: $\hat{H}_{\text{SUV}}=\hat{H} + i \epsilon \hat{G}$. Here, $\hat{H}$ is the Hamiltonian for a system with a spontaneously broken regular symmetry, $\epsilon$ is the strength of the non-unitary perturbation, and $\hat{G}$ is a Hermitian operator coupling to the order parameter of $\hat{H}$. For example, a harmonic crystal could break the unitarity of Schr\"odinger dynamics by evolving according to:
\begin{align}
\hat{H}_{\text{SUV}}=\frac{\hat{P}^2_{\text{tot}}}{2mN} + i \epsilon N \left(\hat{X}_{\text{com}} - x_0\right)^2.
\label{eq:XtalSUVdef}
\end{align}
Here $\epsilon$ is the strength of the non-unitary perturbation, and the factor $N$ arises from the coupling to the order parameter~\cite{vanwezelAmJPhys}.

The unitarity breaking field again causes the emergence of a singular limit, but now in time evolution rather than in the equilibrium ground state:
\begin{align}
\label{eq:XtalSUV}
    \lim_{{N}\to \infty} \lim_{\epsilon\to 0} e^{-\frac{i}{\hbar}t \hat{H}_{\text{SUV}}} \ket{P_{\text{tot}}=0} &= \ket{P_{\text{tot}}=0} \\
    \lim_{\epsilon\to 0} \lim_{{N}\to \infty}  e^{-\frac{i}{\hbar}t \hat{H}_{\text{SUV}}} \ket{P_{\text{tot}}=0} &= \ket{X_{\text{com}}=x_0} ~~~ \forall t>0. \notag
\end{align}
That is, if the unitarity breaking perturbation is strictly zero, the crystal evolves strictly according to Schr\"odinger's equation. If there is even an infinitesimally small (but non-zero) perturbation $\epsilon$, however, the crystal instantaneously collapses into a fully localised in the thermodynamic limit. 

Neither of the limits in Eq.~\eqref{eq:XtalSUV} actually exists, but as before, they indicate a diverging susceptibility of the crystal to unitarity-breaking perturbations. That is, the perturbation required to evolve non-unitarily to a localised state is of the order of $1/N$. In practice, the required perturbation is therefore so weak that we can never detect or control it, and for all practical purposes human-sized harmonic crystals instantaneously localise in the real world. Because the localisation is unavoidable, unpredictable, and uncontrollable, we say the unitarity violation is spontaneous.

Notice that microscopic harmonic crystals consisting of only a few atoms do not spontaneously evolve to a localised state following weak external perturbations, and that the perturbation required to localise a macroscopic crystal will have only a negligible effect on microscopic systems. As with conventional symmetry breaking, unitarity breaking is an emergent effect that becomes unavoidable as the thermodynamic limit is approached. Moreover, as long as it couples to the order parameter, the precise shape and strength of the external perturbation are irrelevant to the final state that the crystal evolves towards.

\subsection*{Quantum measurement}
For spontaneous unitarity violations to explain quantum measurement, perturbations of the type of Eq.~\eqref{eq:XtalSUVdef} are not sufficient. As shown in Refs.~\cite{Wezel10, Mertens_PRA_21, Mertens22,Lotte}, Born's rule can emerge only if the unitarity breaking term is both stochastic and non-linear. In the main text, we therefore consider dynamical equations like Eq.~\eqref{Eq:curlyG2}, which act on the entangled state of a microscopic system with the pointer of a macroscopic measurement device.

Pointers used in any classical measurement device are necessarily symmetry-broken objects, and symmetry broken states are also necessarily pointer states in the decoherence sense~\cite{Zurek_1981}. 
 The result of spontaneous unitarity violation in the dynamics of a superposition over multiple pointer states is a near-instantaneous evolution towards a single pointer state. As in regular spontaneous symmetry breaking, the collapse is not actually spontaneous, but for measurement machines of any realistic size, exceedingly small non-unitary perturbations suffice to cause collapse dynamics that is for all practical purposes unpredictable, inevitable, and instantaneous.

Once again, the emergent dynamics is spontaneous in the sense of being unavoidable and unpredictable, and it is universal in the sense of Born’s rule arising regardless of the precise strength of the unitarity breaking perturbation. Born’s rule emerges rather than being imposed or assumed, in the sense that the stochastic term in the non-unitary dynamics is linear (apart from an overall geometric factor ensuring normalisation, which does not influence the final state statistics) and thus independent of the state being measured.

\section{Von Neumann measurement \label{app:Quantum Strong Measurement}}
In this appendix, we motivate why we restrict attention in the main text to entangled states of system and measurement device, by summarizing the measurement setup originally introduced by Von Neumann~\cite{Von_Neumann2018-bo}. In the projective limit this describes a strong measurement, as opposed to so-called quantum weak measurements~\cite{Tamir_2013,Svensson_2012}. This appendix is based on appendix D of Ref.~\cite{aritro1}, the main points of which are reproduced here for sake of completeness. 

We consider a measurement apparatus $A$, performing a single measurement of the observable $\hat{O}_S$ on the system $S$. We assume quantum theory to apply to the measurement device as well as the system, and consider their joint evolution. The measurement apparatus is assumed to contain a pointer whose position $x$ indicates the measurement outcome. More generally, $x$ represents an eigenvalue of the order parameter describing the broken symmetry of the measurement apparatus~\cite{Wezel_SSBlecturenotes}. The use of pointer states with a spontaneously broken symmetry is necessitated by the fact that only these states are susceptible to spontaneous unitarity violation~\cite{vanwezelprb}. As discussed in the main text, this yields a preferred basis for measurement outcomes, which must be ordered states with a spontaneously broken symmetry.
 
Any states of the measurement apparatus, $\ket{\psi}_A$, may be expressed in a basis of pointer states $\ket{x}_A$ localised at $x$:
\begin{align}
\ket{\psi}_A = \int dx ~ \psi(x) \,\ket{x}_A
\end{align}
Since the pointer is a symmetry-broken object, its state before measurement will be a sharply peaked Gaussian:
\begin{align}
\psi(x) = \left(\frac{1}{2\pi \Delta^2}\right)^{\frac{1}{4}}e^{-x^2/4\Delta^2}.
\label{Eq:pointer}
\end{align}

Pointer states centered at well-separated positions have an exponentially small overlap, and can be used to resolve different measurement outcomes. 
To act as a measurement device, the interaction between system and apparatus should cause the measurement device to display different pointer states for different initial system states $\ket{\sigma}_S$, corresponding to different eigenvalues $\sigma$ of the observable $\hat{O}_S$ to be measured. A generic interaction Hamiltonian yielding this behaviour is given by:
\begin{align}
\hat{H}_{\text{int}} = \gamma \hat{O}_S \otimes \hat{P}_A.
\end{align}
Here, $\gamma$ is the interaction strength and $\hat{P}_A$ is the canonical momentum operator conjugate to the pointer position: $[\hat{X}_A,\hat{P}_A]=i\hbar$. The time evolution generated by this Hamiltonian acts as a shift operator on the pointer position, shifting it in proportion to the eigenvalue of the system observable:
\begin{align}
e^{-\frac{i}{\hbar} t \hat{H}_{\text{int}}} \ket{\sigma}_S \ket{\psi}_A = \int dx ~ \psi(x - \sigma {\gamma t }/{\hbar}) \,\ket{\sigma}_S \ket{x}_A
\end{align}

Generically, the system before measurement will be a superposition of multiple eigenstates of $\hat{O}_S$. The combined initial state of system and apparatus is then given by $\ket{\Psi(t=0)}_{SA} = \sum_\sigma \phi_\sigma \ket{\sigma}_S \ket{\psi}_A$. The unitary time evolution caused by the interaction Hamiltonian then yields macroscopic entanglement: 
\begin{align}
    \ket{\Psi(t)}_{SA}= \sum_\sigma \int dx ~  \phi_\sigma \, \psi(x - \sigma {\gamma t }/{\hbar})  \,\ket{\sigma}_S \ket{x}_A
    \label{Eq:VNS_endstate}
\end{align}
Although the actual interaction between system and apparatus may be more involved than what is considered here, that will not affect the qualitative entanglement between system and apparatus. Although the system and measurement device are entangled from the moment they start interacting, the amount of entanglement grows with time, since pointer states centered at different $x_\sigma(t)$ separate over time.

After some time, the entangled states consists of well-separated components 
(i.e. $\langle x_\sigma\ket{x_{\sigma'}}_A \approx \delta_{\sigma,\sigma'}$), which form the starting point for the discussion in the main text:
\begin{align}
\ket{\Psi}_{SA} &=\sum_\sigma \phi_\sigma \ket{\sigma}_S\ket{x_\sigma}_A, \notag \\
\text{with} ~~~ \ket{x_\sigma}_A &=\int dx\, \psi(x - x_\sigma) \,|x\rangle_A. 
\label{eq:finalvN}
\end{align}

Although the entanglement dynamics is in reality not instantaneous, the evolving overlap during the build-up of entanglement will only affect the speed of the unitarity-breaking dynamics and does not influence the final states obtained in the process, nor their statistics. The SUV mechanism is envisoned to occur for this macroscopic entangled state consisting of a superposition of macroscopic pointers, and possibly a local environment. Any unitarily evolving environment would lead to further entanglement (a von Neumann chain), but cannot yield collapse-like dynamics, unless unitarity is spontaneously broken. This is the premise of Eq.~\eqref{Eq:key} in the main text, where $\ket{\psi_t}$ may be identified with $\ket{\Psi}_{SA}$ in Eq.~\eqref{eq:finalvN}.

\bibliography{biblio_AM}

\begin{thebibliography}{101}%
\makeatletter
\providecommand \@ifxundefined [1]{%
 \@ifx{#1\undefined}
}%
\providecommand \@ifnum [1]{%
 \ifnum #1\expandafter \@firstoftwo
 \else \expandafter \@secondoftwo
 \fi
}%
\providecommand \@ifx [1]{%
 \ifx #1\expandafter \@firstoftwo
 \else \expandafter \@secondoftwo
 \fi
}%
\providecommand \natexlab [1]{#1}%
\providecommand \enquote  [1]{``#1''}%
\providecommand \bibnamefont  [1]{#1}%
\providecommand \bibfnamefont [1]{#1}%
\providecommand \citenamefont [1]{#1}%
\providecommand \href@noop [0]{\@secondoftwo}%
\providecommand \href [0]{\begingroup \@sanitize@url \@href}%
\providecommand \@href[1]{\@@startlink{#1}\@@href}%
\providecommand \@@href[1]{\endgroup#1\@@endlink}%
\providecommand \@sanitize@url [0]{\catcode `\\12\catcode `\$12\catcode
  `\&12\catcode `\#12\catcode `\^12\catcode `\_12\catcode `\%12\relax}%
\providecommand \@@startlink[1]{}%
\providecommand \@@endlink[0]{}%
\providecommand \url  [0]{\begingroup\@sanitize@url \@url }%
\providecommand \@url [1]{\endgroup\@href {#1}{\urlprefix }}%
\providecommand \urlprefix  [0]{URL }%
\providecommand \Eprint [0]{\href }%
\providecommand \doibase [0]{http://dx.doi.org/}%
\providecommand \selectlanguage [0]{\@gobble}%
\providecommand \bibinfo  [0]{\@secondoftwo}%
\providecommand \bibfield  [0]{\@secondoftwo}%
\providecommand \translation [1]{[#1]}%
\providecommand \BibitemOpen [0]{}%
\providecommand \bibitemStop [0]{}%
\providecommand \bibitemNoStop [0]{.\EOS\space}%
\providecommand \EOS [0]{\spacefactor3000\relax}%
\providecommand \BibitemShut  [1]{\csname bibitem#1\endcsname}%
\let\auto@bib@innerbib\@empty
\bibitem [{\citenamefont {Bassi}\ and\ \citenamefont
  {Ghirardi}(2003)}]{Bassi_03_PhyRep}%
  \BibitemOpen
  \bibfield  {author} {\bibinfo {author} {\bibfnamefont {A.}~\bibnamefont
  {Bassi}}\ and\ \bibinfo {author} {\bibfnamefont {G.}~\bibnamefont
  {Ghirardi}},\ }\href {\doibase 10.1016/s0370-1573(03)00103-0} {\bibfield
  {journal} {\bibinfo  {journal} {Physics Reports}\ } (\bibinfo {year}
  {2003}),\ 10.1016/s0370-1573(03)00103-0}\BibitemShut {NoStop}%
\bibitem [{\citenamefont {Leggett}(2005)}]{leggett2005quantum}%
  \BibitemOpen
  \bibfield  {author} {\bibinfo {author} {\bibfnamefont {A.}~\bibnamefont
  {Leggett}},\ }\href {\doibase 10.1126/science.1109541} {\bibfield  {journal}
  {\bibinfo  {journal} {science}\ }\textbf {\bibinfo {volume} {307}},\ \bibinfo
  {pages} {871} (\bibinfo {year} {2005})}\BibitemShut {NoStop}%
\bibitem [{\citenamefont {Bassi}\ \emph
  {et~al.}(2013{\natexlab{a}})\citenamefont {Bassi}, \citenamefont {Lochan},
  \citenamefont {Satin}, \citenamefont {Singh},\ and\ \citenamefont
  {Ulbricht}}]{overview}%
  \BibitemOpen
  \bibfield  {author} {\bibinfo {author} {\bibfnamefont {A.}~\bibnamefont
  {Bassi}}, \bibinfo {author} {\bibfnamefont {K.}~\bibnamefont {Lochan}},
  \bibinfo {author} {\bibfnamefont {S.}~\bibnamefont {Satin}}, \bibinfo
  {author} {\bibfnamefont {T.~P.}\ \bibnamefont {Singh}}, \ and\ \bibinfo
  {author} {\bibfnamefont {H.}~\bibnamefont {Ulbricht}},\ }\href {\doibase
  10.1103/RevModPhys.85.471} {\bibfield  {journal} {\bibinfo  {journal} {Rev.
  of mod. phys}\ }\textbf {\bibinfo {volume} {85}},\ \bibinfo {pages}
  {471–527} (\bibinfo {year} {2013}{\natexlab{a}})}\BibitemShut {NoStop}%
\bibitem [{\citenamefont {Arndt}\ and\ \citenamefont
  {Hornberger}(2014)}]{arndt2014testing}%
  \BibitemOpen
  \bibfield  {author} {\bibinfo {author} {\bibfnamefont {M.}~\bibnamefont
  {Arndt}}\ and\ \bibinfo {author} {\bibfnamefont {K.}~\bibnamefont
  {Hornberger}},\ }\href {\doibase 10.1038/nphys2863} {\bibfield  {journal}
  {\bibinfo  {journal} {Nature Physics}\ }\textbf {\bibinfo {volume} {10}},\
  \bibinfo {pages} {271} (\bibinfo {year} {2014})}\BibitemShut {NoStop}%
\bibitem [{\citenamefont {Carlesso}\ \emph {et~al.}(2022)\citenamefont
  {Carlesso}, \citenamefont {Donadi}, \citenamefont {Ferialdi}, \citenamefont
  {Paternostro}, \citenamefont {Ulbricht},\ and\ \citenamefont
  {Bassi}}]{carlesso2022present}%
  \BibitemOpen
  \bibfield  {author} {\bibinfo {author} {\bibfnamefont {M.}~\bibnamefont
  {Carlesso}}, \bibinfo {author} {\bibfnamefont {S.}~\bibnamefont {Donadi}},
  \bibinfo {author} {\bibfnamefont {L.}~\bibnamefont {Ferialdi}}, \bibinfo
  {author} {\bibfnamefont {M.}~\bibnamefont {Paternostro}}, \bibinfo {author}
  {\bibfnamefont {H.}~\bibnamefont {Ulbricht}}, \ and\ \bibinfo {author}
  {\bibfnamefont {A.}~\bibnamefont {Bassi}},\ }\href {\doibase
  10.1038/s41567-021-01489-5} {\bibfield  {journal} {\bibinfo  {journal}
  {Nature Physics}\ }\textbf {\bibinfo {volume} {18}},\ \bibinfo {pages} {243}
  (\bibinfo {year} {2022})}\BibitemShut {NoStop}%
\bibitem [{\citenamefont {Komar}(1962)}]{Komar62}%
  \BibitemOpen
  \bibfield  {author} {\bibinfo {author} {\bibfnamefont {A.}~\bibnamefont
  {Komar}},\ }\href {\doibase 10.1103/PhysRev.126.365} {\bibfield  {journal}
  {\bibinfo  {journal} {Phys. Rev.}\ }\textbf {\bibinfo {volume} {126}},\
  \bibinfo {pages} {365} (\bibinfo {year} {1962})}\BibitemShut {NoStop}%
\bibitem [{\citenamefont {Wigner}(1995)}]{Wigner95}%
  \BibitemOpen
  \bibfield  {author} {\bibinfo {author} {\bibfnamefont {E.~P.}\ \bibnamefont
  {Wigner}},\ }\enquote {\bibinfo {title} {Review of the quantum-mechanical
  measurement problem},}\ in\ \href {\doibase 10.1007/978-3-642-78374-6_19}
  {\emph {\bibinfo {booktitle} {Philosophical Reflections and Syntheses}}},\
  \bibinfo {editor} {edited by\ \bibinfo {editor} {\bibfnamefont
  {J.}~\bibnamefont {Mehra}}}\ (\bibinfo  {publisher} {Springer Berlin
  Heidelberg},\ \bibinfo {address} {Berlin, Heidelberg},\ \bibinfo {year}
  {1995})\ pp.\ \bibinfo {pages} {225--244}\BibitemShut {NoStop}%
\bibitem [{\citenamefont {Van~Wezel}(2010)}]{Wezel10}%
  \BibitemOpen
  \bibfield  {author} {\bibinfo {author} {\bibfnamefont {J.}~\bibnamefont
  {Van~Wezel}},\ }\href {\doibase 10.3390/sym2020582} {\bibfield  {journal}
  {\bibinfo  {journal} {Symmetry}\ }\textbf {\bibinfo {volume} {2}},\ \bibinfo
  {pages} {582} (\bibinfo {year} {2010})}\BibitemShut {NoStop}%
\bibitem [{\citenamefont {Mukherjee}\ \emph {et~al.}(2024)\citenamefont
  {Mukherjee}, \citenamefont {Gotur}, \citenamefont {Aalberts}, \citenamefont
  {van~den Ende}, \citenamefont {Mertens},\ and\ \citenamefont {van
  Wezel}}]{aritro1}%
  \BibitemOpen
  \bibfield  {author} {\bibinfo {author} {\bibfnamefont {A.}~\bibnamefont
  {Mukherjee}}, \bibinfo {author} {\bibfnamefont {S.}~\bibnamefont {Gotur}},
  \bibinfo {author} {\bibfnamefont {J.}~\bibnamefont {Aalberts}}, \bibinfo
  {author} {\bibfnamefont {R.}~\bibnamefont {van~den Ende}}, \bibinfo {author}
  {\bibfnamefont {L.}~\bibnamefont {Mertens}}, \ and\ \bibinfo {author}
  {\bibfnamefont {J.}~\bibnamefont {van Wezel}},\ }\href {\doibase
  10.3390/e26020131} {\bibfield  {journal} {\bibinfo  {journal} {Entropy}\
  }\textbf {\bibinfo {volume} {26}} (\bibinfo {year} {2024}),\
  10.3390/e26020131}\BibitemShut {NoStop}%
\bibitem [{\citenamefont {Zurek}(1982)}]{Zurek_1982}%
  \BibitemOpen
  \bibfield  {author} {\bibinfo {author} {\bibfnamefont {W.~H.}\ \bibnamefont
  {Zurek}},\ }\href {\doibase 10.1103/physrevd.26.1862} {\bibfield  {journal}
  {\bibinfo  {journal} {Phys. Rev. D}\ } (\bibinfo {year} {1982}),\
  10.1103/physrevd.26.1862}\BibitemShut {NoStop}%
\bibitem [{\citenamefont {Schlosshauer}(2005)}]{Schlosshauer_2005}%
  \BibitemOpen
  \bibfield  {author} {\bibinfo {author} {\bibfnamefont {M.}~\bibnamefont
  {Schlosshauer}},\ }\href {\doibase 10.1103/revmodphys.76.1267} {\bibfield
  {journal} {\bibinfo  {journal} {Reviews of Modern Physics}\ } (\bibinfo
  {year} {2005}),\ 10.1103/revmodphys.76.1267}\BibitemShut {NoStop}%
\bibitem [{\citenamefont {Zurek}(2009)}]{Zurek2009}%
  \BibitemOpen
  \bibfield  {author} {\bibinfo {author} {\bibfnamefont {W.~H.}\ \bibnamefont
  {Zurek}},\ }\href {\doibase 10.1038/nphys1202} {\bibfield  {journal}
  {\bibinfo  {journal} {Nature Physics}\ }\textbf {\bibinfo {volume} {5}},\
  \bibinfo {pages} {181} (\bibinfo {year} {2009})}\BibitemShut {NoStop}%
\bibitem [{\citenamefont {Allahverdyan}\ \emph {et~al.}(2013)\citenamefont
  {Allahverdyan}, \citenamefont {Balian},\ and\ \citenamefont
  {Nieuwenhuizen}}]{theo13}%
  \BibitemOpen
  \bibfield  {author} {\bibinfo {author} {\bibfnamefont {A.~E.}\ \bibnamefont
  {Allahverdyan}}, \bibinfo {author} {\bibfnamefont {R.}~\bibnamefont
  {Balian}}, \ and\ \bibinfo {author} {\bibfnamefont {T.~M.}\ \bibnamefont
  {Nieuwenhuizen}},\ }\href {\doibase
  https://doi.org/10.1016/j.physrep.2012.11.001} {\bibfield  {journal}
  {\bibinfo  {journal} {Physics Reports}\ }\textbf {\bibinfo {volume} {525}},\
  \bibinfo {pages} {1} (\bibinfo {year} {2013})}\BibitemShut {NoStop}%
\bibitem [{\citenamefont {Dieks}(1989)}]{Dieks_1989}%
  \BibitemOpen
  \bibfield  {author} {\bibinfo {author} {\bibfnamefont {D.}~\bibnamefont
  {Dieks}},\ }\href {\doibase 10.1016/0375-9601(89)90510-0} {\bibfield
  {journal} {\bibinfo  {journal} {Physics Lett. A}\ } (\bibinfo {year}
  {1989}),\ 10.1016/0375-9601(89)90510-0}\BibitemShut {NoStop}%
\bibitem [{\citenamefont {Adler}(2003)}]{Adler_2003}%
  \BibitemOpen
  \bibfield  {author} {\bibinfo {author} {\bibfnamefont {S.~L.}\ \bibnamefont
  {Adler}},\ }\href {\doibase 10.1016/S1355-2198(02)00086-2} {\bibfield
  {journal} {\bibinfo  {journal} {Stud. Hist. Phil. Science B}\ }\textbf
  {\bibinfo {volume} {34}},\ \bibinfo {pages} {135} (\bibinfo {year}
  {2003})}\BibitemShut {NoStop}%
\bibitem [{\citenamefont {von Stillfried}(2008)}]{Stillfried_2008}%
  \BibitemOpen
  \bibfield  {author} {\bibinfo {author} {\bibfnamefont {N.}~\bibnamefont {von
  Stillfried}},\ }\href {\doibase 10.1038/453978c} {\bibfield  {journal}
  {\bibinfo  {journal} {Nature}\ } (\bibinfo {year} {2008}),\
  10.1038/453978c}\BibitemShut {NoStop}%
\bibitem [{\citenamefont {Fortin}\ and\ \citenamefont
  {Lombardi}(2014)}]{Fortin2014}%
  \BibitemOpen
  \bibfield  {author} {\bibinfo {author} {\bibfnamefont {S.}~\bibnamefont
  {Fortin}}\ and\ \bibinfo {author} {\bibfnamefont {O.}~\bibnamefont
  {Lombardi}},\ }\href {\doibase 10.1007/s10701-014-9791-3} {\bibfield
  {journal} {\bibinfo  {journal} {Foundations of Physics}\ }\textbf {\bibinfo
  {volume} {44}},\ \bibinfo {pages} {426} (\bibinfo {year} {2014})}\BibitemShut
  {NoStop}%
\bibitem [{\citenamefont {Everett}(1957)}]{Everett_1957}%
  \BibitemOpen
  \bibfield  {author} {\bibinfo {author} {\bibfnamefont {H.}~\bibnamefont
  {Everett}},\ }\href {\doibase 10.1103/revmodphys.29.454} {\bibfield
  {journal} {\bibinfo  {journal} {Reviews of Modern Physics}\ } (\bibinfo
  {year} {1957}),\ 10.1103/revmodphys.29.454}\BibitemShut {NoStop}%
\bibitem [{\citenamefont {Bohm}(1952{\natexlab{a}})}]{Bohm52A}%
  \BibitemOpen
  \bibfield  {author} {\bibinfo {author} {\bibfnamefont {D.}~\bibnamefont
  {Bohm}},\ }\href {\doibase 10.1103/PhysRev.85.166} {\bibfield  {journal}
  {\bibinfo  {journal} {Phys. Rev.}\ }\textbf {\bibinfo {volume} {85}},\
  \bibinfo {pages} {166} (\bibinfo {year} {1952}{\natexlab{a}})}\BibitemShut
  {NoStop}%
\bibitem [{\citenamefont {Bohm}(1952{\natexlab{b}})}]{Bohm52B}%
  \BibitemOpen
  \bibfield  {author} {\bibinfo {author} {\bibfnamefont {D.}~\bibnamefont
  {Bohm}},\ }\href {\doibase 10.1103/PhysRev.85.180} {\bibfield  {journal}
  {\bibinfo  {journal} {Phys. Rev.}\ }\textbf {\bibinfo {volume} {85}},\
  \bibinfo {pages} {180} (\bibinfo {year} {1952}{\natexlab{b}})}\BibitemShut
  {NoStop}%
\bibitem [{\citenamefont {Rovelli}(1996)}]{rovelli1996relational}%
  \BibitemOpen
  \bibfield  {author} {\bibinfo {author} {\bibfnamefont {C.}~\bibnamefont
  {Rovelli}},\ }\href {\doibase 10.1007/BF02302261} {\bibfield  {journal}
  {\bibinfo  {journal} {International Journal of Theoretical Physics}\ }\textbf
  {\bibinfo {volume} {35}},\ \bibinfo {pages} {1637} (\bibinfo {year}
  {1996})}\BibitemShut {NoStop}%
\bibitem [{\citenamefont {Fuchs}\ \emph {et~al.}(2014)\citenamefont {Fuchs},
  \citenamefont {Mermin}, \citenamefont {Schack},\ and\ \citenamefont
  {Schack}}]{Fuchs_2014}%
  \BibitemOpen
  \bibfield  {author} {\bibinfo {author} {\bibfnamefont {C.~A.}\ \bibnamefont
  {Fuchs}}, \bibinfo {author} {\bibfnamefont {N.~D.}\ \bibnamefont {Mermin}},
  \bibinfo {author} {\bibfnamefont {R.}~\bibnamefont {Schack}}, \ and\ \bibinfo
  {author} {\bibfnamefont {R.}~\bibnamefont {Schack}},\ }\href {\doibase
  10.1119/1.4874855} {\bibfield  {journal} {\bibinfo  {journal} {American
  Journal of Physics}\ } (\bibinfo {year} {2014}),\
  10.1119/1.4874855}\BibitemShut {NoStop}%
\bibitem [{\citenamefont {Bohm}\ and\ \citenamefont
  {Bub}(1966)}]{BohmBub_66_RevModPhys}%
  \BibitemOpen
  \bibfield  {author} {\bibinfo {author} {\bibfnamefont {D.}~\bibnamefont
  {Bohm}}\ and\ \bibinfo {author} {\bibfnamefont {J.}~\bibnamefont {Bub}},\
  }\href {\doibase 10.1103/RevModPhys.38.453} {\bibfield  {journal} {\bibinfo
  {journal} {Rev. Mod. Phys.}\ }\textbf {\bibinfo {volume} {38}},\ \bibinfo
  {pages} {453} (\bibinfo {year} {1966})}\BibitemShut {NoStop}%
\bibitem [{\citenamefont {Pearle}(1976)}]{Pearle_76}%
  \BibitemOpen
  \bibfield  {author} {\bibinfo {author} {\bibfnamefont {P.}~\bibnamefont
  {Pearle}},\ }\href {\doibase 10.1103/physrevd.13.857} {\bibfield  {journal}
  {\bibinfo  {journal} {Phys. Rev. D}\ } (\bibinfo {year} {1976}),\
  10.1103/physrevd.13.857}\BibitemShut {NoStop}%
\bibitem [{\citenamefont {Pearle}(1989)}]{Pearle_89_PRA}%
  \BibitemOpen
  \bibfield  {author} {\bibinfo {author} {\bibfnamefont {P.}~\bibnamefont
  {Pearle}},\ }\href {\doibase 10.1103/physreva.39.2277} {\bibfield  {journal}
  {\bibinfo  {journal} {Phys. Rev. A}\ } (\bibinfo {year} {1989}),\
  10.1103/physreva.39.2277}\BibitemShut {NoStop}%
\bibitem [{\citenamefont {Gisin}(1984)}]{Gisin84}%
  \BibitemOpen
  \bibfield  {author} {\bibinfo {author} {\bibfnamefont {N.}~\bibnamefont
  {Gisin}},\ }\href {\doibase 10.1103/PhysRevLett.52.1657} {\bibfield
  {journal} {\bibinfo  {journal} {Phys. Rev. Lett.}\ }\textbf {\bibinfo
  {volume} {52}},\ \bibinfo {pages} {1657} (\bibinfo {year}
  {1984})}\BibitemShut {NoStop}%
\bibitem [{\citenamefont {Ghirardi}\ \emph {et~al.}(1986)\citenamefont
  {Ghirardi}, \citenamefont {Rimini},\ and\ \citenamefont
  {Weber}}]{Ghirardi_1986}%
  \BibitemOpen
  \bibfield  {author} {\bibinfo {author} {\bibfnamefont {G.}~\bibnamefont
  {Ghirardi}}, \bibinfo {author} {\bibfnamefont {A.}~\bibnamefont {Rimini}}, \
  and\ \bibinfo {author} {\bibfnamefont {T.}~\bibnamefont {Weber}},\ }\href
  {\doibase 10.1103/physrevd.34.470} {\bibfield  {journal} {\bibinfo  {journal}
  {Phys. Rev. D}\ } (\bibinfo {year} {1986}),\
  10.1103/physrevd.34.470}\BibitemShut {NoStop}%
\bibitem [{\citenamefont {Diósi}(1987)}]{Diosi_87_PLA}%
  \BibitemOpen
  \bibfield  {author} {\bibinfo {author} {\bibfnamefont {L.}~\bibnamefont
  {Diósi}},\ }\href {\doibase 10.1016/0375-9601(87)90681-5} {\bibfield
  {journal} {\bibinfo  {journal} {Physics Lett. A}\ } (\bibinfo {year}
  {1987}),\ 10.1016/0375-9601(87)90681-5}\BibitemShut {NoStop}%
\bibitem [{\citenamefont {Ghirardi}\ \emph {et~al.}(1990)\citenamefont
  {Ghirardi}, \citenamefont {Pearle},\ and\ \citenamefont
  {Rimini}}]{Ghirardi_90_PRA}%
  \BibitemOpen
  \bibfield  {author} {\bibinfo {author} {\bibfnamefont {G.}~\bibnamefont
  {Ghirardi}}, \bibinfo {author} {\bibfnamefont {P.}~\bibnamefont {Pearle}}, \
  and\ \bibinfo {author} {\bibfnamefont {A.}~\bibnamefont {Rimini}},\ }\href
  {\doibase 10.1103/physreva.42.78} {\bibfield  {journal} {\bibinfo  {journal}
  {Phys. Rev. A}\ } (\bibinfo {year} {1990}),\
  10.1103/physreva.42.78}\BibitemShut {NoStop}%
\bibitem [{\citenamefont {Percival}(1995)}]{percival95}%
  \BibitemOpen
  \bibfield  {author} {\bibinfo {author} {\bibfnamefont {I.~C.}\ \bibnamefont
  {Percival}},\ }\href {\doibase 10.1098/rspa.1995.0139} {\bibfield  {journal}
  {\bibinfo  {journal} {Proceedings of the Royal Society of London. Series A:
  Mathematical and Physical Sciences}\ }\textbf {\bibinfo {volume} {451}},\
  \bibinfo {pages} {503} (\bibinfo {year} {1995})}\BibitemShut {NoStop}%
\bibitem [{\citenamefont {Penrose}(1996)}]{Penrose_96}%
  \BibitemOpen
  \bibfield  {author} {\bibinfo {author} {\bibfnamefont {R.}~\bibnamefont
  {Penrose}},\ }\href {\doibase 10.1007/bf02105068} {\bibfield  {journal}
  {\bibinfo  {journal} {General Relativity and Gravitation}\ } (\bibinfo {year}
  {1996}),\ 10.1007/bf02105068}\BibitemShut {NoStop}%
\bibitem [{\citenamefont {Di\'osi}(1989)}]{diosi_1989}%
  \BibitemOpen
  \bibfield  {author} {\bibinfo {author} {\bibfnamefont {L.}~\bibnamefont
  {Di\'osi}},\ }\href {\doibase 10.1103/physreva.40.1165} {\bibfield  {journal}
  {\bibinfo  {journal} {Phys. Rev. A}\ }\textbf {\bibinfo {volume} {40}},\
  \bibinfo {pages} {1165–1174} (\bibinfo {year} {1989})}\BibitemShut
  {NoStop}%
\bibitem [{\citenamefont {Marshall}\ \emph {et~al.}(2003)\citenamefont
  {Marshall}, \citenamefont {Simon}, \citenamefont {Penrose},\ and\
  \citenamefont {Bouwmeester}}]{Bouwmeester_2003}%
  \BibitemOpen
  \bibfield  {author} {\bibinfo {author} {\bibfnamefont {W.}~\bibnamefont
  {Marshall}}, \bibinfo {author} {\bibfnamefont {C.}~\bibnamefont {Simon}},
  \bibinfo {author} {\bibfnamefont {R.}~\bibnamefont {Penrose}}, \ and\
  \bibinfo {author} {\bibfnamefont {D.}~\bibnamefont {Bouwmeester}},\ }\href
  {\doibase 10.1103/PhysRevLett.91.130401} {\bibfield  {journal} {\bibinfo
  {journal} {Phys. Rev. Lett.}\ }\textbf {\bibinfo {volume} {91}},\ \bibinfo
  {pages} {130401} (\bibinfo {year} {2003})}\BibitemShut {NoStop}%
\bibitem [{\citenamefont {Donadi}\ \emph {et~al.}(2020)\citenamefont {Donadi},
  \citenamefont {Piscicchia}, \citenamefont {Curceanu}, \citenamefont
  {Di\'{o}si}, \citenamefont {Laubenstein},\ and\ \citenamefont
  {Bassi}}]{underground}%
  \BibitemOpen
  \bibfield  {author} {\bibinfo {author} {\bibfnamefont {S.}~\bibnamefont
  {Donadi}}, \bibinfo {author} {\bibfnamefont {K.}~\bibnamefont {Piscicchia}},
  \bibinfo {author} {\bibfnamefont {C.}~\bibnamefont {Curceanu}}, \bibinfo
  {author} {\bibfnamefont {L.}~\bibnamefont {Di\'{o}si}}, \bibinfo {author}
  {\bibfnamefont {M.}~\bibnamefont {Laubenstein}}, \ and\ \bibinfo {author}
  {\bibfnamefont {A.}~\bibnamefont {Bassi}},\ }\href {\doibase
  10.1038/s41567-020-1008-4} {\bibfield  {journal} {\bibinfo  {journal} {Nature
  Physics}\ } (\bibinfo {year} {2020}),\ 10.1038/s41567-020-1008-4}\BibitemShut
  {NoStop}%
\bibitem [{\citenamefont {Vinante}\ \emph {et~al.}(2017)\citenamefont
  {Vinante}, \citenamefont {Mezzena}, \citenamefont {Falferi}, \citenamefont
  {Carlesso},\ and\ \citenamefont
  {Bassi}}]{vinante_mezzena_falferi_carlesso_bassi_2017}%
  \BibitemOpen
  \bibfield  {author} {\bibinfo {author} {\bibfnamefont {A.}~\bibnamefont
  {Vinante}}, \bibinfo {author} {\bibfnamefont {R.}~\bibnamefont {Mezzena}},
  \bibinfo {author} {\bibfnamefont {P.}~\bibnamefont {Falferi}}, \bibinfo
  {author} {\bibfnamefont {M.}~\bibnamefont {Carlesso}}, \ and\ \bibinfo
  {author} {\bibfnamefont {A.}~\bibnamefont {Bassi}},\ }\href {\doibase
  10.1103/physrevlett.119.110401} {\bibfield  {journal} {\bibinfo  {journal}
  {Phys. Rev. Lett.}\ }\textbf {\bibinfo {volume} {119}} (\bibinfo {year}
  {2017}),\ 10.1103/physrevlett.119.110401}\BibitemShut {NoStop}%
\bibitem [{\citenamefont {Carlesso}\ \emph {et~al.}(2016)\citenamefont
  {Carlesso}, \citenamefont {Bassi}, \citenamefont {Falferi},\ and\
  \citenamefont {Vinante}}]{carlesso_bassi_falferi_vinante_2016}%
  \BibitemOpen
  \bibfield  {author} {\bibinfo {author} {\bibfnamefont {M.}~\bibnamefont
  {Carlesso}}, \bibinfo {author} {\bibfnamefont {A.}~\bibnamefont {Bassi}},
  \bibinfo {author} {\bibfnamefont {P.}~\bibnamefont {Falferi}}, \ and\
  \bibinfo {author} {\bibfnamefont {A.}~\bibnamefont {Vinante}},\ }\href
  {\doibase 10.1103/physrevd.94.124036} {\bibfield  {journal} {\bibinfo
  {journal} {Phys. Rev. D}\ }\textbf {\bibinfo {volume} {94}} (\bibinfo {year}
  {2016}),\ 10.1103/physrevd.94.124036}\BibitemShut {NoStop}%
\bibitem [{\citenamefont {Adler}(2007)}]{Adler_2007_experiment}%
  \BibitemOpen
  \bibfield  {author} {\bibinfo {author} {\bibfnamefont {S.~L.}\ \bibnamefont
  {Adler}},\ }\href {\doibase 10.1088/1751-8113/40/12/S03} {\bibfield
  {journal} {\bibinfo  {journal} {Journal of Physics A: Mathematical and
  Theoretical}\ }\textbf {\bibinfo {volume} {40}},\ \bibinfo {pages} {2935}
  (\bibinfo {year} {2007})}\BibitemShut {NoStop}%
\bibitem [{\citenamefont {Mertens}\ \emph {et~al.}(2021)\citenamefont
  {Mertens}, \citenamefont {Wesseling}, \citenamefont {Vercauteren},
  \citenamefont {Corrales-Salazar},\ and\ \citenamefont {van
  Wezel}}]{Mertens_PRA_21}%
  \BibitemOpen
  \bibfield  {author} {\bibinfo {author} {\bibfnamefont {L.}~\bibnamefont
  {Mertens}}, \bibinfo {author} {\bibfnamefont {M.}~\bibnamefont {Wesseling}},
  \bibinfo {author} {\bibfnamefont {N.}~\bibnamefont {Vercauteren}}, \bibinfo
  {author} {\bibfnamefont {A.}~\bibnamefont {Corrales-Salazar}}, \ and\
  \bibinfo {author} {\bibfnamefont {J.}~\bibnamefont {van Wezel}},\ }\href
  {\doibase 10.1103/PhysRevA.104.052224} {\bibfield  {journal} {\bibinfo
  {journal} {Phys. Rev. A}\ }\textbf {\bibinfo {volume} {104}},\ \bibinfo
  {pages} {052224} (\bibinfo {year} {2021})}\BibitemShut {NoStop}%
\bibitem [{\citenamefont {Mertens}\ \emph {et~al.}(2023)\citenamefont
  {Mertens}, \citenamefont {Wesseling},\ and\ \citenamefont {van
  Wezel}}]{Mertens22}%
  \BibitemOpen
  \bibfield  {author} {\bibinfo {author} {\bibfnamefont {L.}~\bibnamefont
  {Mertens}}, \bibinfo {author} {\bibfnamefont {M.}~\bibnamefont {Wesseling}},
  \ and\ \bibinfo {author} {\bibfnamefont {J.}~\bibnamefont {van Wezel}},\
  }\href {\doibase 10.21468/SciPostPhys.14.5.114} {\bibfield  {journal}
  {\bibinfo  {journal} {SciPost Phys.}\ }\textbf {\bibinfo {volume} {14}},\
  \bibinfo {pages} {114} (\bibinfo {year} {2023})}\BibitemShut {NoStop}%
\bibitem [{\citenamefont {van Wezel}(2022)}]{WezelBerry}%
  \BibitemOpen
  \bibfield  {author} {\bibinfo {author} {\bibfnamefont {J.}~\bibnamefont {van
  Wezel}},\ }\href {\doibase 10.1088/1751-8121/ac9163} {\bibfield  {journal}
  {\bibinfo  {journal} {J. Phys. A: Math. Theor.}\ }\textbf {\bibinfo {volume}
  {55}},\ \bibinfo {pages} {401001} (\bibinfo {year} {2022})}\BibitemShut
  {NoStop}%
\bibitem [{Note1()}]{Note1}%
  \BibitemOpen
  \bibinfo {note} {Notice the difference between the time-inversion symmetry
  discussed here, and the time-reversal symmetry that is spontaneously broken
  in for example a magnet. The magnetized equilibrium configuration is static
  and evolves the same way under time evolution forwards and backwards in time.
  That is, its dynamics still has time inversion symmetry.}\BibitemShut {Stop}%
\bibitem [{\citenamefont {van Wezel}(2008{\natexlab{a}})}]{Wezel_2008}%
  \BibitemOpen
  \bibfield  {author} {\bibinfo {author} {\bibfnamefont {J.}~\bibnamefont {van
  Wezel}},\ }\href {\doibase 10.1103/physrevb.78.054301} {\bibfield  {journal}
  {\bibinfo  {journal} {Phys. Rev. B}\ }\textbf {\bibinfo {volume} {78}},\
  \bibinfo {pages} {054301} (\bibinfo {year} {2008}{\natexlab{a}})}\BibitemShut
  {NoStop}%
\bibitem [{\citenamefont {Beekman}\ \emph {et~al.}(2019)\citenamefont
  {Beekman}, \citenamefont {Rademaker},\ and\ \citenamefont {van
  Wezel}}]{Wezel_SSBlecturenotes}%
  \BibitemOpen
  \bibfield  {author} {\bibinfo {author} {\bibfnamefont {A.~J.}\ \bibnamefont
  {Beekman}}, \bibinfo {author} {\bibfnamefont {L.}~\bibnamefont {Rademaker}},
  \ and\ \bibinfo {author} {\bibfnamefont {J.}~\bibnamefont {van Wezel}},\
  }\href {\doibase 10.21468/SciPostPhysLectNotes.11} {\bibfield  {journal}
  {\bibinfo  {journal} {SciPost Phys. Lect. Notes}\ }\textbf {\bibinfo {volume}
  {11}} (\bibinfo {year} {2019}),\
  10.21468/SciPostPhysLectNotes.11}\BibitemShut {NoStop}%
\bibitem [{\citenamefont {Zurek}(1981)}]{Zurek_1981}%
  \BibitemOpen
  \bibfield  {author} {\bibinfo {author} {\bibfnamefont {W.~H.}\ \bibnamefont
  {Zurek}},\ }\href {\doibase 10.1103/physrevd.24.1516} {\bibfield  {journal}
  {\bibinfo  {journal} {Phys. Rev. D}\ }\textbf {\bibinfo {volume} {24}},\
  \bibinfo {pages} {1516–1525} (\bibinfo {year} {1981})}\BibitemShut
  {NoStop}%
\bibitem [{\citenamefont {Bassi}\ \emph
  {et~al.}(2013{\natexlab{b}})\citenamefont {Bassi}, \citenamefont {Lochan},
  \citenamefont {Satin}, \citenamefont {Singh},\ and\ \citenamefont
  {Ulbricht}}]{Bassi2013Review}%
  \BibitemOpen
  \bibfield  {author} {\bibinfo {author} {\bibfnamefont {A.}~\bibnamefont
  {Bassi}}, \bibinfo {author} {\bibfnamefont {K.}~\bibnamefont {Lochan}},
  \bibinfo {author} {\bibfnamefont {S.}~\bibnamefont {Satin}}, \bibinfo
  {author} {\bibfnamefont {T.~P.}\ \bibnamefont {Singh}}, \ and\ \bibinfo
  {author} {\bibfnamefont {H.}~\bibnamefont {Ulbricht}},\ }\href {\doibase
  10.1103/RevModPhys.85.471} {\bibfield  {journal} {\bibinfo  {journal} {Rev.
  Mod. Phys.}\ }\textbf {\bibinfo {volume} {85}},\ \bibinfo {pages} {471}
  (\bibinfo {year} {2013}{\natexlab{b}})}\BibitemShut {NoStop}%
\bibitem [{\citenamefont {Adler}\ and\ \citenamefont
  {Bassi}(2007)}]{Adler2007_col0}%
  \BibitemOpen
  \bibfield  {author} {\bibinfo {author} {\bibfnamefont {S.~L.}\ \bibnamefont
  {Adler}}\ and\ \bibinfo {author} {\bibfnamefont {A.}~\bibnamefont {Bassi}},\
  }\href {\doibase 10.1088/1751-8113/40/50/012} {\bibfield  {journal} {\bibinfo
   {journal} {Journal of Physics A: Mathematical and Theoretical}\ }\textbf
  {\bibinfo {volume} {40}},\ \bibinfo {pages} {15083} (\bibinfo {year}
  {2007})}\BibitemShut {NoStop}%
\bibitem [{\citenamefont {Adler}\ and\ \citenamefont
  {Bassi}(2008)}]{Adler_2008_col1}%
  \BibitemOpen
  \bibfield  {author} {\bibinfo {author} {\bibfnamefont {S.~L.}\ \bibnamefont
  {Adler}}\ and\ \bibinfo {author} {\bibfnamefont {A.}~\bibnamefont {Bassi}},\
  }\href {\doibase 10.1088/1751-8113/41/39/395308} {\bibfield  {journal}
  {\bibinfo  {journal} {Journal of Physics A: Mathematical and Theoretical}\
  }\textbf {\bibinfo {volume} {41}},\ \bibinfo {pages} {395308} (\bibinfo
  {year} {2008})}\BibitemShut {NoStop}%
\bibitem [{\citenamefont {Bassi}\ and\ \citenamefont
  {Ghirardi}(2002)}]{Bassi_2002_col4}%
  \BibitemOpen
  \bibfield  {author} {\bibinfo {author} {\bibfnamefont {A.}~\bibnamefont
  {Bassi}}\ and\ \bibinfo {author} {\bibfnamefont {G.}~\bibnamefont
  {Ghirardi}},\ }\href {\doibase 10.1103/physreva.65.042114} {\bibfield
  {journal} {\bibinfo  {journal} {Physical Review A}\ } (\bibinfo {year}
  {2002}),\ 10.1103/physreva.65.042114}\BibitemShut {NoStop}%
\bibitem [{\citenamefont {Pearle}(1996)}]{Pearle_1996_col6}%
  \BibitemOpen
  \bibfield  {author} {\bibinfo {author} {\bibfnamefont {P.}~\bibnamefont
  {Pearle}},\ }\enquote {\bibinfo {title} {Wavefunction collapse models with
  nonwhite noise},}\ in\ \href {\doibase 10.1007/978-94-015-8656-6_8} {\emph
  {\bibinfo {booktitle} {Perspectives on Quantum Reality: Non-Relativistic,
  Relativistic, and Field-Theoretic}}}\ (\bibinfo  {publisher} {Springer
  Netherlands},\ \bibinfo {address} {Dordrecht},\ \bibinfo {year} {1996})\ pp.\
  \bibinfo {pages} {93--109}\BibitemShut {NoStop}%
\bibitem [{\citenamefont {Ferialdi}\ and\ \citenamefont
  {Bassi}(2012)}]{Ferialdi_2012_col7}%
  \BibitemOpen
  \bibfield  {author} {\bibinfo {author} {\bibfnamefont {L.}~\bibnamefont
  {Ferialdi}}\ and\ \bibinfo {author} {\bibfnamefont {A.}~\bibnamefont
  {Bassi}},\ }\href {\doibase 10.1103/physreva.86.022108} {\bibfield  {journal}
  {\bibinfo  {journal} {Physical Review A}\ } (\bibinfo {year} {2012}),\
  10.1103/physreva.86.022108}\BibitemShut {NoStop}%
\bibitem [{\citenamefont {von Neumann}(2018)}]{Von_Neumann2018-bo}%
  \BibitemOpen
  \bibfield  {author} {\bibinfo {author} {\bibfnamefont {J.}~\bibnamefont {von
  Neumann}},\ }in\ \href@noop {} {\emph {\bibinfo {booktitle} {Mathematical
  Foundations of Quantum Mechanics}}},\ \bibinfo {editor} {edited by\ \bibinfo
  {editor} {\bibfnamefont {N.~A.}\ \bibnamefont {Wheeler}}}\ (\bibinfo
  {publisher} {Princeton University Press},\ \bibinfo {year}
  {2018})\BibitemShut {NoStop}%
\bibitem [{\citenamefont {Di\'{o}si}(1987)}]{diosi_1987}%
  \BibitemOpen
  \bibfield  {author} {\bibinfo {author} {\bibfnamefont {L.}~\bibnamefont
  {Di\'{o}si}},\ }\href {\doibase 10.1016/0375-9601(87)90681-5} {\bibfield
  {journal} {\bibinfo  {journal} {Physics Lett. A}\ }\textbf {\bibinfo {volume}
  {120}},\ \bibinfo {pages} {377–381} (\bibinfo {year} {1987})}\BibitemShut
  {NoStop}%
\bibitem [{\citenamefont {Penrose}(2014)}]{penrose14}%
  \BibitemOpen
  \bibfield  {author} {\bibinfo {author} {\bibfnamefont {R.}~\bibnamefont
  {Penrose}},\ }\href {\doibase 10.1007/s10701-013-9770-0} {\bibfield
  {journal} {\bibinfo  {journal} {Found. Phys.}\ }\textbf {\bibinfo {volume}
  {44}},\ \bibinfo {pages} {557} (\bibinfo {year} {2014})}\BibitemShut
  {NoStop}%
\bibitem [{\citenamefont {Karolyhazy}(1966)}]{Karolyhazy_66}%
  \BibitemOpen
  \bibfield  {author} {\bibinfo {author} {\bibfnamefont {F.}~\bibnamefont
  {Karolyhazy}},\ }\href {\doibase 10.1007/bf02717926} {\bibfield  {journal}
  {\bibinfo  {journal} {Nuovo Cimento Della Societa Italiana Di Fisica A-nuclei
  Particles and Fields}\ } (\bibinfo {year} {1966}),\
  10.1007/bf02717926}\BibitemShut {NoStop}%
\bibitem [{\citenamefont {Christian}(2005)}]{christian_2005}%
  \BibitemOpen
  \bibfield  {author} {\bibinfo {author} {\bibfnamefont {J.}~\bibnamefont
  {Christian}},\ }\href {\doibase 10.1103/physrevlett.95.160403} {\bibfield
  {journal} {\bibinfo  {journal} {Phys. Rev. Lett.}\ }\textbf {\bibinfo
  {volume} {95}} (\bibinfo {year} {2005}),\
  10.1103/physrevlett.95.160403}\BibitemShut {NoStop}%
\bibitem [{\citenamefont {Adler}(2014)}]{Adler_2014_grav0}%
  \BibitemOpen
  \bibfield  {author} {\bibinfo {author} {\bibfnamefont {S.~L.}\ \bibnamefont
  {Adler}},\ }\href {\doibase null} {\bibfield  {journal} {\bibinfo  {journal}
  {arXiv: General Relativity and Quantum Cosmology}\ } (\bibinfo {year}
  {2014}),\ null}\BibitemShut {NoStop}%
\bibitem [{\citenamefont {Singh}(2015)}]{Singh_2015_grav1}%
  \BibitemOpen
  \bibfield  {author} {\bibinfo {author} {\bibfnamefont {T.~P.}\ \bibnamefont
  {Singh}},\ }\href {\doibase null} {\bibfield  {journal} {\bibinfo  {journal}
  {arXiv: Quantum Physics}\ } (\bibinfo {year} {2015}),\ null}\BibitemShut
  {NoStop}%
\bibitem [{\citenamefont {Diósi}\ and\ \citenamefont
  {Papp}(2009)}]{Diósi_2009_grav2}%
  \BibitemOpen
  \bibfield  {author} {\bibinfo {author} {\bibfnamefont {L.}~\bibnamefont
  {Diósi}}\ and\ \bibinfo {author} {\bibfnamefont {T.~N.}\ \bibnamefont
  {Papp}},\ }\href {\doibase 10.1016/j.physleta.2009.07.020} {\bibfield
  {journal} {\bibinfo  {journal} {Physics Letters A}\ } (\bibinfo {year}
  {2009}),\ 10.1016/j.physleta.2009.07.020}\BibitemShut {NoStop}%
\bibitem [{\citenamefont {Gasbarri}\ \emph {et~al.}(2017)\citenamefont
  {Gasbarri}, \citenamefont {Toroš}, \citenamefont {Donadi},\ and\
  \citenamefont {Bassi}}]{Gasbarri_2017_grav3}%
  \BibitemOpen
  \bibfield  {author} {\bibinfo {author} {\bibfnamefont {G.}~\bibnamefont
  {Gasbarri}}, \bibinfo {author} {\bibfnamefont {M.}~\bibnamefont {Toroš}},
  \bibinfo {author} {\bibfnamefont {S.}~\bibnamefont {Donadi}}, \ and\ \bibinfo
  {author} {\bibfnamefont {A.}~\bibnamefont {Bassi}},\ }\href {\doibase
  10.1103/physrevd.96.104013} {\bibfield  {journal} {\bibinfo  {journal}
  {Physical Review D}\ } (\bibinfo {year} {2017}),\
  10.1103/physrevd.96.104013}\BibitemShut {NoStop}%
\bibitem [{\citenamefont {Lindblad}(1976)}]{Lindblad1976}%
  \BibitemOpen
  \bibfield  {author} {\bibinfo {author} {\bibfnamefont {G.}~\bibnamefont
  {Lindblad}},\ }\href {\doibase 10.1007/BF01608499} {\bibfield  {journal}
  {\bibinfo  {journal} {Commun. Math. Phys.}\ }\textbf {\bibinfo {volume}
  {48}},\ \bibinfo {pages} {119} (\bibinfo {year} {1976})}\BibitemShut
  {NoStop}%
\bibitem [{\citenamefont {Gorini}\ \emph {et~al.}(1976)\citenamefont {Gorini},
  \citenamefont {Kossakowski},\ and\ \citenamefont {Sudarshan}}]{GKS76}%
  \BibitemOpen
  \bibfield  {author} {\bibinfo {author} {\bibfnamefont {V.}~\bibnamefont
  {Gorini}}, \bibinfo {author} {\bibfnamefont {A.}~\bibnamefont {Kossakowski}},
  \ and\ \bibinfo {author} {\bibfnamefont {E.~C.~G.}\ \bibnamefont
  {Sudarshan}},\ }\href {\doibase 10.1063/1.522979} {\bibfield  {journal}
  {\bibinfo  {journal} {J. Math. Phys.}\ }\textbf {\bibinfo {volume} {17}},\
  \bibinfo {pages} {821} (\bibinfo {year} {1976})}\BibitemShut {NoStop}%
\bibitem [{\citenamefont {Hänggi}\ and\ \citenamefont
  {Jung}(1994)}]{Hanggi94}%
  \BibitemOpen
  \bibfield  {author} {\bibinfo {author} {\bibfnamefont {P.}~\bibnamefont
  {Hänggi}}\ and\ \bibinfo {author} {\bibfnamefont {P.}~\bibnamefont {Jung}},\
  }\bibfield  {booktitle} {\emph {\bibinfo {booktitle} {Advances in Chemical
  Physics}},\ }\href {\doibase 10.1002/9780470141489.ch4} {\ ,\ \bibinfo
  {pages} {239} (\bibinfo {year} {1994})}\BibitemShut {NoStop}%
\bibitem [{\citenamefont {{Ł}uczka}(2005)}]{Luczka2005}%
  \BibitemOpen
  \bibfield  {author} {\bibinfo {author} {\bibfnamefont {J.}~\bibnamefont
  {{Ł}uczka}},\ }\href {\doibase 10.1063/1.1860471} {\bibfield  {journal}
  {\bibinfo  {journal} {Chaos}\ }\textbf {\bibinfo {volume} {15}},\ \bibinfo
  {pages} {026107} (\bibinfo {year} {2005})}\BibitemShut {NoStop}%
\bibitem [{\citenamefont {Revuz}\ and\ \citenamefont
  {Yor}(1999)}]{revuz1999continuous}%
  \BibitemOpen
  \bibfield  {author} {\bibinfo {author} {\bibfnamefont {D.}~\bibnamefont
  {Revuz}}\ and\ \bibinfo {author} {\bibfnamefont {M.}~\bibnamefont {Yor}},\
  }\href@noop {} {\emph {\bibinfo {title} {Continuous martingales and Brownian
  motion}}}\ (\bibinfo  {publisher} {Springer},\ \bibinfo {year}
  {1999})\BibitemShut {NoStop}%
\bibitem [{\citenamefont {Øksendal}(2003)}]{oksendal2003stochastic}%
  \BibitemOpen
  \bibfield  {author} {\bibinfo {author} {\bibfnamefont {B.}~\bibnamefont
  {Øksendal}},\ }\href@noop {} {\emph {\bibinfo {title} {Stochastic
  differential equations: An introduction with applications}}}\ (\bibinfo
  {publisher} {Springer},\ \bibinfo {year} {2003})\BibitemShut {NoStop}%
\bibitem [{\citenamefont {Gardiner}(2004)}]{gardiner2004handbook}%
  \BibitemOpen
  \bibfield  {author} {\bibinfo {author} {\bibfnamefont {C.~W.}\ \bibnamefont
  {Gardiner}},\ }\href@noop {} {\emph {\bibinfo {title} {Handbook of stochastic
  methods for physics, chemistry and the natural sciences}}},\ \bibinfo
  {edition} {3rd}\ ed.,\ \bibinfo {series} {Springer Series in Synergetics},
  Vol.~\bibinfo {volume} {13}\ (\bibinfo  {publisher} {Springer-Verlag},\
  \bibinfo {address} {Berlin},\ \bibinfo {year} {2004})\BibitemShut {NoStop}%
\bibitem [{\citenamefont {Uhlenbeck}\ and\ \citenamefont
  {Ornstein}(1930)}]{OU_OG1930}%
  \BibitemOpen
  \bibfield  {author} {\bibinfo {author} {\bibfnamefont {G.~E.}\ \bibnamefont
  {Uhlenbeck}}\ and\ \bibinfo {author} {\bibfnamefont {L.~S.}\ \bibnamefont
  {Ornstein}},\ }\href {\doibase 10.1103/PhysRev.36.823} {\bibfield  {journal}
  {\bibinfo  {journal} {Phys. Rev.}\ }\textbf {\bibinfo {volume} {36}},\
  \bibinfo {pages} {823} (\bibinfo {year} {1930})}\BibitemShut {NoStop}%
\bibitem [{\citenamefont {Doob}(1942)}]{Doob42}%
  \BibitemOpen
  \bibfield  {author} {\bibinfo {author} {\bibfnamefont {J.~L.}\ \bibnamefont
  {Doob}},\ }\href {http://www.jstor.org/stable/1968873} {\bibfield  {journal}
  {\bibinfo  {journal} {Annals of Mathematics}\ }\textbf {\bibinfo {volume}
  {43}},\ \bibinfo {pages} {351} (\bibinfo {year} {1942})}\BibitemShut
  {NoStop}%
\bibitem [{\citenamefont {Risken}(1996)}]{Risken1996}%
  \BibitemOpen
  \bibfield  {author} {\bibinfo {author} {\bibfnamefont {H.}~\bibnamefont
  {Risken}},\ }\href {\doibase 10.1007/978-3-642-61544-3} {\emph {\bibinfo
  {title} {The Fokker-Planck Equation}}}\ (\bibinfo  {publisher} {Springer
  Berlin Heidelberg},\ \bibinfo {year} {1996})\BibitemShut {NoStop}%
\bibitem [{\citenamefont {Pavliotis}\ and\ \citenamefont
  {Stuart}(2008)}]{Pavliotis2008}%
  \BibitemOpen
  \bibfield  {author} {\bibinfo {author} {\bibfnamefont {G.}~\bibnamefont
  {Pavliotis}}\ and\ \bibinfo {author} {\bibfnamefont {A.}~\bibnamefont
  {Stuart}},\ }\href {\doibase 10.1007/978-0-387-73829-1} {\emph {\bibinfo
  {title} {Multiscale Methods}}}\ (\bibinfo  {publisher} {Springer New York},\
  \bibinfo {year} {2008})\BibitemShut {NoStop}%
\bibitem [{\citenamefont {Zwanzig}(2001)}]{Zwanzig2001}%
  \BibitemOpen
  \bibfield  {author} {\bibinfo {author} {\bibfnamefont {R.~W.}\ \bibnamefont
  {Zwanzig}},\ }\href@noop {} {\emph {\bibinfo {title} {Nonequilibrium
  Statistical Mechanics}}}\ (\bibinfo  {publisher} {Oxford University Press},\
  \bibinfo {address} {New York, NY},\ \bibinfo {year} {2001})\BibitemShut
  {NoStop}%
\bibitem [{\citenamefont {Forman}\ and\ \citenamefont
  {Sørensen}(2008)}]{CNOSPearsonDiffPaper2008}%
  \BibitemOpen
  \bibfield  {author} {\bibinfo {author} {\bibfnamefont {J.}~\bibnamefont
  {Forman}}\ and\ \bibinfo {author} {\bibfnamefont {M.}~\bibnamefont
  {Sørensen}},\ }\href {\doibase
  https://doi.org/10.1111/j.1467-9469.2007.00592.x} {\bibfield  {journal}
  {\bibinfo  {journal} {Scandinavian Journal of Statistics}\ }\textbf {\bibinfo
  {volume} {35}},\ \bibinfo {pages} {438} (\bibinfo {year} {2008})}\BibitemShut
  {NoStop}%
\bibitem [{\citenamefont {Ascione}\ \emph {et~al.}(2021)\citenamefont
  {Ascione}, \citenamefont {Leonenko},\ and\ \citenamefont
  {Pirozzi}}]{CNOStimelocalPearsontoJacobiAscione2021}%
  \BibitemOpen
  \bibfield  {author} {\bibinfo {author} {\bibfnamefont {G.}~\bibnamefont
  {Ascione}}, \bibinfo {author} {\bibfnamefont {N.}~\bibnamefont {Leonenko}}, \
  and\ \bibinfo {author} {\bibfnamefont {E.}~\bibnamefont {Pirozzi}},\ }\href
  {https://doi.org/10.1007/s10955-021-02786-2} {\bibfield  {journal} {\bibinfo
  {journal} {Journal of Statistical Physics}\ }\textbf {\bibinfo {volume}
  {183}} (\bibinfo {year} {2021})}\BibitemShut {NoStop}%
\bibitem [{\citenamefont {Wong}(1964)}]{CNOSWong1964}%
  \BibitemOpen
  \bibfield  {author} {\bibinfo {author} {\bibfnamefont {E.}~\bibnamefont
  {Wong}},\ }in\ \href
  {https://people.eecs.berkeley.edu/~wong/wong_pubs/wong10.pdf} {\emph
  {\bibinfo {booktitle} {Proc. {S}ympos. {A}ppl. {M}ath., {V}ol. {XVI}}}}\
  (\bibinfo  {publisher} {Amer. Math. Soc., Providence, R.I.},\ \bibinfo {year}
  {1964})\ pp.\ \bibinfo {pages} {264--276}\BibitemShut {NoStop}%
\bibitem [{\citenamefont {Horsthemke}\ and\ \citenamefont
  {Lefever}(2006)}]{HorsthemkeBook2006}%
  \BibitemOpen
  \bibfield  {author} {\bibinfo {author} {\bibfnamefont {W.}~\bibnamefont
  {Horsthemke}}\ and\ \bibinfo {author} {\bibfnamefont {R.}~\bibnamefont
  {Lefever}},\ }\href {\doibase 10.1007/3-540-36852-3} {\emph {\bibinfo {title}
  {Noise-Induced Transitions: Theory and Applications in Physics}}},\ \bibinfo
  {series} {Springer Series in Synergetics}, Vol.~\bibinfo {volume} {15}\
  (\bibinfo  {publisher} {Springer-Verlag Berlin Heidelberg},\ \bibinfo {year}
  {2006})\BibitemShut {NoStop}%
\bibitem [{\citenamefont {Bonnin}\ \emph {et~al.}(2019)\citenamefont {Bonnin},
  \citenamefont {Traversa},\ and\ \citenamefont
  {Bonani}}]{BoninTraversaSmallTau}%
  \BibitemOpen
  \bibfield  {author} {\bibinfo {author} {\bibfnamefont {M.}~\bibnamefont
  {Bonnin}}, \bibinfo {author} {\bibfnamefont {F.~L.}\ \bibnamefont
  {Traversa}}, \ and\ \bibinfo {author} {\bibfnamefont {F.}~\bibnamefont
  {Bonani}},\ }\href {\doibase 10.1109/TCSI.2019.2914398} {\bibfield  {journal}
  {\bibinfo  {journal} {IEEE Transactions on Circuits and Systems I: Regular
  Papers}\ }\textbf {\bibinfo {volume} {66}},\ \bibinfo {pages} {3917}
  (\bibinfo {year} {2019})}\BibitemShut {NoStop}%
\bibitem [{\citenamefont {Wong}\ and\ \citenamefont
  {Zakai}(1965{\natexlab{a}})}]{WongZakai1965relation}%
  \BibitemOpen
  \bibfield  {author} {\bibinfo {author} {\bibfnamefont {E.}~\bibnamefont
  {Wong}}\ and\ \bibinfo {author} {\bibfnamefont {M.}~\bibnamefont {Zakai}},\
  }\href {\doibase https://doi.org/10.1016/0020-7225(65)90045-5} {\bibfield
  {journal} {\bibinfo  {journal} {International Journal of Engineering
  Science}\ }\textbf {\bibinfo {volume} {3}},\ \bibinfo {pages} {213} (\bibinfo
  {year} {1965}{\natexlab{a}})}\BibitemShut {NoStop}%
\bibitem [{\citenamefont {Wong}\ and\ \citenamefont
  {Zakai}(1965{\natexlab{b}})}]{wongZakai1965convergence}%
  \BibitemOpen
  \bibfield  {author} {\bibinfo {author} {\bibfnamefont {E.}~\bibnamefont
  {Wong}}\ and\ \bibinfo {author} {\bibfnamefont {M.}~\bibnamefont {Zakai}},\
  }\href@noop {} {\bibfield  {journal} {\bibinfo  {journal} {The Annals of
  Mathematical Statistics}\ }\textbf {\bibinfo {volume} {36}},\ \bibinfo
  {pages} {1560} (\bibinfo {year} {1965}{\natexlab{b}})}\BibitemShut {NoStop}%
\bibitem [{\citenamefont {Wong}\ and\ \citenamefont
  {Zakai}(1969)}]{WongZakai1969}%
  \BibitemOpen
  \bibfield  {author} {\bibinfo {author} {\bibfnamefont {E.}~\bibnamefont
  {Wong}}\ and\ \bibinfo {author} {\bibfnamefont {M.}~\bibnamefont {Zakai}},\
  }\href {\doibase 10.1007/bf00531642} {\bibfield  {journal} {\bibinfo
  {journal} {Zeitschrift f{\"u}r Wahrscheinlichkeitstheorie und Verwandte
  Gebiete}\ }\textbf {\bibinfo {volume} {12}},\ \bibinfo {pages} {87} (\bibinfo
  {year} {1969})}\BibitemShut {NoStop}%
\bibitem [{\citenamefont {Twardowska}(1996)}]{WongZakaiReview}%
  \BibitemOpen
  \bibfield  {author} {\bibinfo {author} {\bibfnamefont {K.}~\bibnamefont
  {Twardowska}},\ }\href {\doibase 10.1007/bf00047670} {\bibfield  {journal}
  {\bibinfo  {journal} {Acta Applicandae Mathematicae}\ }\textbf {\bibinfo
  {volume} {43}},\ \bibinfo {pages} {317} (\bibinfo {year} {1996})}\BibitemShut
  {NoStop}%
\bibitem [{\citenamefont {Pearle}(1984)}]{Pearle1984_bookRef}%
  \BibitemOpen
  \bibfield  {author} {\bibinfo {author} {\bibfnamefont {P.}~\bibnamefont
  {Pearle}},\ }\enquote {\bibinfo {title} {Dynamics of the reduction of the
  statevector},}\ in\ \href {\doibase 10.1007/978-94-009-6286-6_26} {\emph
  {\bibinfo {booktitle} {The Wave-Particle Dualism: A Tribute to Louis de
  Broglie on his 90th Birthday}}},\ \bibinfo {editor} {edited by\ \bibinfo
  {editor} {\bibfnamefont {S.}~\bibnamefont {Diner}}, \bibinfo {editor}
  {\bibfnamefont {D.}~\bibnamefont {Fargue}}, \bibinfo {editor} {\bibfnamefont
  {G.}~\bibnamefont {Lochak}}, \ and\ \bibinfo {editor} {\bibfnamefont
  {F.}~\bibnamefont {Selleri}}}\ (\bibinfo  {publisher} {Springer
  Netherlands},\ \bibinfo {address} {Dordrecht},\ \bibinfo {year} {1984})\ pp.\
  \bibinfo {pages} {457--483}\BibitemShut {NoStop}%
\bibitem [{\citenamefont {Einstein}(1905)}]{Einstein1905}%
  \BibitemOpen
  \bibfield  {author} {\bibinfo {author} {\bibfnamefont {A.}~\bibnamefont
  {Einstein}},\ }\href {\doibase https://doi.org/10.1002/andp.19053220806}
  {\bibfield  {journal} {\bibinfo  {journal} {Annalen der Physik}\ }\textbf
  {\bibinfo {volume} {322}},\ \bibinfo {pages} {549} (\bibinfo {year}
  {1905})}\BibitemShut {NoStop}%
\bibitem [{\citenamefont {von Smoluchowski}(1906)}]{Smoluchowski1906}%
  \BibitemOpen
  \bibfield  {author} {\bibinfo {author} {\bibfnamefont {M.}~\bibnamefont {von
  Smoluchowski}},\ }\href {\doibase https://doi.org/10.1002/andp.19063261405}
  {\bibfield  {journal} {\bibinfo  {journal} {Annalen der Physik}\ }\textbf
  {\bibinfo {volume} {326}},\ \bibinfo {pages} {756} (\bibinfo {year}
  {1906})}\BibitemShut {NoStop}%
\bibitem [{\citenamefont {Kubo}(1966)}]{Kubo_1966}%
  \BibitemOpen
  \bibfield  {author} {\bibinfo {author} {\bibfnamefont {R.}~\bibnamefont
  {Kubo}},\ }\href {\doibase 10.1088/0034-4885/29/1/306} {\bibfield  {journal}
  {\bibinfo  {journal} {Reports on Progress in Physics}\ }\textbf {\bibinfo
  {volume} {29}},\ \bibinfo {pages} {255} (\bibinfo {year} {1966})}\BibitemShut
  {NoStop}%
\bibitem [{\citenamefont {Gisin}(1989)}]{Gisin:1989sx}%
  \BibitemOpen
  \bibfield  {author} {\bibinfo {author} {\bibfnamefont {N.}~\bibnamefont
  {Gisin}},\ }\href@noop {} {\bibfield  {journal} {\bibinfo  {journal} {Helv.
  Phys. Acta}\ }\textbf {\bibinfo {volume} {62}},\ \bibinfo {pages} {363}
  (\bibinfo {year} {1989})}\BibitemShut {NoStop}%
\bibitem [{\citenamefont {Bassi}\ and\ \citenamefont
  {Hejazi}(2015)}]{Bassi2015}%
  \BibitemOpen
  \bibfield  {author} {\bibinfo {author} {\bibfnamefont {A.}~\bibnamefont
  {Bassi}}\ and\ \bibinfo {author} {\bibfnamefont {K.}~\bibnamefont {Hejazi}},\
  }\href {\doibase 10.1088/0143-0807/36/5/055027} {\bibfield  {journal}
  {\bibinfo  {journal} {European Journal of Physics}\ }\textbf {\bibinfo
  {volume} {36}},\ \bibinfo {pages} {055027} (\bibinfo {year}
  {2015})}\BibitemShut {NoStop}%
\bibitem [{\citenamefont {Breuer}\ and\ \citenamefont
  {Petruccione}(2002)}]{Breur_Petr02}%
  \BibitemOpen
  \bibfield  {author} {\bibinfo {author} {\bibfnamefont {H.~P.}\ \bibnamefont
  {Breuer}}\ and\ \bibinfo {author} {\bibfnamefont {F.}~\bibnamefont
  {Petruccione}},\ }\href@noop {} {\emph {\bibinfo {title} {The theory of open
  quantum systems}}}\ (\bibinfo  {publisher} {Oxford University Press},\
  \bibinfo {address} {Great Clarendon Street},\ \bibinfo {year}
  {2002})\BibitemShut {NoStop}%
\bibitem [{\citenamefont {van Wezel}\ and\ \citenamefont
  {Oosterkamp}(2012)}]{Wezel_2012}%
  \BibitemOpen
  \bibfield  {author} {\bibinfo {author} {\bibfnamefont {J.}~\bibnamefont {van
  Wezel}}\ and\ \bibinfo {author} {\bibfnamefont {T.~H.}\ \bibnamefont
  {Oosterkamp}},\ }\href {\doibase 10.1098/rspa.2011.0201} {\bibfield
  {journal} {\bibinfo  {journal} {Proceedings of The Royal Society A:
  Mathematical, Physical and Engineering Sciences}\ } (\bibinfo {year}
  {2012}),\ 10.1098/rspa.2011.0201}\BibitemShut {NoStop}%
\bibitem [{\citenamefont {Landi}\ and\ \citenamefont
  {Paternostro}(2021)}]{Entropy_production_article}%
  \BibitemOpen
  \bibfield  {author} {\bibinfo {author} {\bibfnamefont {G.~T.}\ \bibnamefont
  {Landi}}\ and\ \bibinfo {author} {\bibfnamefont {M.}~\bibnamefont
  {Paternostro}},\ }\href {\doibase 10.1103/RevModPhys.93.035008} {\bibfield
  {journal} {\bibinfo  {journal} {Rev. Mod. Phys.}\ }\textbf {\bibinfo {volume}
  {93}},\ \bibinfo {pages} {035008} (\bibinfo {year} {2021})}\BibitemShut
  {NoStop}%
\bibitem [{\citenamefont {Bishop}(2004)}]{BISHOP20041_AB0}%
  \BibitemOpen
  \bibfield  {author} {\bibinfo {author} {\bibfnamefont {R.~C.}\ \bibnamefont
  {Bishop}},\ }\href {\doibase https://doi.org/10.1016/j.shpsb.2001.11.001}
  {\bibfield  {journal} {\bibinfo  {journal} {Studies in History and Philosophy
  of Science Part B: Studies in History and Philosophy of Modern Physics}\
  }\textbf {\bibinfo {volume} {35}},\ \bibinfo {pages} {1} (\bibinfo {year}
  {2004})}\BibitemShut {NoStop}%
\bibitem [{\citenamefont {Prigogine}(1999)}]{PRIGOGINE1999_AB1}%
  \BibitemOpen
  \bibfield  {author} {\bibinfo {author} {\bibfnamefont {I.}~\bibnamefont
  {Prigogine}},\ }\href {\doibase
  https://doi.org/10.1016/S0378-4371(98)00527-5} {\bibfield  {journal}
  {\bibinfo  {journal} {Physica A: Statistical Mechanics and its Applications}\
  }\textbf {\bibinfo {volume} {263}},\ \bibinfo {pages} {528} (\bibinfo {year}
  {1999})},\ \bibinfo {note} {proceedings of the 20th IUPAP International
  Conference on Statistical Physics}\BibitemShut {NoStop}%
\bibitem [{\citenamefont {Morozov}\ and\ \citenamefont
  {Ropke}(1998)}]{Morozov1998_Zubarev_AB2}%
  \BibitemOpen
  \bibfield  {author} {\bibinfo {author} {\bibnamefont {Morozov}}\ and\
  \bibinfo {author} {\bibnamefont {Ropke}},\ }\href {\doibase
  10.5488/cmp.1.4.673} {\bibfield  {journal} {\bibinfo  {journal} {Condensed
  Matter Physics}\ }\textbf {\bibinfo {volume} {1}},\ \bibinfo {pages} {673}
  (\bibinfo {year} {1998})}\BibitemShut {NoStop}%
\bibitem [{\citenamefont {Jou}\ \emph {et~al.}(1996)\citenamefont {Jou},
  \citenamefont {Casas-Vázquez},\ and\ \citenamefont
  {Lebon}}]{Jou1996_zubarev_AB3}%
  \BibitemOpen
  \bibfield  {author} {\bibinfo {author} {\bibfnamefont {D.}~\bibnamefont
  {Jou}}, \bibinfo {author} {\bibfnamefont {J.}~\bibnamefont {Casas-Vázquez}},
  \ and\ \bibinfo {author} {\bibfnamefont {G.}~\bibnamefont {Lebon}},\ }\href
  {\doibase 10.1007/978-3-642-97671-1} {\emph {\bibinfo {title} {Extended
  Irreversible Thermodynamics}}}\ (\bibinfo  {publisher} {Springer Berlin
  Heidelberg},\ \bibinfo {year} {1996})\BibitemShut {NoStop}%
\bibitem [{\citenamefont {Luzzi}\ \emph {et~al.}(2006)\citenamefont {Luzzi},
  \citenamefont {Vasconcellos},\ and\ \citenamefont
  {Ramos}}]{LuzziVasconcellosRamos2006_AB4}%
  \BibitemOpen
  \bibfield  {author} {\bibinfo {author} {\bibfnamefont {R.}~\bibnamefont
  {Luzzi}}, \bibinfo {author} {\bibfnamefont {A.}~\bibnamefont {Vasconcellos}},
  \ and\ \bibinfo {author} {\bibfnamefont {J.}~\bibnamefont {Ramos}},\ }\href
  {\doibase 10.1393/ncr/i2006-10009-1} {\bibfield  {journal} {\bibinfo
  {journal} {La Rivista del Nuovo Cimento}\ }\textbf {\bibinfo {volume} {29}},\
  \bibinfo {pages} {1–82} (\bibinfo {year} {2006})}\BibitemShut {NoStop}%
\bibitem [{\citenamefont {Eu}(1998)}]{Eu1998_AB5}%
  \BibitemOpen
  \bibfield  {author} {\bibinfo {author} {\bibfnamefont {B.~C.}\ \bibnamefont
  {Eu}},\ }\href {\doibase 10.1007/978-94-017-2438-8} {\emph {\bibinfo {title}
  {Nonequilibrium Statistical Mechanics}}}\ (\bibinfo  {publisher} {Springer
  Netherlands},\ \bibinfo {year} {1998})\BibitemShut {NoStop}%
\bibitem [{\citenamefont {Demni}\ and\ \citenamefont
  {Zani}(2009)}]{CNOSJacobiDEMNI2009518}%
  \BibitemOpen
  \bibfield  {author} {\bibinfo {author} {\bibfnamefont {N.}~\bibnamefont
  {Demni}}\ and\ \bibinfo {author} {\bibfnamefont {M.}~\bibnamefont {Zani}},\
  }\href {\doibase https://doi.org/10.1016/j.spa.2008.02.015} {\bibfield
  {journal} {\bibinfo  {journal} {Stochastic Processes and their Applications}\
  }\textbf {\bibinfo {volume} {119}},\ \bibinfo {pages} {518} (\bibinfo {year}
  {2009})}\BibitemShut {NoStop}%
\bibitem [{\citenamefont {van Wezel}\ and\ \citenamefont {van~den
  Brink}(2007)}]{vanwezelAmJPhys}%
  \BibitemOpen
  \bibfield  {author} {\bibinfo {author} {\bibfnamefont {J.}~\bibnamefont {van
  Wezel}}\ and\ \bibinfo {author} {\bibfnamefont {J.}~\bibnamefont {van~den
  Brink}},\ }\href {\doibase 10.1119/1.2730839} {\bibfield  {journal} {\bibinfo
   {journal} {Am. J. Phys.}\ }\textbf {\bibinfo {volume} {75}},\ \bibinfo
  {pages} {635} (\bibinfo {year} {2007})}\BibitemShut {NoStop}%
\bibitem [{\citenamefont {van Wezel}(2008{\natexlab{b}})}]{vanwezelprb}%
  \BibitemOpen
  \bibfield  {author} {\bibinfo {author} {\bibfnamefont {J.}~\bibnamefont {van
  Wezel}},\ }\href {\doibase 10.1103/PhysRevB.78.054301} {\bibfield  {journal}
  {\bibinfo  {journal} {Phys. Rev. B}\ }\textbf {\bibinfo {volume} {78}},\
  \bibinfo {pages} {054301} (\bibinfo {year} {2008}{\natexlab{b}})}\BibitemShut
  {NoStop}%
\bibitem [{\citenamefont {Mertens}(2020)}]{Lotte}%
  \BibitemOpen
  \bibfield  {author} {\bibinfo {author} {\bibfnamefont {L.}~\bibnamefont
  {Mertens}},\ }\emph {\bibinfo {title} {Spontaneous Unitary Violations and
  Effective Non-linearity in Relation to Quantum State Reduction}},\ \href@noop
  {} {Master's thesis},\ \bibinfo  {school} {University of Amsterdam} (\bibinfo
  {year} {2020})\BibitemShut {NoStop}%
\bibitem [{\citenamefont {Tamir}\ and\ \citenamefont
  {Cohen}(2013)}]{Tamir_2013}%
  \BibitemOpen
  \bibfield  {author} {\bibinfo {author} {\bibfnamefont {B.}~\bibnamefont
  {Tamir}}\ and\ \bibinfo {author} {\bibfnamefont {E.}~\bibnamefont {Cohen}},\
  }\href {\doibase 10.12743/quanta.v2i1.14} {\bibfield  {journal} {\bibinfo
  {journal} {Quanta}\ } (\bibinfo {year} {2013}),\
  10.12743/quanta.v2i1.14}\BibitemShut {NoStop}%
\bibitem [{\citenamefont {Svensson}(2012)}]{Svensson_2012}%
  \BibitemOpen
  \bibfield  {author} {\bibinfo {author} {\bibfnamefont {B.~E.~Y.}\
  \bibnamefont {Svensson}},\ }\href {\doibase 10.12743/quanta.v2i1.12}
  {\bibfield  {journal} {\bibinfo  {journal} {arXiv: Quantum Physics}\ }
  (\bibinfo {year} {2012}),\ 10.12743/quanta.v2i1.12}\BibitemShut {NoStop}%
\end{thebibliography}%


\end{document}